\documentclass[sn-mathphys]{sn-jnl}

\usepackage{caption}
\usepackage{subcaption}
\usepackage{siunitx}
\usepackage{url}

\jyear{2021}%

\theoremstyle{thmstyleone}%
\newtheorem{theorem}{Theorem}
\theoremstyle{thmstyletwo}%

\theoremstyle{thmstylethree}%

\begin{document}

	\title[An age-structured epidemic model with vaccination]{An age-structured epidemic model with vaccination}


	\author[1]{\fnm{Ruiyang} \sur{Zhou}}\email{ruiyangzhou@outlook.com}

	\author*[1,2]{\fnm{Fengying} \sur{Wei}}\email{weifengying@fzu.edu.cn}

	\affil[1]{\orgdiv{College of Mathematics and Computer Science}, \orgname{Fuzhou University}, \orgaddress{\street{Wulong River North Street 2}, \city{Fuzhou}, \postcode{350116}, \state{Fujian Province}, \country{PR China}}}

	\affil[2]{\orgdiv{Key Laboratory of Operations Research and Control of Universities in Fujian}, \orgname{Fuzhou University}, \orgaddress{\street{Wulong River North Street 2}, \city{Fuzhou}, \postcode{350116}, \state{Fujian Province}, \country{PR China}}}


	\abstract{In this article, we construct an age-structured model for COVID-19 with vaccination and analyze it from multiple perspectives. We derive the unique disease-free equilibrium point and the basic reproduction number $ \mathscr{R}_0 $, then we show that the disease-free equilibrium is locally asymptotically stable when $    \mathscr{R}_0 < 1 $, while is unstable when $ \mathscr{R}_0 > 1 $. We also work out endemic equilibrium points and reveal the stability. We use sensitivity analysis to explore how parameters influence $ \mathscr{R}_0 $. Sensitivity analysis helps us develop more targeted strategies to control epidemics. Finally, this model is used to discuss the cases in Shijiazhuang, Hebei Province at the beginning of 2021. We compare reported cases with the simulation to evaluate the measures taken by Shijiazhuang government. Our study shows how age structure, vaccination and drastic containment measures can affect the epidemic.
	}




	\maketitle

	\section{Introduction}\label{sec:introduction}


	COVID-19 is an infectious disease caused by the SARS-CoV-2 virus.
	The main clinical characteristics have been figured out by \cite{Characteristics01}.
	Many people infected with the virus will
	experience mild to moderate respiratory illness and recover
	without requiring special treatment. However, some will become
	seriously ill and require medical attention. Older people and
	those with underlying medical conditions like cardiovascular
	disease, diabetes, chronic respiratory disease, or cancer are more
	likely to develop serious illnesses. Anyone can get sick with
	COVID-19 and become seriously ill or die at any age.


	According to WHO, on November 27, 2021, there are \SI{259 502 031}
	confirmed cases and \SI{5 183 003} confirmed deaths over the world. At
	the same time, vaccine doses administered \SI{7 702 859 718} on
	November 25, 2021. According to National Health Commission of the
	People's Republic of China, China has reported \SI{98 631} confirmed
	cases and \SI{4636} confirmed deaths on November 27, 2021.

	There exist various kinds of mathematical models to study
	epidemics, such as SIR, SIRS, SEIR, and SEIRS on the transmission
	of diseases. Some of them is age-structured model, such as in
	COVID-19 epidemic\cite{Age-structured1, Age-structured2},
	tuberculosis transmission \cite{Age-structured3}, and measles
	epidemics\cite{Age-structured4}. For COVID-19 epidemic, people in
	different ages have different clinical characteristics, so it's
	necessary to include age-structure into the model.

	COVID-19 vaccines can produce protection against the disease.
	Developing immunity through vaccination means there is a reduced
	risk of developing the illness and its consequences. This immunity
	helps people fight the virus if exposed. Getting vaccinated can
	also protect people around. This is particularly important to
	protect people at increased risk for severe illness from COVID-19,
	such as healthcare providers, older or elderly adults, and people
	with other medical conditions. Some articles have studied the
	importance and the impact of COVID-19 vaccination, such as in
	\cite{vaccination1, vaccination2}.


	In this article, we study the dynamics of an age-structured
	epidemic model with vaccination. In this model, individuals are
	distinguished both by age and the stage of the disease. The
	formulation of the model is addressed in Section~\ref{sec:model}.
	We calculate $ \mathscr{R}_0 $ in Subsection \ref{sec:brn}.
	We discuss the Properties of the model near the disease-free
	equilibrium point in Subsection~\ref{sec:dfe}, including the
	disease-free equilibrium point, the basic reproduction number, and
	the stability of disease-free equilibrium. We study the endemic
	equilibrium in Subsection~\ref{sec:eep}. We find the endemic
	equilibrium points and their stability. Sensitivity analysis of
	the model parameters is given in Section~\ref{sec:5} to show the
	importance of vaccination, city lock-down, and other factors. In
	Section~\ref{sec:6}, we apply our model to analyze the cases in
	Shijiazhuang, Hebei Province. Finally, we make some discussion in
	Section~\ref{sec:discussion} and the conclusion is drawn in
	Section~\ref{sec:conclusion}.

	\section{Model}\label{sec:model}

\begin{figure}[h]
		\centering
		\includegraphics[width=0.8\textwidth]{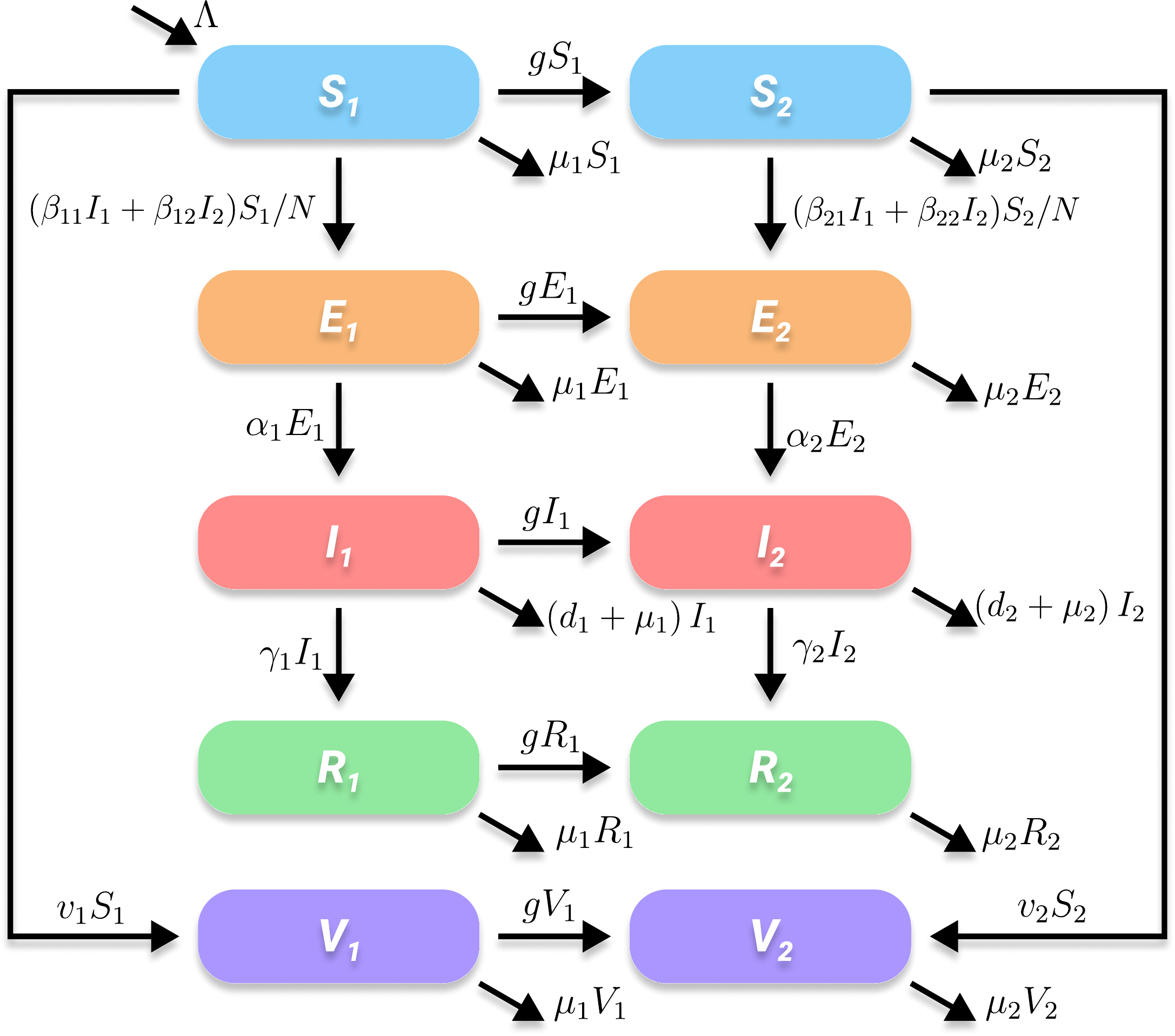}
		\caption{Diagram for conversions between the two age group in compartments.}
		\label{fig:progression}
	\end{figure}

	In this section, we consider an age-structured epidemic model in
	which individuals are distinguishable both by their age and the
	stage of the disease. In the model, the total size of the
	population $ N(t) $ contains $ 2 $ age groups, we use subscripts $
	i=1, 2 $ to stand for age group $\left( < 60 \text{ yr} \right)$ and
	age group $\left( \geqslant 60 \text{ yr} \right)$ respectively. The population
	in the same age group is divided into 5 clusters, including the
	susceptible individuals ($ S_i $), the latent individuals ($ E_i
	$), individuals with infectious ($ I_i $), the recovered
	individuals ($ R_i $) and those with vaccination ($ V_i $). The
	latent class consist of individuals infected, but without an
	infectious status, while the infectious class consists of those
	with infectious status. We divide vaccinated individuals
	separately for figuring out the effects of vaccination in an
	age-structured population. So the solution as $ X=\left( S_1, S_2, E_1, E_2, I_1, I_2,
	R_1, R_2, V_1, V_2 \right)^\mathrm{T} $, and the total population
	size is $ N = S_1+S_2+E_1+E_2+I_1+I_2+R_1+R_2+V_1+V_2 $. The transmission diagram is given in
	Figure~\ref{fig:progression}.

	Our model is derived as follows:
	\begin{equation}
		\label{eq:model}
		\begin{array}{l}
			\vspace{1ex}
			\displaystyle \dot{S}_1=\Lambda-(\beta_{11} I_1+\beta_{12} I_2) \frac{S_1}{N} -(v_1+\mu_1+g) S_1,            \\ \vspace{1ex}
			\displaystyle \dot{S}_2=g S_1-(\beta_{21} I_1+\beta_{22} I_2) \frac{S_2}{N}-\left(v_2+\mu_2\right) S_2,      \\ \vspace{1ex}
			\displaystyle \dot{E}_1=(\beta_{11} I_1+\beta_{12} I_2) \frac{S_1}{N}-\left(\alpha_1+\mu_1+g\right) E_1,     \\ \vspace{1ex}
			\displaystyle \dot{E}_2=g E_1+(\beta_{21} I_1+\beta_{22} I_2) \frac{S_2}{N}-\left(\alpha_2+\mu_2\right) E_2, \\ \vspace{1ex}
			\displaystyle \dot{I}_1=\alpha_1 E_1-\left(\gamma_1+d_1+\mu_1+g\right) I_1,                                  \\ \vspace{1ex}
			\displaystyle \dot{I}_2=gI_1+\alpha_2 E_2-\left(\gamma_2+d_2+\mu_2\right) I_2,                               \\ \vspace{1ex}
			\displaystyle \dot{R}_1=\gamma_1 I_1-\left(\mu_1+g\right) R_1,                                               \\ \vspace{1ex}
			\displaystyle \dot{R}_2=g R_1+\gamma_2 I_2-\mu_2 R_2,                                                        \\ \vspace{1ex}
			\displaystyle \dot{V}_1=v_1 S_1-\left(\mu_1+g\right) V_1,                                                    \\ \vspace{1ex}
			\displaystyle \dot{V}_2=g V_1+v_2 S_2-\mu_2 V_2.
		\end{array}
	\end{equation}
	All of the parameters are assumed to be constants. We assume that
	$ \mu_i>0 ~ (i = 1, 2) $, while the others are nonnative. The
	description of the parameters used in the model \eqref{eq:model} is
	shown on Table \ref{tab:parameters_description}.

	\begin{table}[h]
		\begin{center}
			\begin{minipage}{1\textwidth}
				\caption{Parameters and the corresponding descriptions to model \eqref{eq:model}}\label{tab:parameters_description}%
				\begin{tabular}{@{}cllll@{}}
					\toprule
					Parameters & Descriptions                           & Value                       & Reference value                        & Sources                                                    \\
					\midrule
					$\Delta T$      & Awareness delay                        & 21                          & $( 17, 23 )$                           & \cite{Data_AD}                                             \\
					$N$             & Shijiazhuang population                & $ 1039.42\times {10^{4}} $  & $ 1039.42\times {10^{4}} $                        & \cite{Data_people}   \\
					$\Lambda$       & Recruitment rate\tnote{1}              & 300.82                      & 300.82                                 & \cite{Data_age_rate}                                       \\
					$g$             & Ageing rate from G1 to G2\tnote{2}     & $ 4.57 \times {10^{ - 5}} $ & -                                      & Assumed                                                    \\
					$\beta_{11}$    & Infection rate led by $ I_1 $\tnote{1} & (0.0014, 0.288)             & $( 0, 0.5 )$                           & \cite{Data_AD}                     \\
					$\beta_{12}$    & Infection rate led by $ I_2 $\tnote{1} & (0.0015, 0.300)             & $( 0, 0.5 )$                           & \cite{Data_AD}                    \\
					$\beta_{21}$    & Infection rate led by $ I_1 $\tnote{2} & (0.0022, 0.440)             & $( 0, 0.5 )$                           & \cite{Data_AD}                     \\
					$\beta_{22}$    & Infection rate led by $ I_2 $\tnote{2} & (0.0020, 0.390)             & $( 0, 0.5 )$                           & \cite{Data_AD}                     \\
					$\alpha_{1}$    & Average conversation period\tnote{1}   & 4                           & $(3.53, 6.6)$                          & \cite{Data_incubation,Data_incubation_1,Data_incubation_2} \\
					$\alpha_{2}$    & Average conversation period\tnote{2}   & 3                           & $(3.53, 6.6)$                          & \cite{Data_incubation,Data_incubation_1,Data_incubation_2} \\
					$\gamma_{1}$    & Average recovery period\tnote{1}       & 28                          & $( 13.6, 33.4 )$                       & \cite{Data_gamma_0,Data_gamma_1}                           \\
					$\gamma_{2}$    & Average recovery period\tnote{2}       & 35                          & $( 13.6, 33.4 )$                       & \cite{Data_gamma_0,Data_gamma_1}                           \\
					$v_{1}$         & Vaccination rate\tnote{1}              & 0                           & -                                      & Assumed                                                    \\
					$v_{2}$         & Vaccination rate\tnote{2}              & 0                           & -                                      & Assumed                                                    \\
					$d_{1}$         & SARS-CoV-2 death rate\tnote{1}         & $ 2.37 \times {10^{ - 7}} $ & $(1 \times 10^{-7}, 3 \times 10^{-7})$ & \cite{Data_clinic,Data_death,Data_death1}                    \\
					$d_{2}$         & SARS-CoV-2 death rate\tnote{2}         & $ 1.42 \times {10^{ - 5}} $ & $(1 \times 10^{-7}, 4 \times 10^{-7})$ & \cite{Data_clinic,Data_death,Data_death1}                     \\
					$\mu_{1}$       & Natural death rate\tnote{1}            & $ 2.37 \times {10^{ - 7}} $ & $ 2.37 \times {10^{ - 7}} $                       & \cite{Data_mu}                     \\
					$\mu_{2}$       & Natural death rate\tnote{2}            & $ 1.42 \times {10^{ - 5}} $ & $ 1.42 \times {10^{ - 5}} $                       & \cite{Data_mu}                    \\
					$p$             & Population ageing rate\tnote{2}        & 19.38\%                     & 19.38\%                                & \cite{Data_age_rate}                                       \\
					\botrule
				\end{tabular}
				\footnotetext[1]{G1 means the individuals who are under 60 years old, short for G1 $( < 60 \text{ yr})$;}
				\footnotetext[2]{G2 means the individuals who are over 60 years old, short for G2 $(\geqslant 60 \text{ yr} )$.}
			\end{minipage}
		\end{center}
	\end{table}

	\section{Long-term Dynamics of the Model}

	In this section, we give the expression of the basic reproduction number $ \mathscr{R}_0 $,
	and discuss the properties of model \eqref{eq:model} around
	the disease-free equilibrium point and the endemic equilibrium
	point and their stabilities.

	\subsection{Basic Reproduction Number $\mathscr{R}_0$} \label{sec:brn}

	The basic reproduction number $ \mathscr{R}_0 $ describes the
	expected number of secondary case from primary case during the
	infectious period of the primary case, which is an important
	threshold parameter in studying the epidemics according to
	\cite{reproduction1}. We use the method provided in
	\cite{reproduction1} to calculate the basic reproduction number $
	\mathscr{R}_0 $.

	Let $ I_1 = I_2 = E_1 = E_2 = R_1 = R_2 = 0 $. Then, model \eqref{eq:model}
	can be simplified as follows:
	\begin{equation}
		\label{eq:DFEeqation}
		\begin{array}{lll}
			\Lambda-(v_1+\mu_1+g) S_1^0     =0, & \quad
			g S_1^0-\left(v_2+\mu_2\right) S_2^0  =0, \\
			v_1 S_1^0-\left(\mu_1+g\right) V_1^0    =0, & \quad
			g V_1^0+v_2 S_2^0-\mu_2 V_2^0            =0.
		\end{array}
	\end{equation}

	The equations \eqref{eq:DFEeqation} are linear with respect
	to $ S_1^0 $, $ S_2^0 $, $ V_1^0 $ and $ V_2^0 $, so we derive
	\begin{equation}
		\label{eq:DFE}
		\begin{array}{lll}
			\vspace{1ex}
			\displaystyle S_1^0=\frac{\Lambda}{g+\mu_1 +v_1 }, & \quad
			\displaystyle S_2^0=\frac{\Lambda g}{\left(\mu_{2}+v_{2}\right)\left(g+\mu_{1}+v_{1}\right)}, \\ \vspace{1ex}
			\displaystyle V_1^0=\frac{\Lambda \,v_1 }{{\left(g+\mu_1 \right)}\,{\left(g+\mu_1 +v_1 \right)}}, & \quad
			\displaystyle V_2^0=\frac{\Lambda \,g\,{\left(g\,v_2 +\mu_1 \,v_2 +\mu_2 \,v_1 +v_1 \,v_2 \right)}}
			{\mu_2 \,{\left(g+\mu_1 \right)}\,{\left(\mu_2 +v_2 \right)}\,{\left(g+\mu_1 +v_1
			\right)}},
		\end{array}
	\end{equation}
	which gives a unique disease-free equilibrium point $ X^0=\left(
	S_1^0, S_2^0, 0, 0, 0, 0, 0, 0, V_1^0, V_2^0\right)^\mathrm{T} $
	to model (\ref{eq:model}). Further, the initial total population size
	is $ N^0 = S_1^0+S_2^0+V_1^0+V_2^0 $.

	Firstly, the appearance of new infective individuals in compartments $
	I $ and $ E $ is written as:
	\begin{equation}
		\mathscr{F}_{1\times 10} =
		\left(
		\left(\beta_{11} I_1+\beta_{12} I_2\right) \displaystyle\frac{S_1}{N},
		\left(\beta_{21} I_1+\beta_{22} I_2\right) \displaystyle\frac{S_2}{N},
		0                                                                    ,
		0                                                                    ,
		\cdots                                                               ,
		0                                                                       \\
		\right)
		\text{.}
	\end{equation}

	Let $\mathscr{V}=\mathscr{V}^- -\mathscr{V}^+
	=(Q_1, Q_2, Q_3, Q_4, Q_5)^\mathrm{T}-(P_1, P_2, P_3, P_4, P_5)^\mathrm{T}$,
	where $\mathscr{V}^-$ stands for the transfer
	of the individuals out of compartment $ I $ and $\mathscr{V}^+$ stands for the transfer of the individuals into compartment $ I $ by all
	other means (except for the appearance of new infective individuals). The components
	of are respectively
	\begin{equation}
		\begin{array}{lll}
			P_1 = \left(\left( \alpha_1+g+\mu_1 \right)E_1    , \left( \alpha_2+\mu_2 \right)E_2 \right),    & \quad
			P_2 = \left( \left(\gamma_1+d_1+\mu_1+g\right) I_1 , \left(\gamma_2+d_2+\mu_2\right) I_2  \right),       \\
			P_3 =\left(  \left(v_1+\mu_1+g\right) S_1              , \left(v_2+\mu_2\right) S_2     \right), & \quad
			P_4 = \left( \left(\mu_1+g\right) R_1              , \mu_2 R_2
			\right), \\
			P_5 =\left( \left(\mu_1+g\right) V_1              , \mu_2 V_2
			\right), & \quad
			Q_1 = \left(  0            , g E_1  \right), \\
			Q_2 = \left(    \alpha_1 E_1 , g I_1+\alpha_2 E_2  \right), & \quad
			Q_3 = \left( \Lambda      , g S_1     \right), \\
			Q_4 = \left(    \gamma_1 I_1 , g R_1+\gamma_2 I_2  \right), & \quad
			Q_5 = \left( v_1 S_1            , g V_1+v_2 S_2
			\right).
		\end{array}
	\end{equation}

	We denote $Y =\left(E_1, E_2, I_1, I_2, S_1, S_2, R_1, R_2, V_1, V_2\right)$,
	for $i, j = 1, 2, 3, 4$ and $Y^0=(0, 0, 0, 0, S_1^0, S_2^0,
	0, 0, V_1^0, V_2^0)$, by Lemma 1 in \cite{reproduction1}, we obtain
	\begin{equation}
		F
		:=
		\left(\displaystyle\frac{\partial \mathscr{F}_i}{\partial Y_j}(Y^0)
		\right)_{4\times 4}
		=
		\left(\begin{array}{cccc}
				  \vspace{1ex}
				  0 & 0 & \displaystyle\frac{S_{1}^0 \beta_{11}}{N^0} & \displaystyle\frac{S_{1}^0 \beta_{12}}{N^0} \\ \vspace{1ex}
				  0     & 0 & \displaystyle\frac{S_{2}^0 \beta_{21}}{N^0} & \displaystyle\frac{S_{2}^0 \beta_{22}}{N^0} \\ \vspace{1ex}
				  0     & 0 & 0                                           & 0                                           \\ \vspace{1ex}
				  0     & 0 & 0                                           & 0
		\end{array}\right)
		\text{, }
	\end{equation}
	\begin{equation}
		V
		:=
		\left(
		\displaystyle \frac{\partial \mathscr{V}_i}{\partial Y_j}(Y^0)
		\right)_{4\times 4}
		=
		\left(
		\begin{array}{cccc}
			\alpha_{1}+\mu_{1}+g & 0                  & 0                          & 0                        \\
			-g                   & \alpha_{2}+\mu_{2} & 0                          & 0                        \\
			-\alpha_{1}          & 0                  & \gamma_{1}+d_{1}+\mu_{1}+g & 0                        \\
			0                    & -\alpha_{2}        & -g                         & \gamma_{2}+d_{2}+\mu_{2}
		\end{array}
		\right)
		\text{,}
	\end{equation}
	which then yields that
	\begin{equation}
		\label{eq:kij}
		\begin{array}{llllll}
			FV^{-1}
			& = &
			\left(\begin{array}{cccc}
					  \vspace{1ex}
					  \displaystyle\frac{S_1^0 \Sigma_{11} }{\sigma_1 } & \displaystyle\frac{S_1^0 \,\alpha_2 \,\beta_{12} }{\sigma_2 } & \displaystyle\frac{S_1^0 \,\Sigma_{13} }{\sigma_3 } & \displaystyle\frac{S_1^0 \,\beta_{12} }{\sigma_4} \\\vspace{1ex}
					  \displaystyle\frac{S_2^0 \Sigma_{21} }{\sigma_1 }      & \displaystyle\frac{S_2^0 \,\alpha_2 \,\beta_{22} }{\sigma_2 } & \displaystyle\frac{S_2^0 \,\Sigma_{23} }{\sigma_3 } & \displaystyle\frac{S_2^0 \,\beta_{22} }{\sigma_4} \\\vspace{1ex}
					  0                                                      & 0                                                             & 0                                                   & 0                                                 \\\vspace{1ex}
					  0                                                      & 0                                                             & 0                                                   & 0
			\end{array}\right)
			& := &
			\left(\begin{array}{cccc}
					  k_{11} & k_{12} & k_{13} & k_{14} \\
					  k_{21} & k_{22} & k_{23} & k_{24} \\
					  0      & 0      & 0      & 0      \\
					  0      & 0      & 0      & 0
			\end{array}\right)
			>0
			\text {,}
		\end{array}
	\end{equation}
	where
	\begin{equation}
		\begin{array}{ccl}
			\Sigma_{11} & = & \beta_{11} \left( \alpha_{1} \mu_{2}^{2}+\alpha_{1} \alpha_{2} d_{2} +\alpha_{1} \alpha_{2} \gamma_{2}+\alpha_{1} \alpha_{2} \mu_{2}+\alpha_{1} d_{2} \mu_{2}+\alpha_{1} \gamma_{2} \mu_{2}\right)   \\
			&   & +\beta_{12} \left( \alpha_{2} g^{2} +\alpha_{1} \alpha_{2} g+\alpha_{2} d_{1} g+\alpha_{2} g \gamma_{1}+\alpha_{1} g \mu_{2}+\alpha_{2} g \mu_{1}\right) ,                                           \\
			\Sigma_{21} & = & \beta_{21} \left( \alpha_{1} \mu_{2}^{2} +\alpha_{1} \alpha_{2} d_{2}+\alpha_{1} \alpha_{2} \gamma_{2}+\alpha_{1} \alpha_{2} \mu_{2}+\alpha_{1} d_{2} \mu_{2} +\alpha_{1} \gamma_{2} \mu_{2} \right) \\
			&   & +\beta_{22} \left( \alpha_{2} g^{2}+\alpha_{1} \alpha_{2} g +\alpha_{2} d_{1} g+\alpha_{2} g \gamma_{1}+\alpha_{1} g \mu_{2}+\alpha_{2} g \mu_{1} \right) ,                                          \\
			\Sigma_{13} & = & \beta_{11} \left( d_{2}+ \gamma_{2} + \mu_{2} \right) +\beta_{12} g,                                                                                                                                 \\
			\Sigma_{23} & = & \beta_{21} \left( d_{2} + \mu_{2} \right) +\beta_{22} \left( g + \gamma_{2} \right),                                                                                                                 \\
			\sigma_1    & = & N^0\,{\left(\alpha_1 +g+\mu_1 \right)}\,{\left(d_1 +g+\gamma_1 +\mu_1 \right)},                                                                                                                      \\
			\sigma_2    & = & N^0\,{\left(\alpha_2 +\mu_2 \right)}\,{\left(d_2 +\gamma_2 +\mu_2 \right)},                                                                                                                          \\
			\sigma_3    & = & N^0\,{\left(d_1 +g+\gamma_1 +\mu_1 \right)},                                                                                                                                                         \\
			\sigma_4    & = & N^0\,{\left(d_2 +\gamma_2 +\mu_2 \right)}.
		\end{array}
	\end{equation}

	The characteristic equation of matrix $ FV^{-1} $ is followed
	\begin{equation}
		\label{EigenvalueCharpoly}
		\lambda^2 \left[\left(\lambda-k_{11}\right)\left(\lambda-k_{22}\right)-k_{12}k_{21}\right]=0
		\text {.}
	\end{equation}
	Then, the eigenvalues of \eqref{EigenvalueCharpoly} are
	respectively
	\begin{equation}
		\begin{array}{lll}
			& \lambda_1  =  0, \quad
			\lambda_2  =  0, \\\vspace{1ex}
			& \lambda_3  =   \displaystyle\frac{1}{2}\left(k_{11}+k_{22}-\sqrt{k_{11}^{2}-2 k_{11} k_{22}+k_{22}^{2}+4 k_{12} k_{21}}\right), \\\vspace{1ex}
			& \lambda_4  =   \displaystyle\frac{1}{2}\left(k_{11}+k_{22}+\sqrt{k_{11}^{2}-2 k_{11} k_{22}+k_{22}^{2}+4 k_{12} k_{21}}\right),
		\end{array}
	\end{equation}
	with
	\begin{equation}
		k_{11}^{2}-2 k_{11} k_{22}+k_{22}^{2}+4 k_{12} k_{21}=\left(k_{11}-k_{22}\right)^2+4 k_{12} k_{21}>0
		\text{.}
	\end{equation}

	By the next generation matrix method, the basic reproduction number $ \mathscr{R}_0 $ is the spectral radius of
	$FV^{-1}$, that is:
	\begin{equation}
		\label{eq:R0}
		\begin{array}{lll}
			\mathscr{R}_0 & = & \rho(FV^{-1})                                                                                                                \\ \vspace{1ex}
			& = & \max \left\{ \lambda : \left\| \lambda E_4-FV^{-1} \right\| =0 \right\}                                                      \\
			& = & \displaystyle\frac{1}{2}\left( k_{11}+k_{22}+\sqrt{ k_{11}^{2}-2 k_{11} k_{22}+k_{22}^{2}+4 k_{12} k_{21} } \right) \text{,}
		\end{array}
	\end{equation}
	where the positive parameters $ k_{ij} $ $ (i,j=1, 2, 3, 4) $ are
	defined in \eqref{eq:kij}.

	\subsection{Disease-free Equilibrium and Stability} \label{sec:dfe}

	The Jacobian matrix $ J $ on Appendix \ref{sec:Block matrices} at the disease-free
	equilibrium point $ X = X^0 $ is
	\begin{equation}
		J_{0}=
		J_{X = X^0}:=
		\left(\begin{array}{cc}
				  M_1 & O   \\
				  M_3 & M_4
		\end{array}\right)
		\text{, }
	\end{equation}
	with
	\begin{equation}
		M_1
		=\left(\begin{array}{cc}
				   J_{EE} & J_{EI} \\
				   J_{IE} & J_{II}
		\end{array}\right)
		\text{, }
		M_3
		=\left(\begin{array}{cc}
				   O & J_{SI} \\
				   O & J_{RI} \\
				   O & O
		\end{array}\right)
		\text{, }
		M_4
		=\left(\begin{array}{ccc}
				   J_{SS} & O      & O      \\
				   O      & J_{RR} & O      \\
				   J_{VS} & O      & J_{VV}
		\end{array}\right)
		\text{. }
	\end{equation}
	Further, by the careful computation, we derive
	\begin{equation}
		\begin{array}{lll}
			M_1 & = &
			\left(
			\begin{array}{cccc}
				\vspace{1ex}
				-\alpha_{1}-g-\mu_{1} & 0                   & \displaystyle \frac{S_1^0 \beta_{11}}{N^0} & \displaystyle \frac{S_1^0 \beta_{12}}{N^0} \\ \vspace{1ex}
				g                         & -\alpha_{2}-\mu_{2} & \displaystyle \frac{S_2^0 \beta_{21}}{N^0} & \displaystyle \frac{S_2^0 \beta_{22}}{N^0} \\\vspace{1ex}
				\alpha_{1}                & 0                   & -d_{1}-g-\gamma_{1}-\mu_{1}                & 0                                          \\\vspace{1ex}
				0                         & \alpha_{2}          & g                                          & -d_{2}-\gamma_{2}-\mu_{2}
			\end{array}\right)
			\vspace{1ex} \\
			& := &
			\left(\begin{array}{cccc}
					  -A_1-g & 0    & B_{11}   & B_{12} \\
					  g      & -A_2 & B_{21}   & B_{22} \\
					  C_1    & 0    & -D_{1}-g & 0      \\
					  0      & C_2  & g        & -D_2
			\end{array}
			\right)
			\text{,}
		\end{array}
	\end{equation}
	and
	\begin{equation}
		M_4=
		-
		\left(
		\begin{array}{cccccc}
			g+\mu_{1}+v_{1} & 0             & 0         & 0       & 0         & 0       \\
			-g              & \mu_{2}+v_{2} & 0         & 0       & 0         & 0       \\
			0               & 0             & g+\mu_{1} & 0       & 0         & 0       \\
			0               & 0             & -g        & \mu_{2} & 0         & 0       \\
			-v_1            & 0             & 0         & 0       & g+\mu_{1} & 0       \\
			0               & -v_2          & 0         & 0       & -g        & \mu_{2}
		\end{array}
		\right)
		\text{.}
	\end{equation}
	The characteristic equation of Jacobian matrix $J_{0}$ is equal to
	the product of the characteristic equations of $M_1$ and
	$M_4$. Because the eigenvalues of $ M_4 $ are all negative, and other four eigenvalues depend on
	the characteristic equation of $ M_1 $
	\begin{equation}
		\label{DFEcharployM1}
		\lambda^4+L_1\lambda^3+L_2\lambda^2+L_3\lambda+L_4=0.
	\end{equation}
	where
	\begin{equation}
		\label{CopolyM1}
		\resizebox{\linewidth}{!}{%
			$
			\begin{array}{lll}
				L_1 & = & A_1 +A_2 +D_1 +D_2
				\text{,}
				\\
				L_2 & = & A_1 A_2+A_1 D_1+A_1 D_2+A_2 D_1+A_2 D_2-B_{11} C_1-B_{22} C_2+D_1 D_2                                                                                    \\
				& = & A_1 A_2+A_1 D_2+A_2 D_1+\left(A_1 D_1-B_{11} C_1\right)+\left(A_2 D_2-B_{22} C_2\right)+D_1 D_2
				\text{,}
				\\
				L_3 & = & A_1 A_2 D_1+A_1 A_2 D_2-A_2 B_{11} C_1-A_1 B_{22} C_2+A_1 D_1 D_2+A_2 D_1 D_2-B_{11} C_1 D_2-B_{22} C_2 D_1                                              \\
				& = & \left(A_1+D_1\right) \left(A_2 D_2 - B_{22} C_2\right) + \left(A_2+D_2\right) \left(A_1 D_1 - B_{11} C_1\right)
				\text{,}
				\\
				L_4 & = & A_1 A_2 D_1 D_2-A_2 B_{11} C_1 D_2-A_1 B_{22} C_2 D_1+B_{11} B_{22} C_1 C_2-B_{12} B_{21} C_1 C_2                                                        \\
				& = & -A_1 A_2 D_1 D_2+\left(B_{11} B_{22} -B_{12} B_{21}\right) C_1 C_2 +A_2 D_2 \left(A_1 D_1 - B_{11} C_1\right) + A_1 D_1 \left(A_2 D_2 -B_{22} C_2\right)
				\text{.}
			\end{array}
			$
		}
	\end{equation}

	According to Descartes' Rule of Signs in \cite{Descartes01,Descartes02,Descartes03}, the number of negative roots of the
	characteristic equation \eqref{DFEcharployM1} is equal to the
	number of variations in the change in the coefficient sign, so
	\eqref{DFEcharployM1} has 4 negative values if
	\begin{equation}
		\label{DFE_L_condition}
		L_1>0,\quad L_2>0, \quad L_3>0, \quad L_4>0.
	\end{equation}
	The conclusions \eqref{DFE_L_condition} are valid provided that
	the parameters in \eqref{CopolyM1} satisfy:
	\begin{equation}
		\label{DFE_coefficients_condition}
		\begin{array}{l}
			A_1 D_1 -B_{11} C_1 >0, \quad
			A_2 D_2 -B_{22} C_2 >0, \\
			-A_1 A_2 D_1 D_2+\left(B_{11} B_{22} -B_{12} B_{21}\right) C_1 C_2>0.
		\end{array}
	\end{equation}

	So, we derive the following result:
	\begin{theorem}[Disease-free equilibrium point stability]
		\label{thm:dfe_stability}
		The disease-free equilibrium point $
		X^0 $ is locally asymptotically stable when $ \mathscr{R}_0 <1
		$, while $X^0$ is unstable when $ \mathscr{R}_0 >1 $, when the
		parameters satisfy equation \eqref{DFE_coefficients_condition}.
	\end{theorem}

	\subsection{Endemic Equilibrium Points and Stability} \label{sec:eep}

	In this section, we consider the properties of model (\ref{eq:model}) near the
	endemic equilibrium point $X^*$.

	Let the right hand side of model (\ref{eq:model}) be zero, that is
	\begin{subequations}
		\begin{align}
			& \Lambda-(\beta_{11} I_1^*+\beta_{12} I_2^*) \frac{S_1^*}{N^*} -(v_1+\mu_1+g) S_1^* =0,    \label{eq:EEPa}              \\
			& g S_1^*-(\beta_{21} I_1^*+\beta_{22} I_2^*) \frac{S_2^*}{N^*}-\left(v_2+\mu_2\right) S_2^*=0, \label{eq:EEPb}          \\
			& (\beta_{11} I_1^*+\beta_{12} I_2^*) \frac{S_1^*}{N^*}-\left(\alpha_1+\mu_1+g\right) E_1^*=0,      \label{eq:EEPc}      \\
			& g E_1^*+(\beta_{21} I_1^*+\beta_{22} I_2^*) \frac{S_2^*}{N^*}-\left(\alpha_2+\mu_2\right) E_2^*  =0,   \label{eq:EEPd} \\
			& \alpha_1 E_1^*-\left(\gamma_1+d_1+\mu_1+g\right) I_1^*              =0,        \label{eq:EEPe}                         \\
			& gI_1^*+\alpha_2 E_2^*-\left(\gamma_2+d_2+\mu_2\right) I_2^*    =0,          \label{eq:EEPf}                            \\
			& \gamma_1 I_1^*-\left(\mu_1+g\right) R_1^*     =0,       \label{eq:EEPg}                                                \\
			& g R_1^*+\gamma_2 I_2^*-\mu_2 R_2^*   =0,       \label{eq:EEPh}                                                         \\
			& v_1 S_1^*-\left(\mu_1+g\right) V_1^*   =0,         \label{eq:EEPi}                                                     \\
			& g V_1^*+v_2 S_2^*-\mu_2 V_2^* =0. \label{eq:EEPj}
		\end{align}
	\end{subequations}

	Taking sum of \eqref{eq:EEPa} and \eqref{eq:EEPc}, \eqref{eq:EEPb} and \eqref{eq:EEPd} respectively give two equalities:
	\begin{equation}
		\begin{array}{l}
			\Lambda -(v_1+\mu_1+g) S_1^* -\left(\alpha_1+\mu_1+g\right) E_1^*=0, \\
			g S_1^*-\left(v_2+\mu_2\right) S_2^*+g E_1^* -\left(\alpha_2+\mu_2\right) E_2^*=0,
		\end{array}
	\end{equation}
	so $ S_1^* $ and $ S_2^* $ are expressed as linear functions of $ E_1^* $ and
	$ E_2^* $. Similarly, $ I_1^*, I_2^*, R_1^*, R_2^*, V_1^*, V_2^* $
	are written as linear functions of $ E_1^* $ and $ E_2^* $ by
	\eqref{eq:EEPe}-\eqref{eq:EEPj}:
	\begin{equation}
		\label{EEPlinear}
		\resizebox{\linewidth}{!}{%
			$
			\begin{array}{lll}
				\vspace{1ex}
				S_1^* & = & \displaystyle  \frac{\Lambda }{v_1+\mu_1+g}  -\frac{\alpha_1+\mu_1+g}{v_1+\mu_1+g}     E_1^* :=  u_{10}+u_{11} E_1^*,                                                                                                                                                          \\
				\vspace{1ex}
				S_2^*     & = & \displaystyle \frac{g \Lambda}{\left(v_1+\mu_1+g\right) \left(v_2+\mu_2\right)} +\frac{g \left(v_1-\alpha_1\right)}{\left(v_1+\mu_1+g\right) \left(v_2+\mu_2\right)} E_1^* -\frac{\alpha_2+\mu_2}{\left(v_2+\mu_2\right)} E_2^* :=  u_{20}+u_{21} E_1^* +u_{22}  E_2^*, \\
				I_1^*     & = & \displaystyle \frac{\alpha_1}{\gamma_1+d_1+\mu_1+g} E_1^* := v_{11} E_1^* ,                                                                                                                                                                                         \\
				\vspace{1ex}
				I_2^*     & = & \displaystyle \frac{g \alpha_1}{\left(\gamma_1+d_1+\mu_1+g\right) \left(\gamma_2+d_2+\mu_2\right)} E_1^*+\frac{\alpha_2}{\gamma_2+d_2+\mu_2}       E_2^* :=v_{21} E_1^* + v_{22} E_2^*,                                                                                    \\
				\vspace{1ex}
				R_1^*     & = & \displaystyle \frac{    \gamma_1 \alpha_1}{\left(\mu_1+g\right) \left(\gamma_1+d_1+\mu_1+g\right)}      E_1^* :=x_{11} E_1^*,                                                                                                                                                 \\
				\vspace{1ex}
				R_2^*     & = & \displaystyle \frac{g \alpha_1}{\mu_2 \left(\gamma_1+d_1+\mu_1+g\right)}  \left(\frac{\gamma_1}{\mu_1+g} + \frac{\gamma_2}{\mu_2}\right) E_1^* +\frac{\gamma_2 \alpha_2}{\mu_2 \left(\gamma_2+d_2+\mu_2\right)}    E_2^* := x_{21} E_1^*+x_{22} E_2^*,                    \\
				\vspace{1ex}
				V_1^*     & = & \displaystyle \frac{v_1\Lambda}{\left(\mu_1+g\right) \left(v_1+\mu_1+g\right)}  -\frac{v_1 \left(\alpha_1+\mu_1+g\right) }{\left(\mu_1+g\right) \left(v_1+\mu_1+g\right)} E_1^* := y_{10}  + y_{11}   E_1^*,                                                               \\
				\vspace{1ex}
				V_2^* & = & \displaystyle \frac{g v_1\Lambda}{\mu_2 \left(v_1+\mu_1+g\right) } \left(\frac{1}{\mu_1+g}+\frac{1}{v_2+\mu_2}\right)
				+\frac{g v_1}{\mu_2 \left(v_1+\mu_1+g\right) } \left(-\frac{\alpha_1+\mu_1+g}{\mu_1+g} + \frac{v_1-\alpha_1}{v_2+\mu_2}\right)E_1^*
				\\
				\vspace{1ex}
				&   & \displaystyle -\frac{v_1 \left(\alpha_2 +\mu_2\right)}{\mu_2 \left(v_2+\mu_2\right)}E_2^*
				:=y_{20}+y_{21} E_1^*+y_{22} E_2^*.
			\end{array}
			$
		}
	\end{equation}

	The total population size $ N^* $ is given
	\begin{equation}
		\label{EEPN}
		\begin{array}{lll}
			N^* & =  & S_1^*+S_2^*+E_1^*+E_2^*+I_1^*+I_2^*+R_1^*+R_2^*+V_1^*+V_2^*                                           \\
			& =  & u_{10}+u_{20}+y_{10}+y_{20}+\left(u_{11}+u_{21}+v_{11}+v_{21}+x_{11}+x_{21}+y_{11}+y_{21}\right)E_1^* \\
			&    & +\left(u_{22}+v_{22}+x_{22}+y_{22}\right)E_2^*                                                        \\
			& := & w_0+w_1 E_1^*+w_2 E_2^*.
		\end{array}
	\end{equation}

	Together with \eqref{EEPlinear} and \eqref{EEPN},  equalities
	\eqref{eq:EEPc} and \eqref{eq:EEPd} can be rewritten as:
	\begin{equation}
		\label{EEPequation}
		\begin{array}{l}
			f_1(E_1^*)=a_{11}{E_1^*}^2+2a_{12} E_1^* E_2^*+2a_1 E_1^*+2 a_2 E_2^*=0, \\
			f_2(E_1^*)=b_{11}{E_1^*}^2+2b_{12} E_1^* E_2^*+b_{22}{E_2^*}^2+2b_1 E_1^*+2 b_2 E_2^*=0,
		\end{array}
	\end{equation}
	with
	\begin{equation}
		\resizebox{\linewidth}{!}{%
			$
			\begin{array}{l}
				\vspace{1ex}
				a_{11}= u_{11} \left(\beta_{11} v_{11}+\beta_{12} v_{21}\right)-\left(\alpha_1+\mu_1+g\right)w_1 ,  \quad \vspace{1ex}
				\displaystyle a_{12}= 2u_{11} \beta_{12} v_{21}-2\left(\alpha_1+\mu_1+g\right)w_2, \\
				\vspace{1ex}
				\displaystyle a_{1}= 2u_{10} \left(\beta_{11} v_{11}+\beta_{12} v_{21}\right)-2\left(\alpha_1+\mu_1+g\right)w_0,    \quad
				\vspace{1ex}
				\displaystyle a_{2}= 2u_{10} \beta_{12} v_{22},                                        \\
				\vspace{1ex}
				b_{11}= u_{21} \left(\beta_{21} v_{11}+\beta_{22} v_{21}\right)-g w_1,      \quad
				b_{22}= u_{22} \beta_{22} v_{22} + \left(\alpha_2+\mu_2\right)w_2,
				\\
				\vspace{1ex}
				\displaystyle b_{12}= 2u_{22} \left(\beta_{21} v_{11}+\beta_{22} v_{21}\right)
				+2u_{21} \beta_{22} v_{22}+2\left(\alpha_2+\mu_2\right)w_1-2g w_2,                     \\
				\vspace{1ex}
				\displaystyle b_{1}= 2u_{20} \left(\beta_{21} v_{11}+\beta_{22} v_{21}\right)-2g w_0,     \quad
				\displaystyle b_{2}= 2u_{20} \beta_{22} v_{22} + 2\left(\alpha_2+\mu_2\right)w_0.
			\end{array}
			$
		}
	\end{equation}
	The endemic equilibrium point $X^*$ is located at the intersection of $f_1(E_1^*)=0$ and $f_2(E_1^*)=0$,
	we discuss equalities \eqref{EEPequation} by cases.

	\textbf{Case 1} ($ a_{11} \neq 0$ and $ b_{11} \neq 0$)
	Equalities \eqref{EEPequation} determine a unique endemic
	equilibrium point as
	\begin{equation}
		\label{Deltaequation}
		\begin{array}{l}
			\vspace{1ex}
			\displaystyle \Delta_1=4(a_{12} E_2^*+a_1)^2-8 a_2 a_{11}  E_2^* =0, \\
			\vspace{1ex}
			\displaystyle \Delta_2=4(b_{12} E_2^*+b_1)^2-4b_{11} (b_{22}E_2^*+2 b_2) E_2^* =0,
		\end{array}
	\end{equation}
	then we figure out
	\begin{equation}
		\label{E1quation}
		\displaystyle \frac{a_{12} E_2^*+a_1}{a_{11}} =\displaystyle \frac{b_{12} E_2^*+b_1}{b_{11}},
	\end{equation}
	due to $f_1(E_1^*)=f_2(E_1^*)=0$, which further implies
	\begin{equation}
		\label{E2epoint}
		E_2^*=\displaystyle \frac{b_{11} a_1 +a_{11} b_1}{a_{11} b_{12} - b_{11}
			a_{12}}, \quad
		E_1^*=\displaystyle \frac{a_{12} b_1-a_1 b_{12}}{a_{11} b_{12} - b_{11}
			a_{12}}.
	\end{equation}
	We denote $ K_1=a_{12} b_1-a_1 b_{12}, ~
	K_2=a_{11} b_{12} - b_{11} a_{12},~
	K_3=b_{11} a_1 + a_{11} b_1, $
	thus $ E_1^* $ and $ E_2^* $ keep positive if and only if
	\begin{equation}
		\begin{array}{l}
			\vspace{1ex}
			K_1K_2=(a_{12} b_1-a_1 b_{12})
			(a_{11} b_{12} - b_{11} a_{12})>0, \\
			\vspace{1ex}
			K_2K_3=(a_{11} b_{12} - b_{11} a_{12})
			(b_{11} a_1 +a_{11}b_1)>0,
		\end{array}
	\end{equation}
	We substitute $ E_2^* $ into equalities \eqref{Deltaequation} such that
	\begin{equation}
		\begin{array}{llll}
			\vspace{1ex}
			& \displaystyle \Delta_1 & = & \displaystyle 4\frac{a_{11}}{K_2^2}
			\left( a_{11} K_1^2 - 2 a_2 K_2 K_3 \right)         =0, \\
			\vspace{1ex}
			& \displaystyle \Delta_2 & = & \displaystyle 4\frac{b_{11}}{K_2^2}
			\left[ b_{11} K_1^2 - \left( b_{22} K_3 + 2 b_2 K_2 \right) K_3 \right]   =0,
		\end{array}
	\end{equation}
	which further gives
	\begin{equation}
		a_{11} b_{22} K_3= 2(a_{12}b_{11}-a_{11}b_2)K_2.
	\end{equation}

	\textbf{Case 2} ($ a_{11} = 0 $ and $ b_{11} \neq0$) Equalities
	\eqref{Deltaequation} become
	\begin{equation}
		\label{EEPNewequation2}
		\begin{array}{l}
			2(a_{12} E_2^*+a_1) E_1^*+2 a_2 E_2^*=0, \\
			b_{11}{E_1^*}^2+2(b_{12} E_2^*+b_1) E_1^*+(b_{22}E_2^*+2 b_2) E_2^*=0.
		\end{array}
	\end{equation}

	As $ \Delta_2 = 0 $, we can work out
	\begin{equation}
		\label{EEPNewequation2E1}
		E_1^*=-\displaystyle \frac{b_{12} E_2^*+b_1}{b_{11}},
	\end{equation}
	which combines with \eqref{EEPNewequation2}, we get:
	\begin{equation}
		a_{12} b_{12} {E_2^*}^2+\left(a_{12} b_1+a_1 b_{12}-a_2 b_{11}\right)E_2^*+a_1 b_1=0.
	\end{equation}

	If $ a_{12} b_{12} \neq 0 $ and
	$ \Delta = \left( a_{12} b_1+a_1 b_{12} - a_2 b_{11} \right)^2-4 a_{12} b_{12} a_1 b_1  =0 $, we obtain
	\begin{equation}
		\vspace{1ex}
		E_1^* = \displaystyle \frac{-a_{12} b_1+a_1 b_{12} +a_2 b_{11}}{2 a_{12} b_{11}},
		\quad
		E_2^* = - \displaystyle \frac{a_{12} b_1+a_1 b_{12}-a_2 b_{11}}{2 a_{12} b_{12}}.
	\end{equation}

	If $ a_{12} b_{12} = 0 $, then
	\begin{equation}
		\vspace{1ex}
		E_1^* = \displaystyle \frac{\left(a_{2} b_{11}-a_{12} b_{1}\right)b_{1}}
		{\left(a_{12} b_{1}+a_{1} b_{12}-a_{2} b_{11}\right)b_{11}},
		\quad
		E_2^* = - \displaystyle \frac{a_1 b_1}{a_{12} b_1+a_1 b_{12}-a_2 b_{11}}.
	\end{equation}

	\textbf{Case 3} ($ a_{11} \neq 0 $ and $b_{11} = 0 $) Equalities
	\eqref{Deltaequation} turn out to be
	\begin{equation}
		\begin{array}{l}
			\label{EEPNewequation3}
			a_{11}{E_1^*}^2+2(a_{12} E_2^*+a_1) E_1^*+2 a_2 E_2^*=0, \\
			2(b_{12} E_2^*+b_1) E_1^*+(b_{22}E_2^*+2 b_2) E_2^*=0.
		\end{array}
	\end{equation}

	As $ \Delta_1 = 0 $, we figure out
	\begin{equation}
		\label{EEPNewequation3E1}
		E_1^*=-\displaystyle \frac{a_{12}E_2^*+a_1}{a_{11}},
	\end{equation}
	which combines with \eqref{EEPNewequation3E1}, we have
	\begin{equation}
		\left(a_{11} b_{22}-2a_{12} b_{12}\right)
		{E_2^*}^2+2\left(a_{11} b_2-a_1 b_{12}-a_{12} b_1\right)E_2^*-2a_1 b_1=0.
	\end{equation}

	If $ a_{11} b_{22}-2a_{12} b_{12} \neq 0 $ and
	$ \displaystyle \frac{\Delta}{4} = \left( a_{11} b_2-a_1 b_{12}-a_{12} b_1 \right)^2
	+2 a_1 b_1 \left(a_{11} b_{22}-2a_{12} b_{12}\right)   =0 $, we can get:
	\begin{equation}
		\begin{array}{l}
			\vspace{1ex}
			E_1^* = -\displaystyle \frac{-a_{11} a_{12} b_2+a_{12}^2 b_1 + a_1 a_{11} b_{22} -a_1 a_{12} b_{12}}{ a_{11}
				\left(a_{11} b_{22}-2a_{12} b_{12}\right) }, \\
			E_2^* = - \displaystyle \frac{-a_{11} b_2+a_1 b_{12}+a_{12} b_1}{a_{11} b_{22}-2a_{12} b_{12}}.
		\end{array}
	\end{equation}

	If $ a_{12} b_{12} = 0 $, then
	\begin{equation}
		\begin{array}{l}
			\vspace{1ex}
			E_1^* = -\displaystyle \frac{a_{1}\left(a_{1} b_{12}-a_{11} b_{2}\right)}{a_{11}\left(a_{1} b_{12}-a_{11} b_{2}+a_{12}
			b_{1}\right)}, \\
			E_2^* = - \displaystyle \frac{a_{1} b_{1}}{a_{1} b_{12}-a_{11} b_{2}+a_{12} b_{1}}.
		\end{array}
	\end{equation}

	\textbf{Case 4} ($ a_{11} = b_{11} =0$) Equalities
	\eqref{EEPequation} can be simplified as follows:
	\begin{equation}
		\label{EEPNewequation4}
		2(a_{12} E_2^*+a_1) E_1^*+2 a_2 E_2^*=0,  \\
		2(b_{12} E_2^*+b_1) E_1^*+(b_{22}E_2^*+2 b_2) E_2^*=0.
	\end{equation}

	The first equality of \eqref{EEPNewequation4} gives
	\begin{equation}
		\label{EEPNewequation4E1}
		E_1^*=-\frac{E_2^* a_{2}}{a_{1}+E_2^* a_{12}},
	\end{equation}
	which combines with  \eqref{EEPNewequation4E1}, we have
	\begin{equation}
		\label{EEPNewequation4E2}
		\left(a_{12} b_{22}\right) {E_{2}^*}^2+\left(2 a_{12} b_{2}-2 a_{2} b_{12}+a_{1} b_{22}\right) E_2^*+2 a_{1} b_{2}-2 a_{2} b_{1}=0.
	\end{equation}

	If $ 2 a_{12} b_{22} \neq 0 $ and
	$ \Delta =\left(2 a_{12} b_{2}-2 a_{2} b_{12}+a_{1} b_{22}\right)^{2}-4 a_{12} b_{22}\left(2 a_{1} b_{2}-2 a_{2} b_{1}\right) =0 $, we have
	\begin{equation}
		\vspace{1ex}
		E_1^*= \displaystyle \frac{a_{2}\left(2 a_{12} b_{2}-2 a_{2} b_{12}+a_{1} b_{22}\right)}
		{a_{12}\left(2 a_{2} b_{12}-2 a_{12} b_{2}+a_{1}
		b_{22}\right)},\quad
		E_2^*=-\displaystyle \frac{2 a_{12} b_{2}-2 a_{2} b_{12}+a_{1} b_{22}}{2 a_{12} b_{22}}.
	\end{equation}

	If $ 2 a_{12} b_{22} = 0 $, then we derive
	\begin{equation}
		\vspace{1ex}
		E_1^*= \displaystyle \frac{2 a_{2}\left(a_{1} b_{2}-a_{2} b_{1}\right)}{b_{22} a_{1}^{2}-2 a_{2} b_{12} a_{1}+2 a_{2} a_{12}
			b_{1}},\quad
		E_2^*= -\displaystyle \frac{2 a_{1} b_{2}-2 a_{2} b_{1}}{2 a_{12} b_{2}-2 a_{2} b_{12}+a_{1} b_{22}}.
	\end{equation}

	\begin{theorem}[Endemic equilibrium points extience and uniqueness]
		\label{thm:eep}
		Under the moderate conditions,
		model \eqref{eq:model} has a unique endemic equilibrium points $ X^* $.
	\end{theorem}

	We construct the Jacobian matrix for
	age group $\left( < 60 \text{ yr} \right)$ and age group $\left( \geqslant 60 \text{ yr} \right)$ respectively as follows:
	\begin{equation}
		\resizebox{\linewidth}{!}{%
			$
			J_a= \left(\begin{array}{ccccc}
						   \vspace{1ex}
						   \displaystyle   -\alpha_1 -g-\mu_1 - \frac{S_1^* \,\eta_1 }{(N^*)^2 }
						   & \displaystyle \frac{S_1^* \,\beta_{11} }{N^*}-\frac{S_1^* \,\eta_1 }{(N^*)^2 }
						   & \displaystyle -\eta_1 \,{\left(\frac{S_1^* }{(N^*)^2 }-\frac{1}{N^*}\right)}
						   & \displaystyle -\frac{\eta_1 }{(N^*)^2 }                                                    & \displaystyle -\frac{\eta_1 }{(N^*)^2 }                       \\
						   \vspace{1ex}
						   \alpha_1 & -d_1 -g-\gamma_1 -\mu_1 & 0   & 0        & 0        \\
						   \vspace{1ex}
						   \displaystyle   -\frac{S_1^* \,\eta_1 }{(N^*)^2 }
						   & \displaystyle \frac{S_1^* \,\eta_1 }{(N^*)^2 } -\frac{S_1^* \,\beta_{11} }{N^*}
						   & \displaystyle \eta_1 \,{\left(\frac{S_1^* }{(N^*)^2 }-\frac{1}{N^*}\right)} -\mu_1 -v_1 -g
						   & \displaystyle \frac{\eta_1 }{(N^*)^2 }                                                     & \displaystyle \frac{\eta_1 }{(N^*)^2 }                        \\
						   0        & \gamma_1                & 0   & -g-\mu_1 & 0        \\
						   0        & 0                       & v_1 & 0        & -g-\mu_1
			\end{array}\right)  \text{,}
			$
		}
	\end{equation}

	\begin{equation}
		\resizebox{\linewidth}{!}{%
			$
			J_e=
			\left(\begin{array}{ccccc}
					  \vspace{1ex}
					  \displaystyle -\alpha_2 -\mu_2 -\frac{S_2^* \,\eta_2 }{(N^*)^2 }
					  & \displaystyle \frac{S_2^* \,\beta_{22} }{N^*}-\frac{S_2^* \,\eta_2 }{(N^*)^2 }
					  & \displaystyle -\eta_2{\left(\frac{S_2^* }{(N^*)^2 }-\frac{1}{N^*}\right)}
					  & \displaystyle -\frac{\eta_2 }{(N^*)^2 }                                              & \displaystyle -\frac{\eta_2 }{(N^*)^2 }                   \\
					  \vspace{1ex}
					  \alpha_2 & -d_2 -\gamma_2 -\mu_2 & 0   & 0      & 0      \\
					  \vspace{1ex}
					  \displaystyle -\frac{S_2^* \,\eta_2 }{(N^*)^2 }
					  & \displaystyle \frac{S_2^* \,\eta_2 }{(N^*)^2 } -\frac{S_2^* \,\beta_{22} }{N^*}
					  & \displaystyle \eta_2{\left(\frac{S_2^* }{(N^*)^2 }-\frac{1}{N^*}\right)} -v_2 -\mu_2
					  & \displaystyle \frac{\eta_2 }{(N^*)^2 }                                               & \displaystyle \frac{\eta_2 }{(N^*)^2 }                    \\
					  0        & \gamma_2              & 0   & -\mu_2 & 0      \\
					  0        & 0                     & v_2 & 0      & -\mu_2
			\end{array}\right)  \text{.}
			$
		}
	\end{equation}

	Because the same structure for elements in $ J_a $ and $ J_e $, we
	denote them as $M$ as follows:
	\begin{equation}
		M=
		\left(\begin{array}{ccccc}
				  -k_{11} & k_{12}  & k_{13}  & -k_{11} & -k_{11} \\
				  k_{21}  & -k_{22} & 0       & 0       & 0       \\
				  -k_{31} & -k_{12} & -k_{33} & k_{11}  & k_{11}  \\
				  0       & k_{42}  & 0       & -k_{44} & 0       \\
				  0       & 0       & k_{53}  & 0       & -k_{44}
		\end{array}\right).
	\end{equation}
	The corresponding characteristic equation of $ M $ is described as
	follows:
	\begin{equation}
		\label{charployM1}
		\lambda^5+L_1\lambda^4+L_2\lambda^3+L_3\lambda^2+L_4\lambda+L_5=0
		\text{,}
	\end{equation}
	with
	\begin{equation}
		\resizebox{\linewidth}{!}{%
			$
			\begin{array}{lll}
				L_1
				& = & k_{11} +k_{22} +k_{33} +2k_{44} >0\text{,}                                                                                                                                                     \\
				L_2
				& = & k_{11} k_{22} -k_{12} k_{21} +k_{11} k_{33} +k_{13} k_{31} +2k_{11} k_{44} +k_{22} k_{33} -k_{11} k_{53} +2k_{22} k_{44} +2k_{33} k_{44} +{k_{44} }^2                                          \\
				& = & \left( k_{11} k_{22} -k_{12} k_{21} \right)   +k_{11} k_{33} +\left(k_{13} k_{31}-k_{11} k_{53}\right) +2k_{11} k_{44} +k_{22} k_{33}  +2k_{22} k_{44} +2k_{33} k_{44} +{k_{44} }^2
				\text{,} \\
				L_3
				& = & k_{11} {k_{44} }^2 -{k_{11} }^2 k_{53} +k_{22} {k_{44} }^2 +k_{33} {k_{44} }^2 +k_{12} k_{13} k_{21} +k_{11} k_{22} k_{33} -k_{12} k_{21} k_{33}                                               \\
				&   & +k_{13} k_{22} k_{31} +k_{11} k_{21} k_{42} +2k_{11} k_{22} k_{44} -2k_{12} k_{21} k_{44} -k_{11} k_{22} k_{53} +2k_{11} k_{33} k_{44}                                                         \\
				&   & +2k_{13} k_{31} k_{44} -k_{11} k_{31} k_{53} +2k_{22} k_{33} k_{44} -k_{11} k_{44} k_{53}                                                                                                      \\
				& = & k_{11} {k_{44} }^2 + k_{11}\left( k_{33} k_{44} -k_{11} k_{53}\right) +k_{22} {k_{44} }^2 +k_{33} {k_{44} }^2 +k_{12} k_{13} k_{21} +k_{33}\left(k_{11} k_{22}  -k_{12} k_{21}\right)            \\
				&   & + k_{22} \left(k_{13}k_{31}-k_{11} k_{53}\right)  +k_{11} k_{21} k_{42} +2k_{44}\left(k_{11} k_{22} - k_{12} k_{21}\right)  + k_{11} k_{33} k_{44} +k_{31}\left(k_{13}  k_{44} -k_{11} k_{53}\right) \\
				& & +2k_{22} k_{33} k_{44} + k_{44} \left(k_{13} k_{31} -k_{11} k_{53}\right)
				\text{,} \\
				L_4
				& = & k_{11} k_{22} {k_{44} }^2 -k_{12} k_{21} {k_{44} }^2 -{k_{11} }^2 k_{22} k_{53} +k_{11} k_{33} {k_{44} }^2 +k_{13} k_{31} {k_{44} }^2 +k_{22} k_{33} {k_{44} }^2 -{k_{11} }^2 k_{44} k_{53}    \\
				&   & -k_{11} k_{13} k_{21} k_{42} +2k_{12} k_{13} k_{21} k_{44} +k_{11} k_{21} k_{33} k_{42} +2k_{11} k_{22} k_{33} k_{44} -2k_{12} k_{21} k_{33} k_{44}                                            \\
				&   & +2k_{13} k_{22} k_{31} k_{44} -k_{11} k_{22} k_{31} k_{53} +k_{11} k_{21} k_{42} k_{44} -k_{11} k_{22} k_{44} k_{53} -k_{11} k_{31} k_{44} k_{53}                                              \\
				& = & {k_{44} }^2 \left(k_{11} k_{22} -k_{12} k_{21}\right) + \left(k_{12} k_{13} k_{21} k_{44} -{k_{11}}^2 k_{22} k_{53}\right) +k_{11}k_{44}\left( k_{33} k_{44}-k_{11} k_{53}\right)              \\
				&   & + k_{31} k_{44} \left(k_{13} k_{44}-k_{11} k_{53} \right) +k_{22} k_{33} {k_{44} }^2 +k_{12} k_{13} k_{21} k_{44} +k_{11} k_{21} \left(k_{33} k_{42}- k_{13} k_{42}\right)                     \\
				& & +2k_{33} k_{44}\left(k_{11} k_{22}  - k_{12} k_{21} \right) +k_{22} k_{31} \left(k_{13} k_{44} -k_{11} k_{53}\right) +k_{11} k_{21} k_{42} k_{44} + k_{22} k_{44} \left(k_{13} k_{31} - k_{11} k_{53}\right)
				\text{,} \\
				L_5
				& = & k_{12} k_{13} k_{21} {k_{44} }^2 +k_{11} k_{22} k_{33} {k_{44} }^2 -k_{12} k_{21} k_{33} {k_{44} }^2 +k_{13} k_{22} k_{31} {k_{44} }^2                                                         \\
				&   & -{k_{11} }^2 k_{22} k_{44} k_{53} -k_{11} k_{13} k_{21} k_{42} k_{44} +k_{11} k_{21} k_{33} k_{42} k_{44} -k_{11} k_{22} k_{31} k_{44} k_{53}                                                  \\
				& = & k_{44}\left(k_{12}k_{13}k_{21}k_{44}-k_{11}^2k_{22}k_{53}\right)+k_{33}k_{44}^2\left(k_{11}k_{22}-k_{12}k_{21}\right)                                                                          \\
				&   & +k_{22}k_{31}k_{44}\left(k_{13}k_{44}-k_{11}k_{53}\right)+k_{11}k_{21}k_{42}k_{44}\left(k_{33}-k_{13}\right)
				\text{.}
			\end{array}
			$
		}
	\end{equation}

	Based on Descartes' Rule of Signs \cite{Descartes01,Descartes02,Descartes03}, the number of negative roots of the
	characteristic equation \eqref{charployM1} is equal to the number
	of variations in the change in the coefficient sign, so
	\eqref{charployM1} has five negative values if
	\begin{equation}
		\label{EEP_L_condition}
		L_1>0,\quad L_2>0,\quad L_3>0,\quad L_4>0,\quad L_5>0.
	\end{equation}
	The conclusions in \eqref{EEP_L_condition} are valid provided that
	the parameters satisfy:
	\begin{equation}
		\label{EEP_constraint}
		\begin{array}{l}
			k_{12}k_{13}k_{21}k_{44}-k_{11}^2k_{22}k_{53} >0, \quad
			k_{33}-k_{13} >0, \quad
			k_{11}k_{22}-k_{12}k_{21}>0, \\
			k_{13}k_{44}-k_{11}k_{53}>0, \quad
			k_{33}k_{44}-k_{11}k_{53}>0.
		\end{array}
	\end{equation}

	Due to
	\begin{equation*}
		k_{33}-k_{13}=\left[ -\eta_2{\left(\frac{S_2^* }{(N^*)^2
		}-\frac{1}{N^*}\right)} +v_2 +\mu_2\right]-\left[
		-\eta_2{\left(\frac{S_2^* }{(N^*)^2
		}-\frac{1}{N^*}\right)}\right]=v_2 +\mu_2>0,
	\end{equation*}
	then conditions \eqref{EEP_constraint} can be simplified as
	\begin{equation}
		\begin{array}{l}
			k_{12}k_{13}k_{21}k_{44}-k_{11}^2k_{22}k_{53} >0, \quad
			k_{11}k_{22}-k_{12}k_{21}>0, \quad
			k_{13}k_{44}-k_{11}k_{53}>0.
		\end{array}
	\end{equation}
	then condition \eqref{EEP_L_condition} is satisfied as well, so $ \lambda_1, \lambda_2, \lambda_3, \lambda_4 \text{ and } \lambda_5 <0 $.

	In short, the characteristic values of the equation system in our model
	are all negative with constraint \eqref{EEP_constraint}, so the endemic equilibrium point is stable global asymptotic.

	\begin{theorem}[Endemic equilibrium points stability]
		\label{thm:eep_stability}
		The endemic equilibrium point $ X^* $
		is locally asymptotically stable as $ \mathscr{R}_0 <1 $, and the
		endemic equilibrium point $ X^* $ is unstable as $ \mathscr{R}_0
		>1 $, under the conditions \eqref{EEP_constraint}.
	\end{theorem}

	\section{Sensitivity Analysis of Parameters} \label{sec:5}

	In this article, we wonder what and how much the parameters affect the epidemic variable we are interested in. For instance, we want to figure out how much some important parameters influence $ \mathscr{R}_0 $, which can show the importance of vaccination, city lock-down, and some other factors.

	To reach the goal, we turn to sensitivity analysis \cite{Age-structured3,sensitivity2,sensitivity3}. We define the sensitivity index $ \Gamma \left( P \right) $ in which the variable $ \mathscr{R}_0 $ depends on the parameter $ P $. When the variable is a differentiable function of the parameter, the sensitivity index may be alternatively defined using partial derivatives as follows:
	\begin{equation}
		\label{sensitivity_index}
		\Gamma \left( P \right) := \displaystyle \frac{\partial \mathscr{R}_0}{\partial P} \cdot \displaystyle  \frac{P}{\mathscr{R}_0} \text{.}
	\end{equation}

	As we have an explicit formula of $ \mathscr{R}_0 $ in \eqref{eq:R0}, we derive an analytical expression for the sensitivity of it, to each of the seventeen different parameters described in Table \ref{tab:parameters_description}. By this way, we can compare the sensitivity indices of basic reproduction number  $ \mathscr{R}_0 $ with respect to some parameters.

	In figure group \ref{fig:Sensitivity} shown on Appendix \ref{sec:simulation result of sensitivity analysis}, we find that $ \mathscr{R}_0 $ is significantly impacted by $ s_1 $, $ s_2 $, $ v_1 $, $ v_2 $, and $ \beta_{ij} (i, j=1, 2)$, while the impact of the other parameters are much smaller. In this sense we can draw the conclusion that the value change of $ \mathscr{R}_0 $ are mainly caused by them.

	According to Figure \ref{fig:Sensitivity_s_1} and \ref{fig:Sensitivity_s_2}, the basic reproduction ratio show different sensitivity in the age group $\left( < 60 \text{ yr} \right)$ and age group $\left( < 60 \text{ yr} \right)$, because of the proportion of age group $\left( < 60 \text{ yr} \right)$ is several times higher than that of age group $\left( \geqslant 60 \text{ yr} \right)$. On the other hand, due to different scales for horizontal axes between Figure \ref{fig:Sensitivity_s_1} and \ref{fig:Sensitivity_s_2}, actually if there is a small change in the proportion of the elderly group, the basic reproduction ratio will have notable change. Considering that the age structure of a region generally does not change in the short term, strengthening the protection of the elderly population will effectively reduce the value of $ \mathscr{R}_0 $.

	According to Figure \ref{fig:Sensitivity_v_1} and \ref{fig:Sensitivity_v_2}, we can find that to increase vaccination rate can decrease $ \mathscr{R}_0 $, but only when the rate is high enough, it can make significant effect on $ \mathscr{R}_0 $. This reveals a situation where the vaccine alone can completely stop the spread of the virus only if the percentage of the immunized population is high enough; otherwise, the authorities should take other measures to control the outbreak.

	From Figure \ref{fig:Sensitivity_beta_{11}}, \ref{fig:Sensitivity_beta_{12}}, \ref{fig:Sensitivity_beta_{21}} and \ref{fig:Sensitivity_beta_{22}}, we can find that for any $ \beta_{ij} \left( i, j = 1, 2 \right) $, they all have positive influence to $ \mathscr{R}_0 $, we can make such explanation to this phenomenon: the higher value of $ \beta $ represents a higher infectivity of the virus and, therefore, a higher value of $ \mathscr{R}_0 $. On the other hand, due to the proportion of the difference of the two age group and population-to-population differential infectivity, their value is different.

	Hence, we can say that our age-structured model provides a realistic description of COVID-19 transmission in different age groups in the population.

	\section{Numerical Simulation of the Cases in Shijiazhuang} \label{sec:6}

	We investigate the reported cases in Shijiazhuang of Hebei
	Province from January 2 to February 28 of the year 2021, and then
	apply our model to analyze the cases in this area.


	\subsection{Cases in Shijiazhuang}

	At the middle of December 2020, a sudden flare-up emerged in
	Shijiazhuang, a rural city and the capital of Hebei province. More
	than 300 individualas were diagnosed with COVID-19 from January 2
	to January 12 of the year 2021. The authorities had taken measures
	to control epidemics to have promptly reduced the spread of
	COVID-19. Some articles have make some statistical analysis for Shijiazhuang. \cite{sjz01, sjz02, Data_incubation_1}

	The control measures of Shijiazhuang government can be divided into 3 stages:
	\begin{enumerate}[1.]
		\item before January 2 of 2021, the local CDC and the local government are out of
		awareness for a new outbreak;
		\item January 7 of 2021, Shijiazhuang announced to lock down the city; \cite{Data_stage2_0,Data_stage2_1}
		\item January 11 of 2021, the isolation of Xiaoguozhuang village
		and other villages nearby are carried out by the local government. \cite{Data_stage3}
	\end{enumerate}
	The COVID-19 in Shijiazhuang was controlled effectively as of
	January 27. \cite{Data_stage4}

\begin{figure}[h]
		\centering
		\begin{subfigure}[b]{0.45\textwidth}
			\centering
			\includegraphics[width=\textwidth]{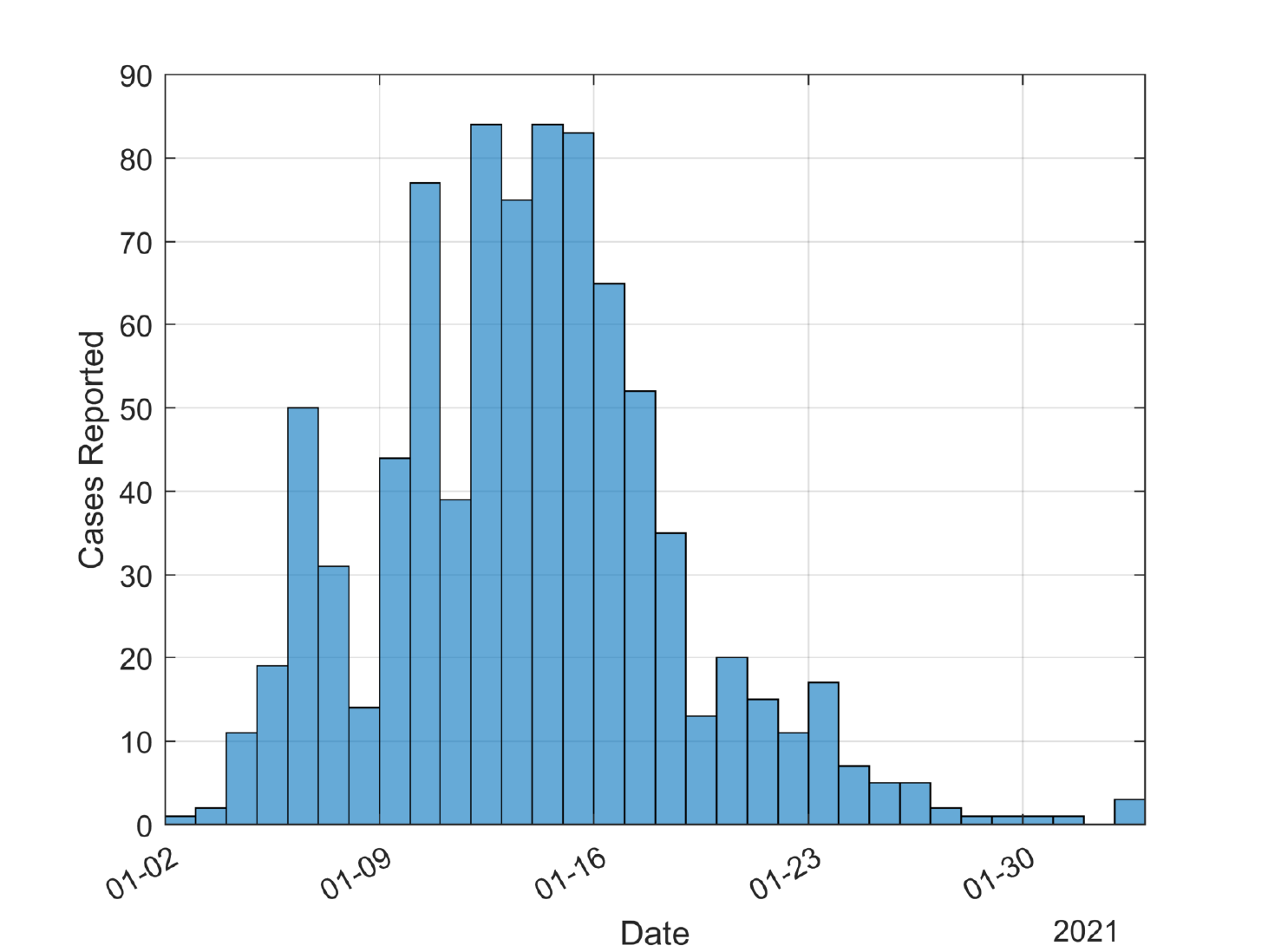}
			\caption{Distribution in time}
			\label{fig:CasesReported}
		\end{subfigure}
		\hfill
		\begin{subfigure}[b]{0.45\textwidth}
			\centering
			\includegraphics[width=\textwidth]{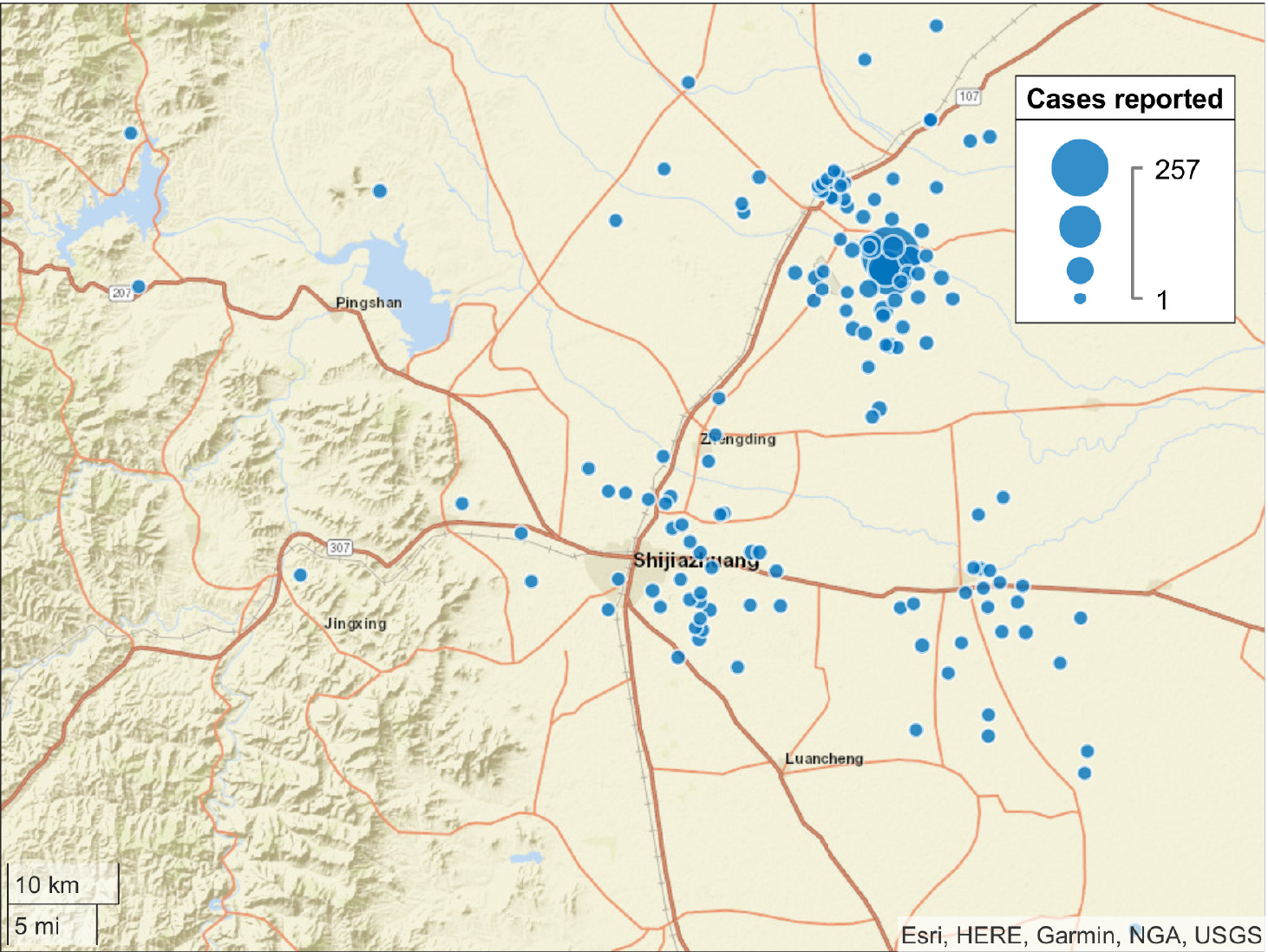}
			\caption{Distribution in space}
			\label{fig:Distribution}
		\end{subfigure}
		\caption{Distributions of the cases reported in time and space}
	\end{figure}

	We collect data from the Health Commission of Hebei Province, and
	then analyze the data in many viewpoints. The number of the daily
	reported cases is plotted in Figure \ref{fig:CasesReported} from
	January 2 to February 3 of the year 2021, and the location where
	the reported cases come from is plotted in Figure
	\ref{fig:Distribution}. It is obvious that the majority of the
	reported cases distributes centered around Xiaoguozhuang village,
	the minor of the reported cases distributes nearby within
	Shijiazhuang city.

\begin{figure}[h]
		\centering
		\begin{subfigure}[b]{0.45\textwidth}
			\centering
			\includegraphics[width=\textwidth]{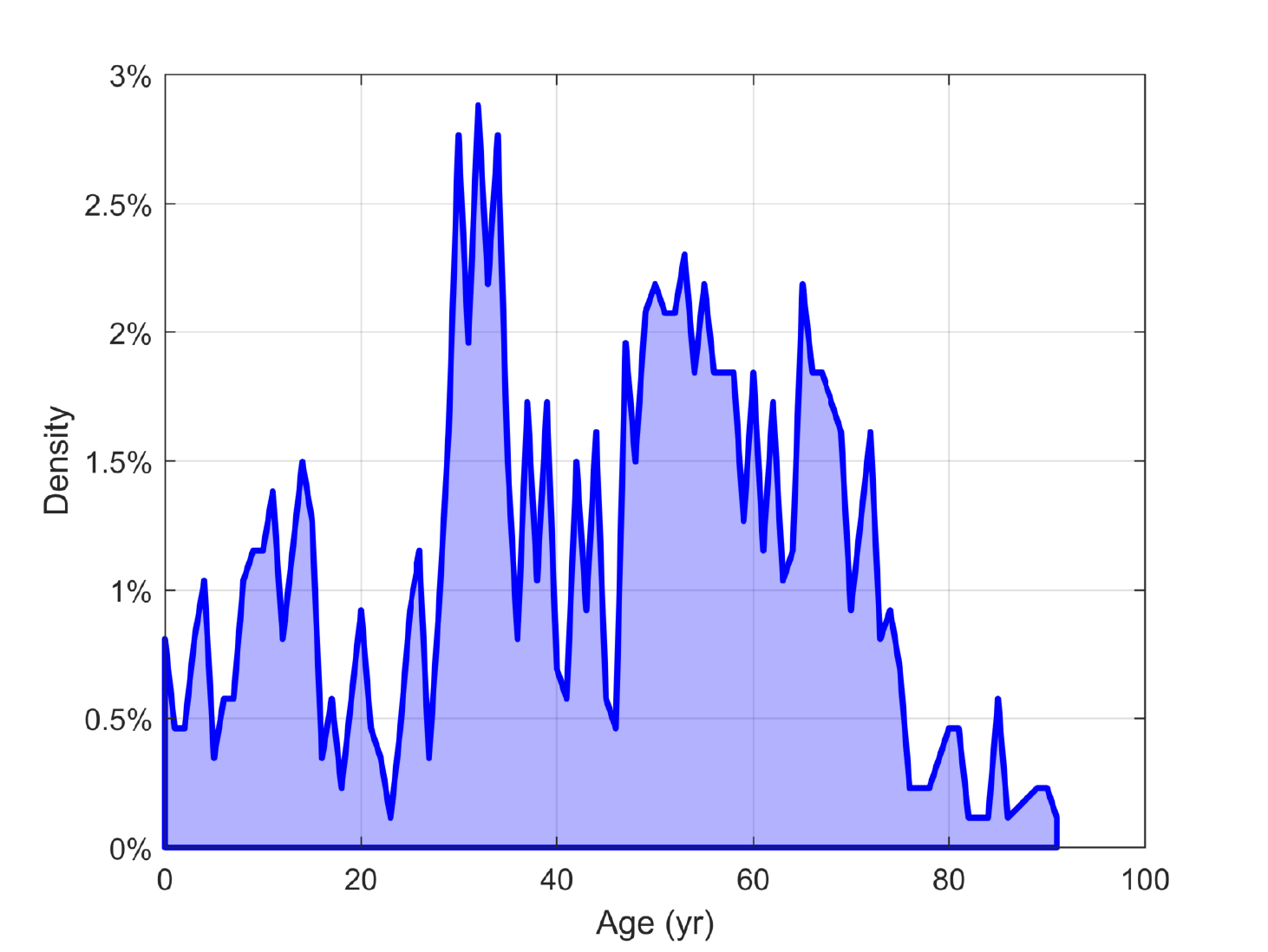}
			\caption{Distribution of ages in year}
			\label{fig:AgeStatistics}
		\end{subfigure}
		\hfill
		\begin{subfigure}[b]{0.45\textwidth}
			\centering
			\includegraphics[width=\textwidth]{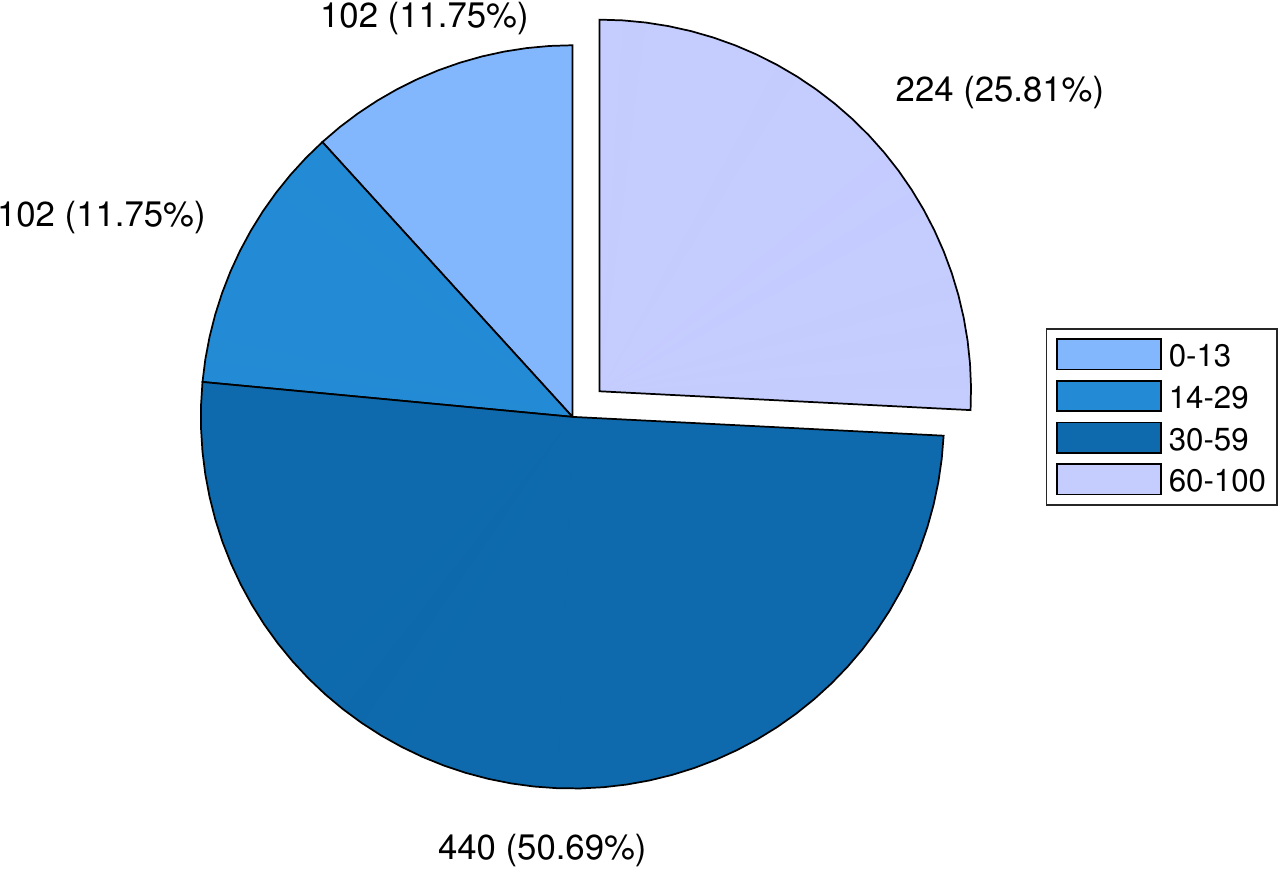}
			\caption{Distribution of ages in group}
			\label{fig:AgeGroup}
		\end{subfigure}
		\caption{Distribution of ages of the cases reported}
	\end{figure}

	We reorganize the reported cases in age in Figure
	\ref{fig:AgeStatistics}. The reported cases over 60 take the
	percentage 25.81\% in Figure \ref{fig:AgeGroup}. We discuss the
	dynamical properties of COVID-19 based on the reported cases in
	Shijiazhuang city. We keep the values of the parameters in
	\ref{tab:parameters_description}, others parameters and their
	corresponding values are presented in Table \ref{tab:parameters_group}.

\begin{figure}[h]
		\centering
		\includegraphics[width=0.9\textwidth]{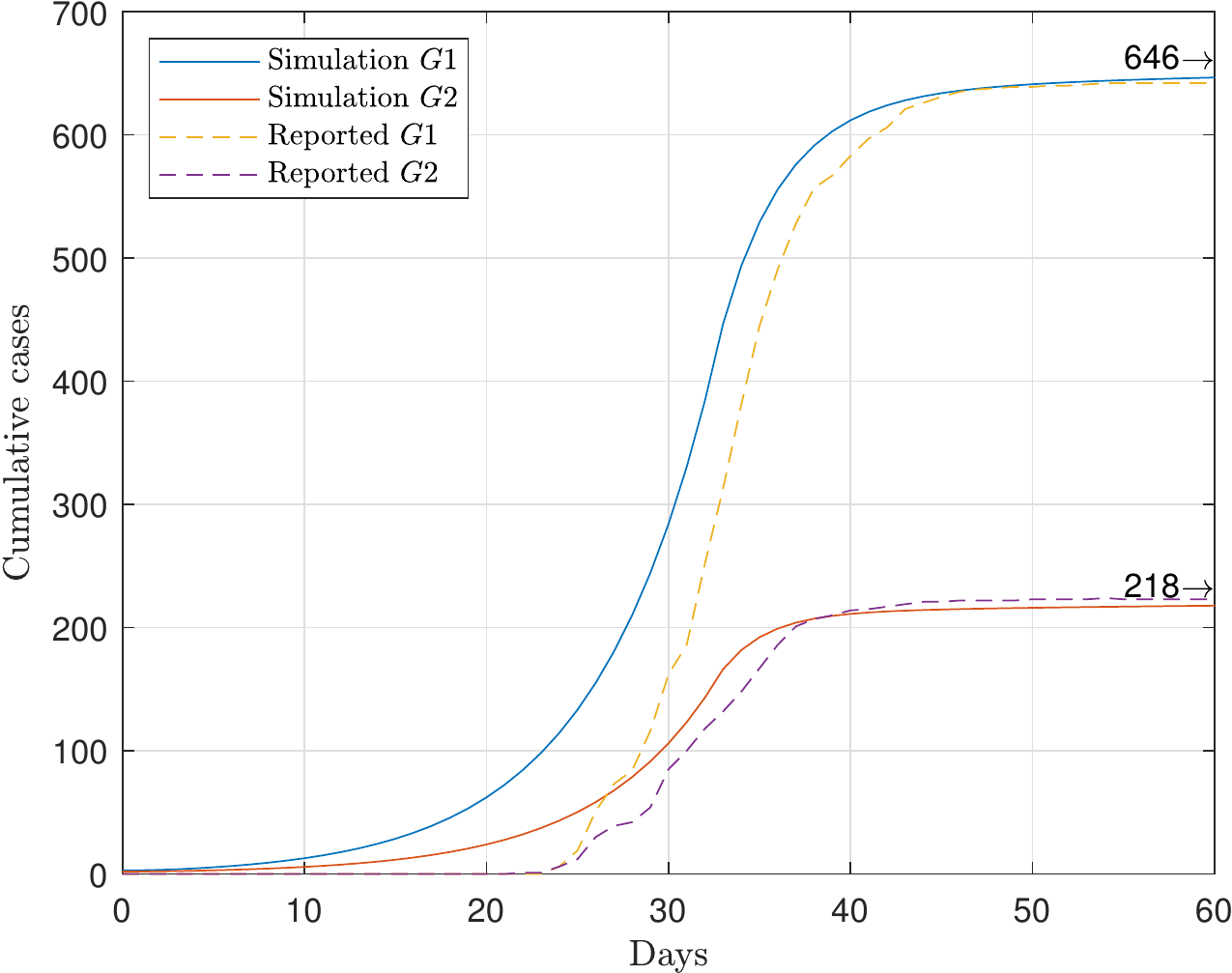}
		\caption{Simulation curves and cumulative reported cases}
		\label{fig:Curve_Fit}
	\end{figure}

	We apply the parameters of Table \ref{tab:parameters_description} to
	simulate the cumulative cases in Shijiazhuang city, where the
	basic production number is $ \mathscr{R}_0 = 8.25 $. Figure
	\ref{fig:Curve_Fit} simulates the number of the cumulative cases
	to fit the real reported confirmed cases for both
	G1 ($<60$yr) and G2 ($\geqslant 60$yr) by days.

	Based on the distributions in time and space and the media reports \cite{Data_stage3}, the individuals in Xiaoguozhuang
	village and villages and towns nearby moved to the remote hotels
	for the centralized quarantine on January 11 of 2021. We assume
	that the infection rates are separated into two-stage step
	functions: the first stage is before January 11, and we take
	$\beta_{11}=0.2880, \beta_{12}=0.3000, \beta_{21}=0.4400,
	\beta_{22}=0.3900$ before the remote centralized quarantine;
	moreover, the second stage is starting from January 12 to February
	6, the chances of contacting with others between the individuals
	are less after January 12, and we take
	$\beta_{11}=0.0014, \beta_{12}=0.0015, \beta_{12}=0.0022, \beta_{22}=0.0020$ for the
	second stage. Further, we provide the daily new cases and the
	current cases for the exposed and the infected, and the cumulative
	cases as well in Figure \ref{fig:Curve_Fit}. 
	On February 6, the new outbreak of COVID-19 in Shijiazhuang
	vanishes \cite{Data_stage4}, the simulations in Figure \ref{fig:Curve_Fit} present the
	coincidence with the control measures of Shijiazhuang government.

	\subsection{More simulations for the dynamical properties of the model}

	Based on the cases in Shijiazhuang city, we adopt more simulations to study the dynamical properties of
	COVID-19 in the rest of this section. The groups of parameters value changed are shown
	in Table \ref{tab:parameters_group}, while others keep the same as Table \ref{tab:parameters_description}.
	\begin{table}[h]
		\begin{center}
			\begin{minipage}{1\textwidth}
				\caption{Parameters and their value used in simulation}\label{tab:parameters_group}%
				\begin{tabular}{@{}lcccc@{}}
					\toprule
					Strategy           & $\beta_{11}$    & $\beta_{12}$  & $\beta_{21}$   & $\beta_{22}$  \\
					\midrule
					Risk-based control      & (0.0288, 0.288) & (0.0300, 0.3) & (0.0440, 0.44) & (0.039, 0.39) \\
					Late quarantine 7 days  & (0.0014, 0.288) & (0.0015, 0.3) & (0.0022, 0.44) & (0.002, 0.39) \\
					High vaccination number & (0.0014, 0.288) & (0.0015, 0.3) & (0.0022, 0.44) & (0.002, 0.39) \\
					Low vaccination number  & (0.0014, 0.288) & (0.0015, 0.3) & (0.0022, 0.44) & (0.002, 0.39) \\
					Lower ageing rate       & (0.0014, 0.288) & (0.0015, 0.3) & (0.0022, 0.44) & (0.002, 0.39) \\
					Higher ageing rate      & (0.0014, 0.288) & (0.0015, 0.3) & (0.0022, 0.44) & (0.002, 0.39) \\
					\botrule
				\end{tabular}
			\end{minipage}

			\addtocounter{table}{-1}
			\begin{minipage}{\textwidth}
				\caption{Parameters and their value used in simulation (\emph{Continued})}
				\begin{tabular*}{\textwidth}{@{\extracolsep{\fill}}lccccc@{}}
					\toprule
					Strategy           & $V_{1}$  & $V_{2}$  & $v_{1}$  & $v_{2}$   & $p$     \\
					\midrule
					Risk-based control      & 0        & 0        & 0        & 0         & 19.38\% \\
					Late quarantine 7 days  & 0        & 0        & 0        & 0         & 19.38\% \\
					High vaccination number & $806200$ & $193800$ & $0.0014$ & $0.00034$ & 19.38\% \\
					Low vaccination number  & $403100$ & $96900$  & $0.0014$ & $0.00034$ & 19.38\% \\
					Low aging rate          & 0        & 0        & 0        & 0         & 14.38\% \\
					High aging rate         & 0        & 0        & 0        & 0         & 24.38\% \\
					\botrule
				\end{tabular*}
			\end{minipage}
		\end{center}
	\end{table}

	\textbf{Variation of infection rate}

	\emph{Scenario A (Risk-based control)}

	We assume that there exist three-level control
	strategy after the remote centralized quarantine on January 11,
	and adopt the decreasing contact rates strategy. For
	instance, we set up a 14-day prevention and control cycle. After
	32 days of free transmission from December 12 of 2020 to January
	11 of 2021 (i.e., 1-32d), then two 14-day control cycles are followed, i.e., from January
	12 of 2021 to January 25 of 2021 (i.e., 33-46d)  and from January
	26 of 2021 to February 8 of 2021 (i.e., 47-60d).
	Finally the transmission of the virus is reduced to the
	minimum level.

	\begin{table}[h]
		\begin{center}
			\begin{minipage}{1\textwidth}
				\caption{Infection rates for medium risk and low risk strategies}\label{tab1}%
				\begin{tabular}{@{}lcccc|lcccc@{}}
					\toprule
					\text{Days} & $\beta_{11}$ & $\beta_{12}$ & $\beta_{21}$ & $\beta_{22}$
					& \text{Days} & $\beta_{11}$ & $\beta_{12}$ & $\beta_{21}$ & $\beta_{22}$ \\
					\midrule
					0-32d & 0.2880 & 0.3000 & 0.4400 & 0.3900
					& 0-32d & 0.2880 & 0.3000 & 0.4400 & 0.3900 \\
					33-46d & 0.1440 & 0.1500 & 0.2200 & 0.1950
					& 33-46d & 0.0720 & 0.0750 & 0.1100 & 0.0975 \\
					47-60d & 0.0288 & 0.0300 & 0.0440 & 0.0390
					& 47-60d & 0.0014 & 0.0015 & 0.0022 & 0.0020 \\
					\botrule
				\end{tabular}
			\end{minipage}
		\end{center}
	\end{table}

	\begin{figure}
		\centering
		\begin{subfigure}[b]{0.45\textwidth}
			\centering
			\includegraphics[width=\textwidth]{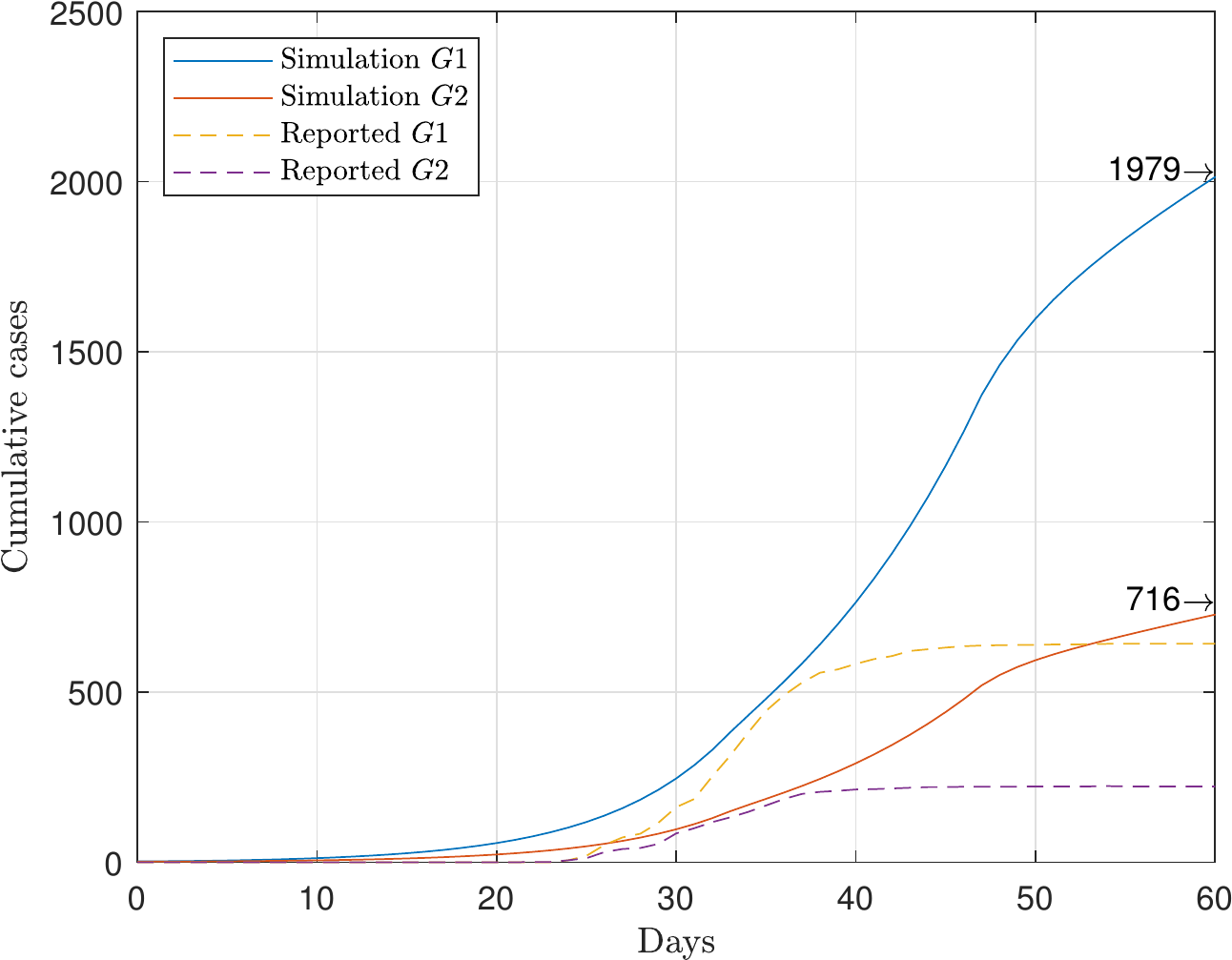}
			\caption{Medium risk level}
			\label{fig:beta_level_1}
		\end{subfigure}
		\hfill
		\begin{subfigure}[b]{0.45\textwidth}
			\centering
			\includegraphics[width=\textwidth]{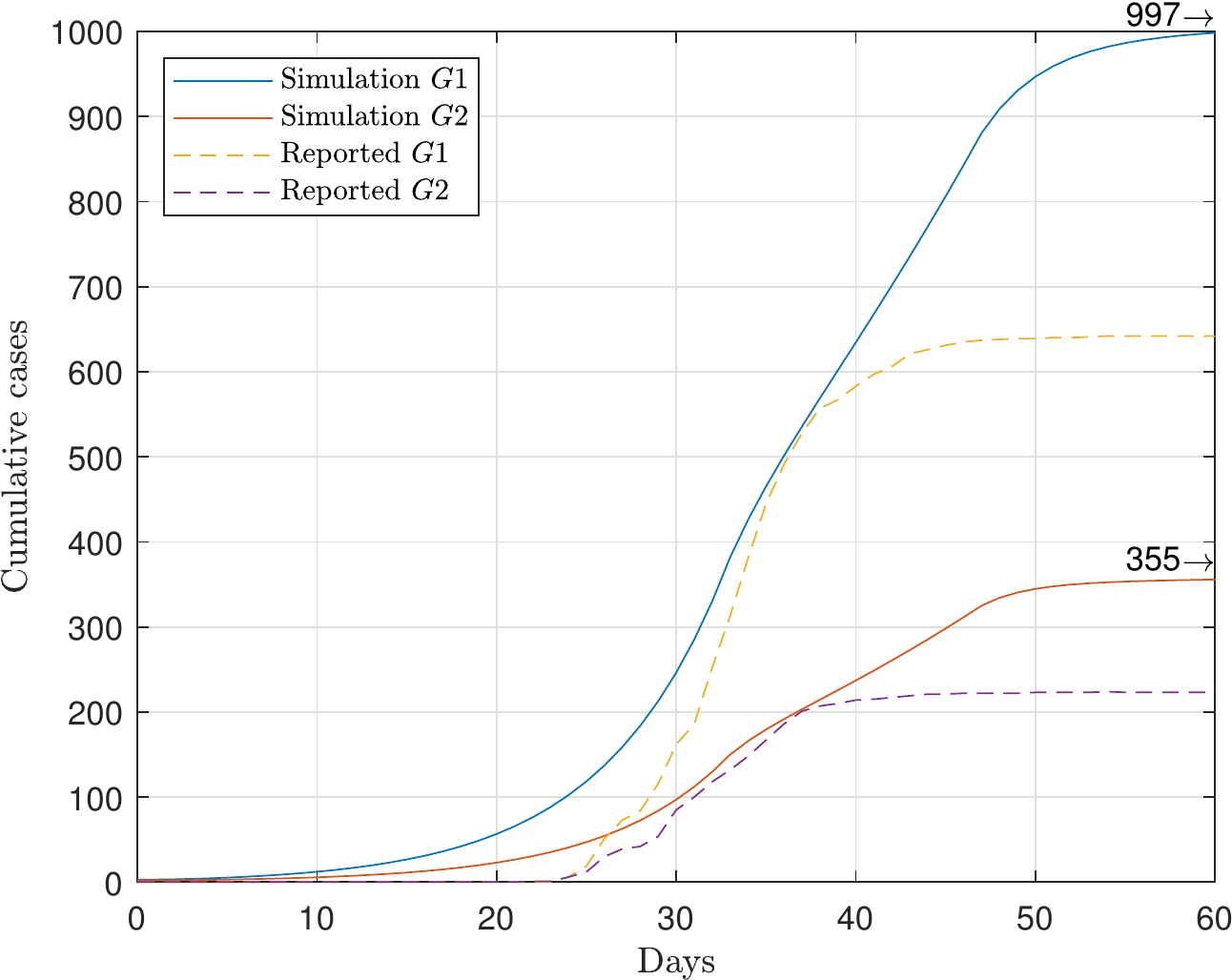}
			\caption{Low risk level}
			\label{fig:beta_level_2}
		\end{subfigure}
		\caption{The impacts of the infection rates with risk-based strategy}
		\label{fig:Change_beta}
	\end{figure}

	\emph{Scenario B (Remote quarantine late strategy)}

\begin{figure}[h]
		\centering
		\begin{subfigure}[b]{0.45\textwidth}
			\centering
			\includegraphics[width=\textwidth]{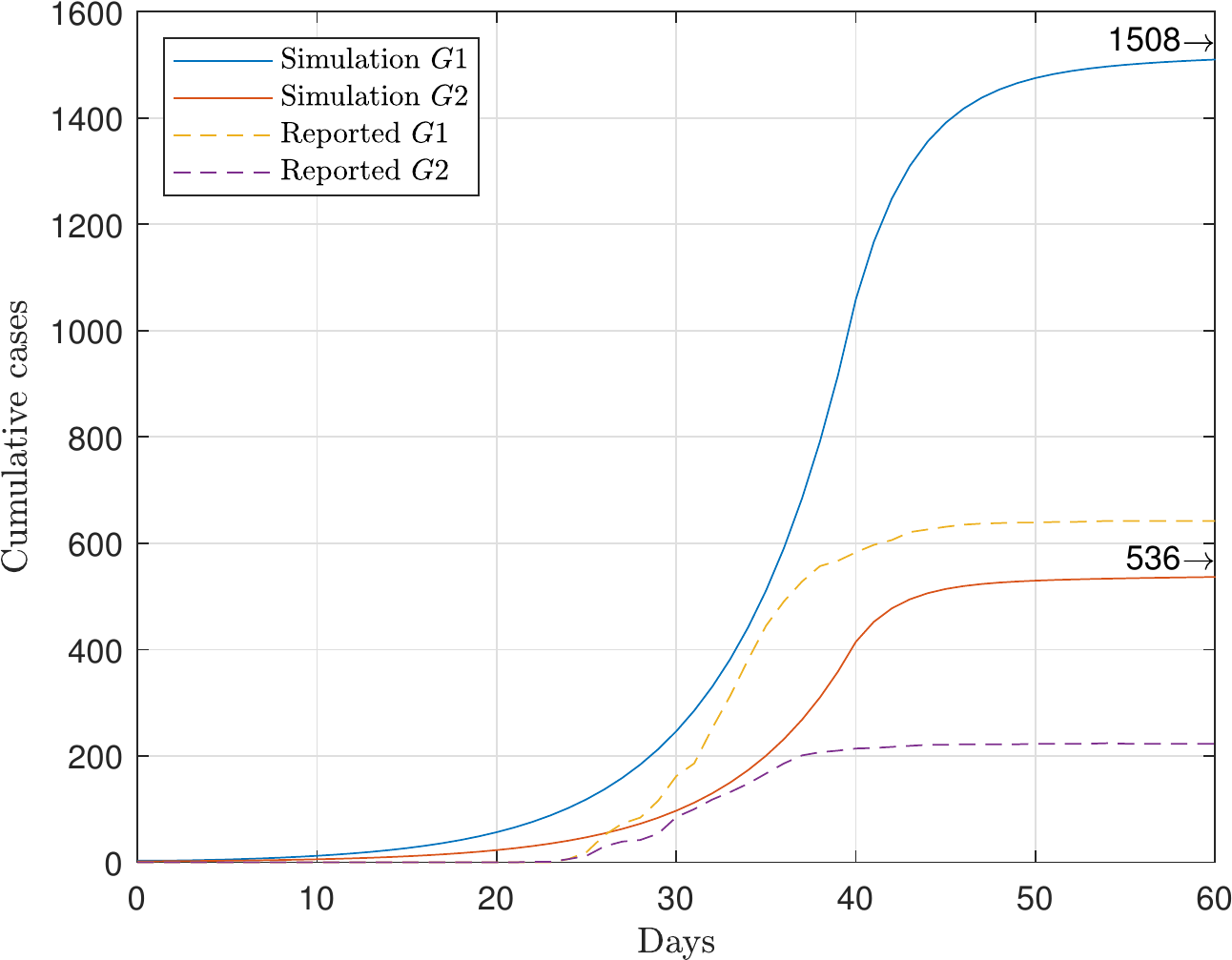}
			\caption{Remote quarantine 7-day late}
			\label{fig:beta_level_3}
		\end{subfigure}
		\hfill
		\begin{subfigure}[b]{0.45\textwidth}
			\centering
			\includegraphics[width=\textwidth]{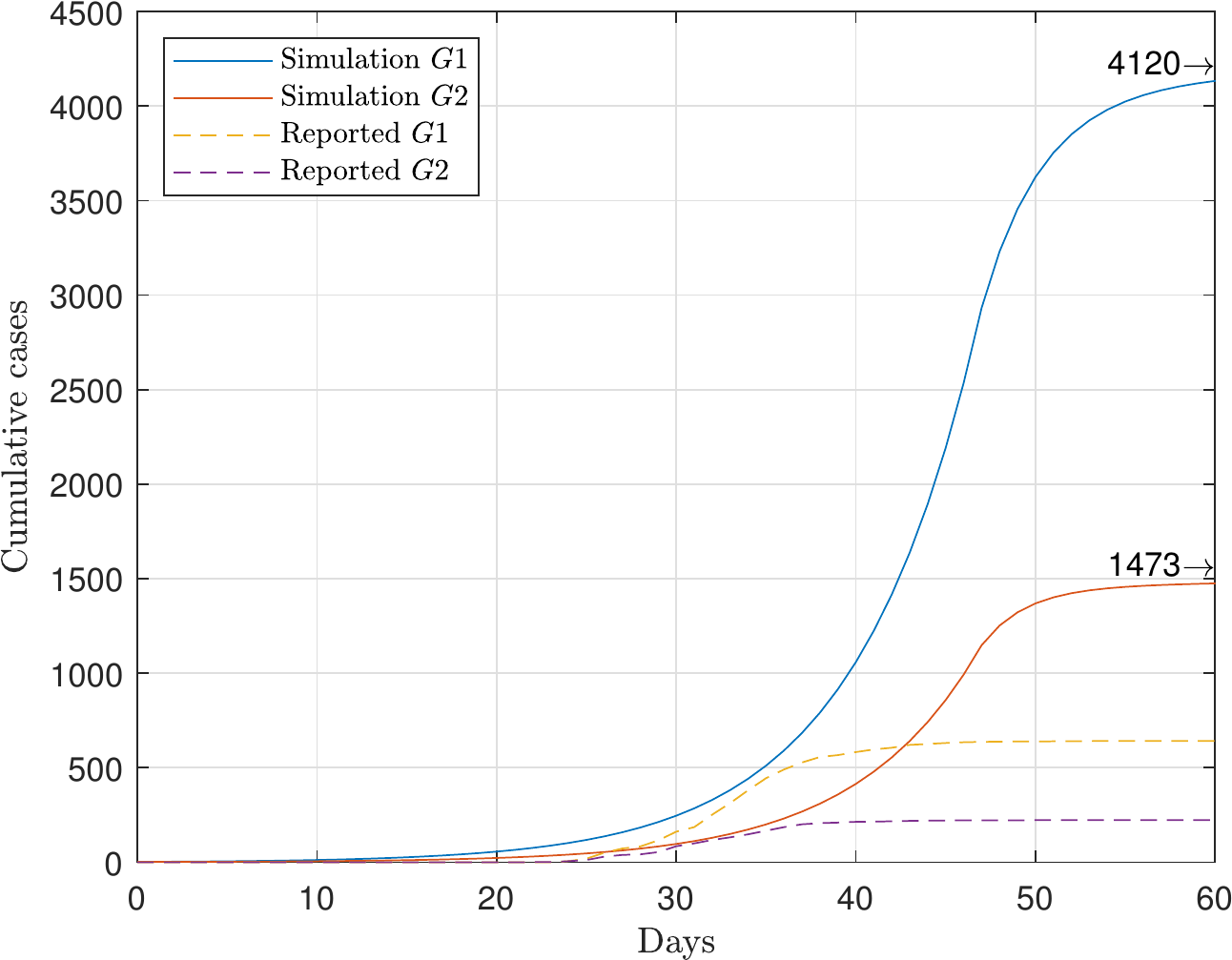}
			\caption{Remote quarantine 14-day late}
			\label{fig:beta_level_4}
		\end{subfigure}
		\caption{The impacts of the infection rates with remote quarantine strategy}
	\end{figure}

	We consider the infection rate in our
	model. Figure \ref{fig:beta_level_4} and Figure \ref{fig:beta_level_2}
	show that the infection rate affect the number of the
	cumulative cases.

	\begin{table}[h]
		\begin{center}
			\begin{minipage}{1\textwidth}
				\caption{Infection rates for remote quarantine 7 days late and 14 days late}
				\begin{tabular}{@{}lcccc|lcccc@{}}
					\toprule
					\text{Days} & $\beta_{11}$ & $\beta_{12}$ & $\beta_{21}$ & $\beta_{22}$
					& \text{Days} & $\beta_{11}$ & $\beta_{12}$ & $\beta_{21}$ & $\beta_{22}$ \\
					\midrule
					0-39d & 0.2880 & 0.3000 & 0.4400 & 0.3900
					& 0-46d & 0.2880 & 0.3000 & 0.4400 & 0.3900 \\
					40-60d & 0.0014 & 0.0015 & 0.0022 & 0.0020
					& 47-60d & 0.0014 & 0.0015 & 0.0022 & 0.0020 \\
					\botrule
				\end{tabular}
			\end{minipage}
		\end{center}
	\end{table}

	\emph{Scenario C (Remote quarantine late and double contact rates strategy)}

	\begin{table}[h]
		\begin{center}
			\begin{minipage}{1\textwidth}
				\caption{Infection rates for remote quarantine and double contact rates}
				\begin{tabular}{@{}lcccc|lcccc@{}}
					\toprule
					\text{Days} & $\beta_{11}$ & $\beta_{12}$ & $\beta_{21}$ & $\beta_{22}$
					& \text{Days} & $\beta_{11}$ & $\beta_{12}$ & $\beta_{21}$ & $\beta_{22}$ \\
					\midrule
					0-39d & 0.5760 & 0.6000 & 0.8800 & 0.7800
					& 0-46d & 0.5760 & 0.6000 & 0.8800 & 0.7800 \\
					40-60d & 0.0014 & 0.0015 & 0.0022 & 0.0020
					& 47-60d & 0.0014 & 0.0015 & 0.0022 & 0.0020 \\
					\botrule
				\end{tabular}
			\end{minipage}
		\end{center}
	\end{table}

	\begin{figure}[h]
		\centering
		\begin{subfigure}[b]{0.45\textwidth}
			\centering
			\includegraphics[width=\textwidth]{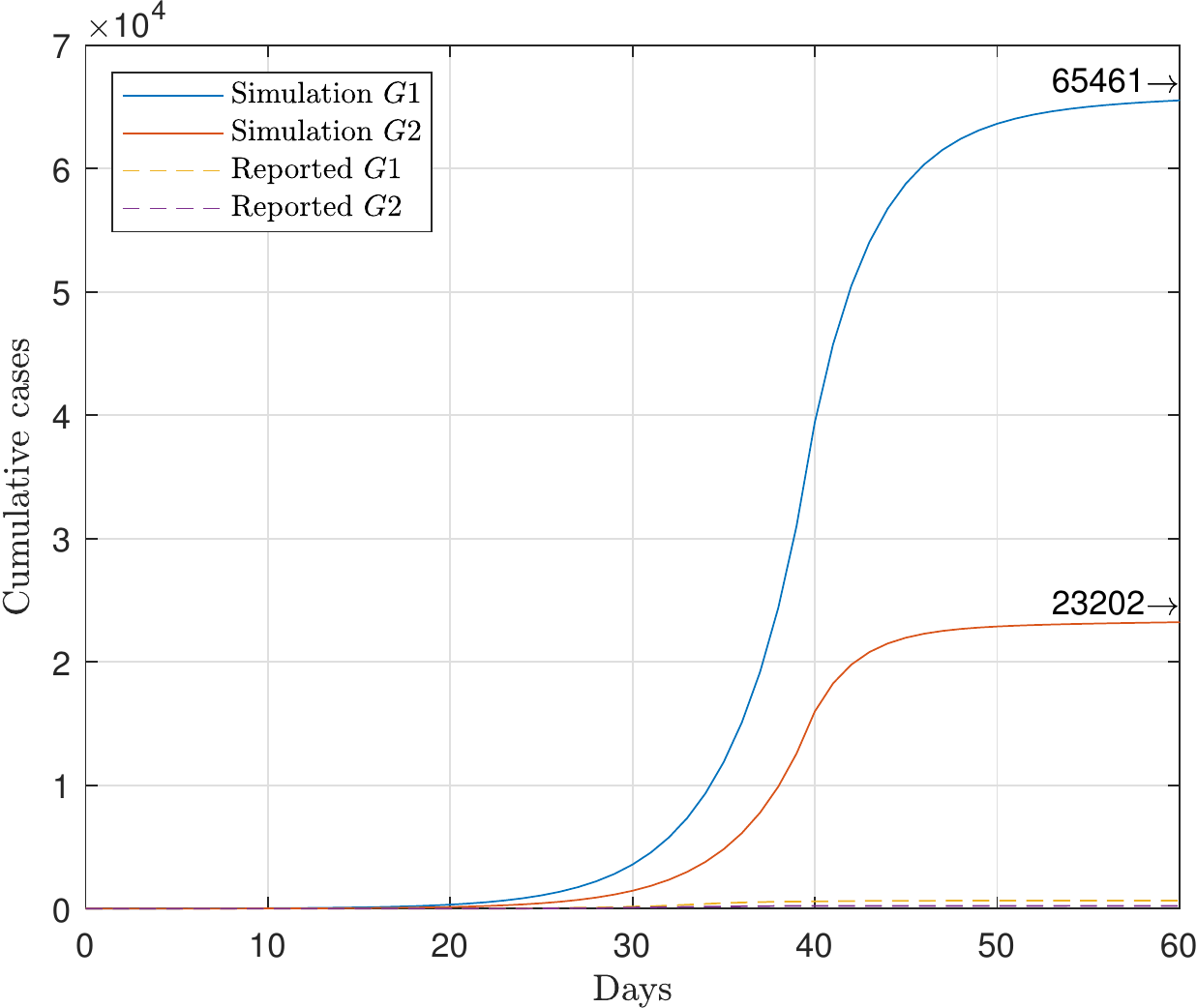}
			\caption{Remote quarantine 7-day late and double contact rates}
			\label{fig:beta_level_5}
		\end{subfigure}
		\hfill
		\begin{subfigure}[b]{0.45\textwidth}
			\centering
			\includegraphics[width=\textwidth]{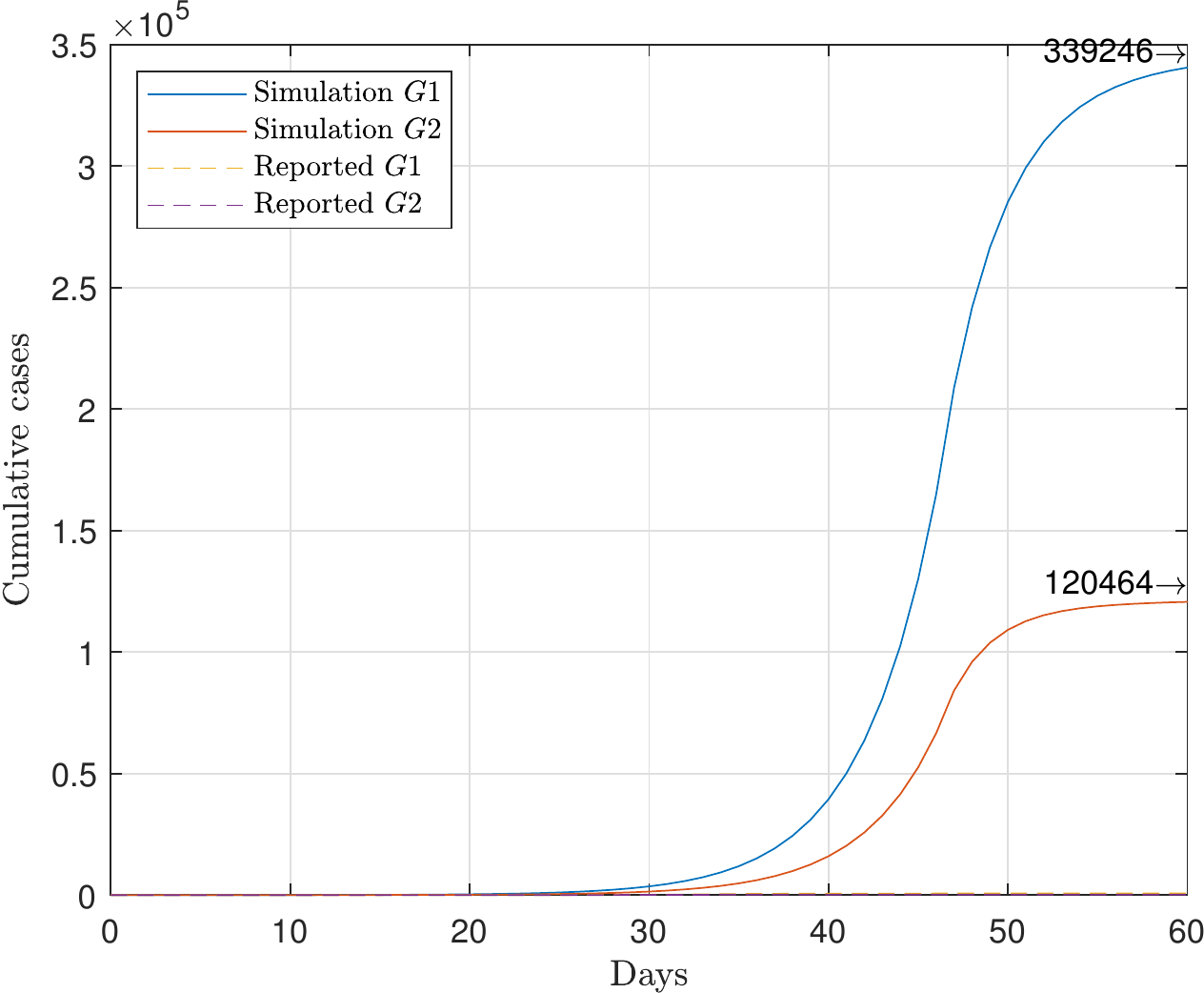}
			\caption{Remote quarantine 14-day late and double contact rates}
			\label{fig:beta_level_6}
		\end{subfigure}
		\caption{The impacts of the infection rates with remote quarantine and double contact rates strategy}
		\label{fig:Change_beta}
	\end{figure}



	\textbf{Variation of vaccination number}

\begin{figure}[h]
		\centering
		\begin{subfigure}[b]{0.45\textwidth}
			\centering
			\includegraphics[width=\textwidth]{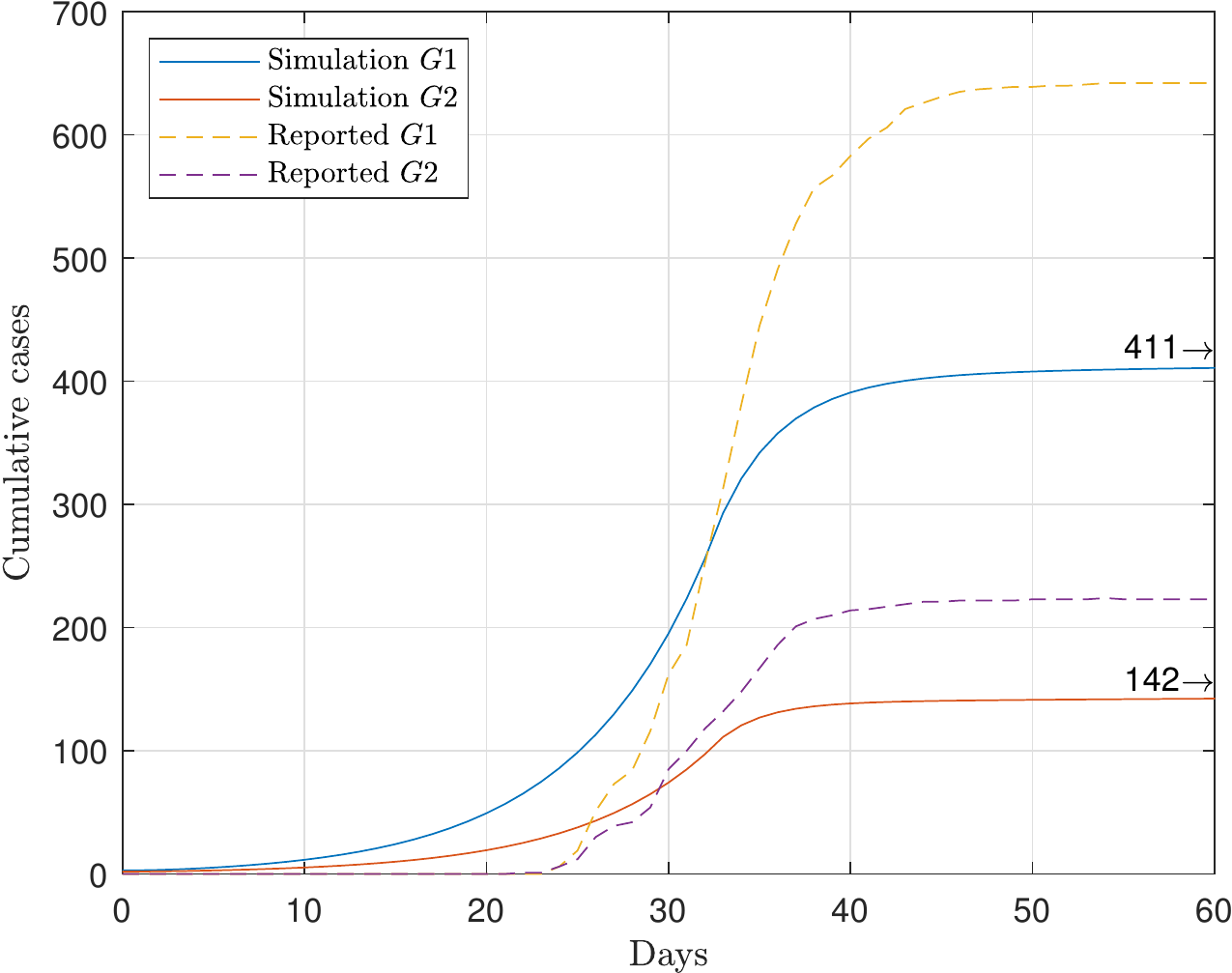}
			\caption{High vaccination number}
			\label{fig:v_high}
		\end{subfigure}
		\hfill
		\begin{subfigure}[b]{0.45\textwidth}
			\centering
			\includegraphics[width=\textwidth]{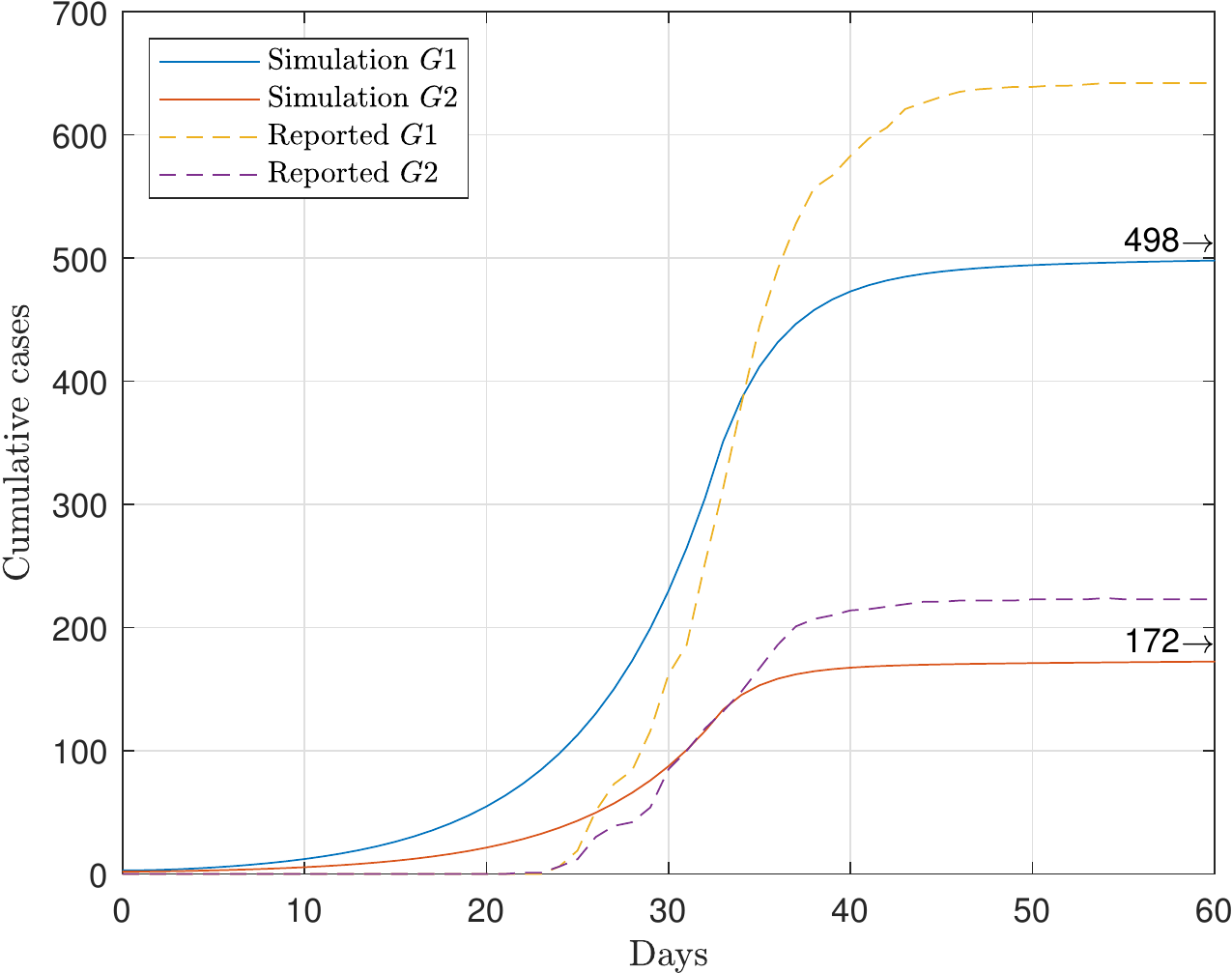}
			\caption{Low vaccination number}
			\label{fig:v_low}
		\end{subfigure}
		\hfill
		\begin{subfigure}[b]{0.45\textwidth}
			\centering
			\includegraphics[width=\textwidth]{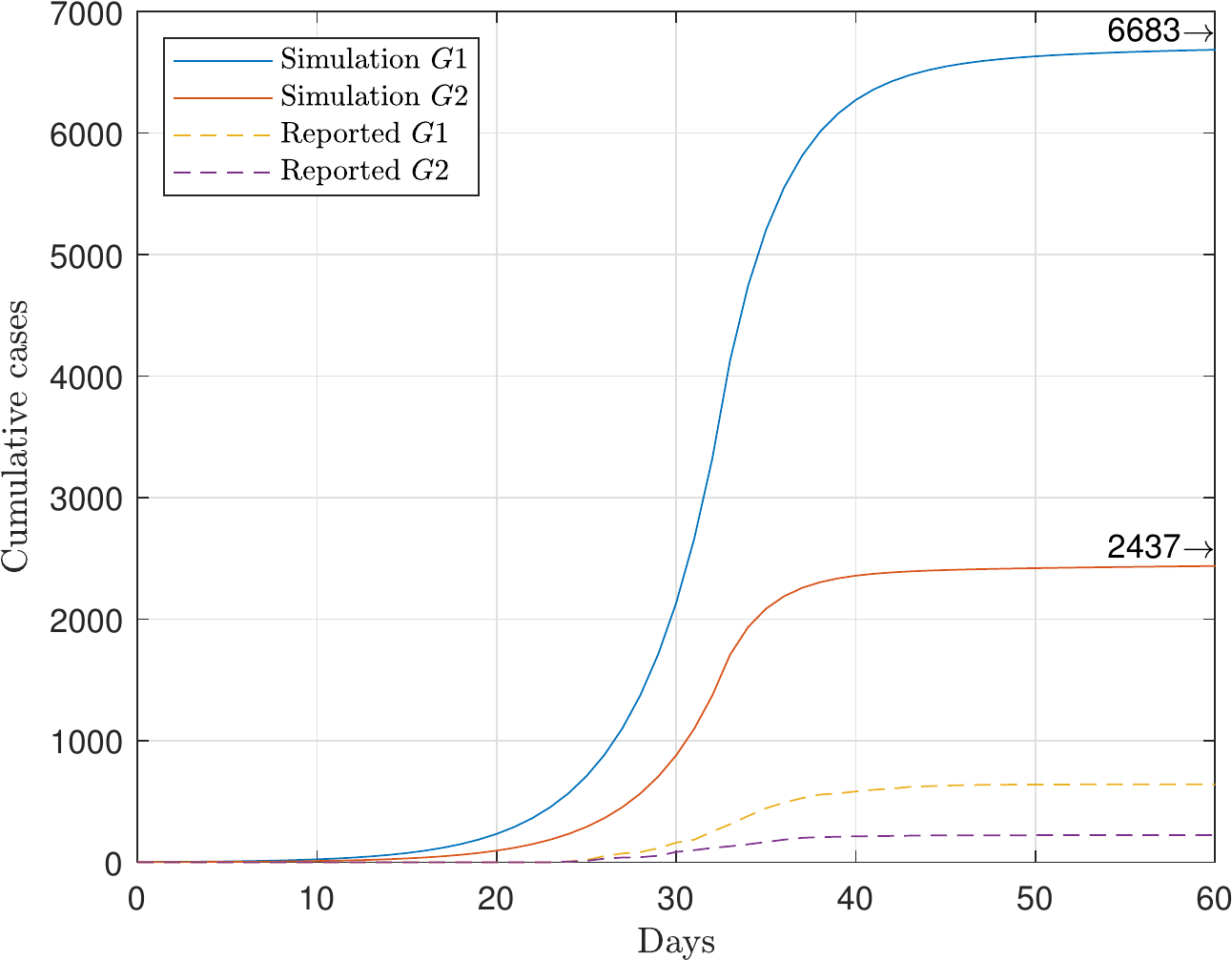}
			\caption{High vaccination number and double contact rates}
			\label{fig:v_high_beta_double}
		\end{subfigure}
		\hfill
		\begin{subfigure}[b]{0.45\textwidth}
			\centering
			\includegraphics[width=\textwidth]{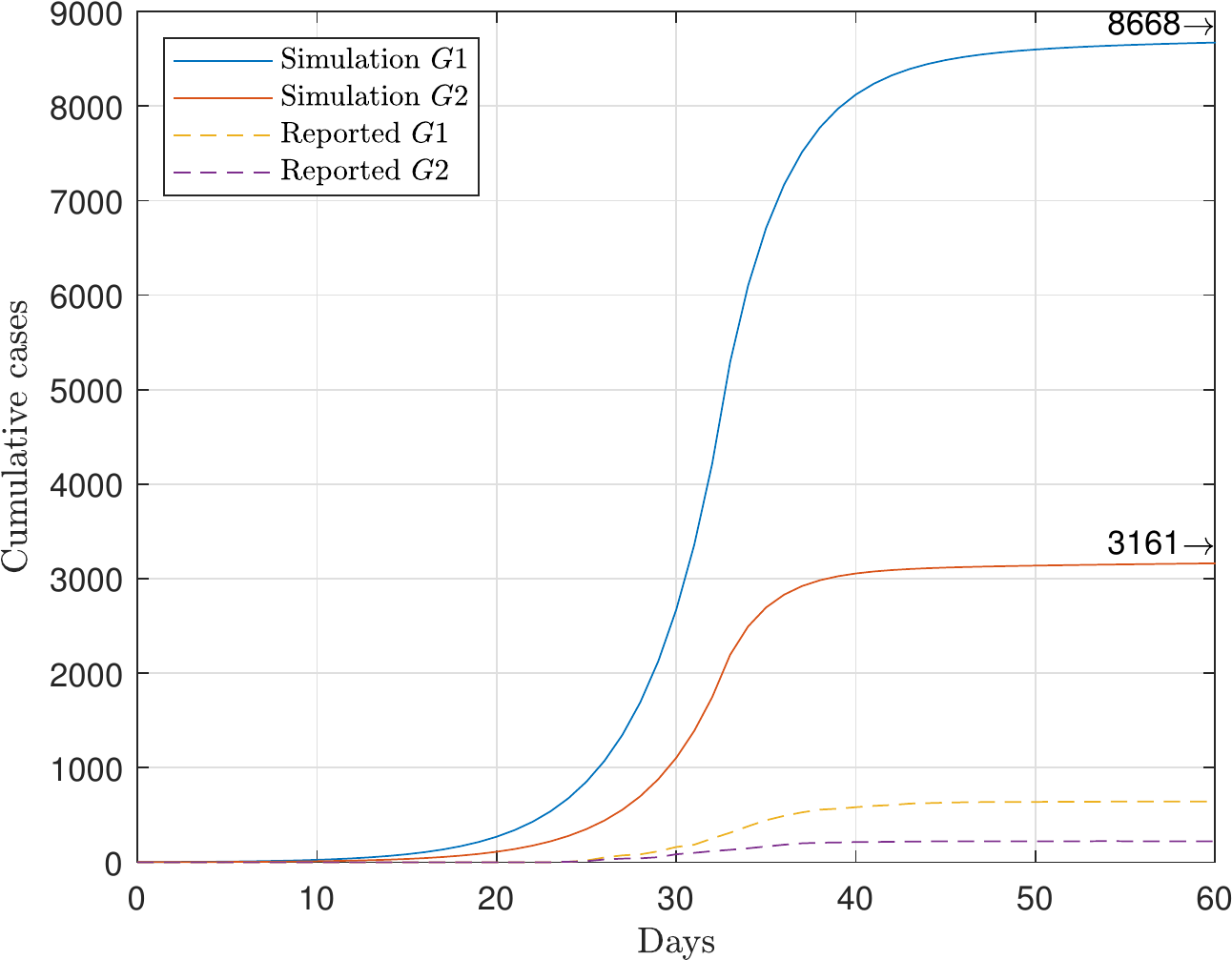}
			\caption{Low vaccination number and double contact rates}
			\label{fig:v_low_beta_double}
		\end{subfigure}
		\caption{The impacts of the vaccination number }
		\label{fig:Change_v}
	\end{figure}

	We consider the vaccination initial values in our model. Figure
	\ref{fig:v_high} shows that vaccination can suppress the the
	number of the cumulative cases. If the vaccination initial value
	is low, the number of the cumulative cases raises at the peak in
	Figure \ref{fig:v_low}.


	\textbf{Variation of population aging rate}

\begin{figure}[h]
		\centering
		\begin{subfigure}[b]{0.45\textwidth}
			\centering
			\includegraphics[width=\textwidth]{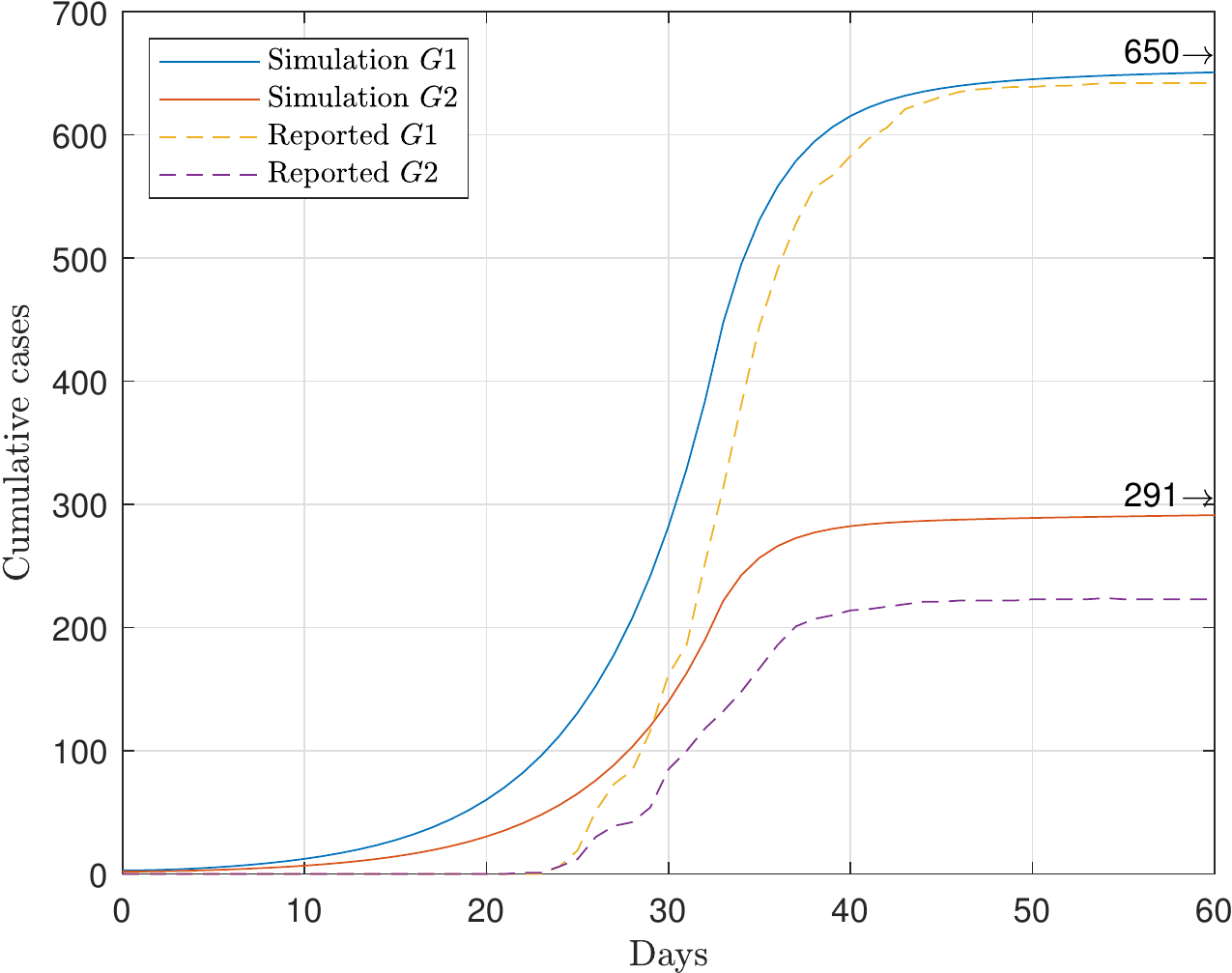}
			\caption{High aging rate 24.38\%}
			\label{fig:p_high}
		\end{subfigure}
		\hfill
		\begin{subfigure}[b]{0.45\textwidth}
			\centering
			\includegraphics[width=\textwidth]{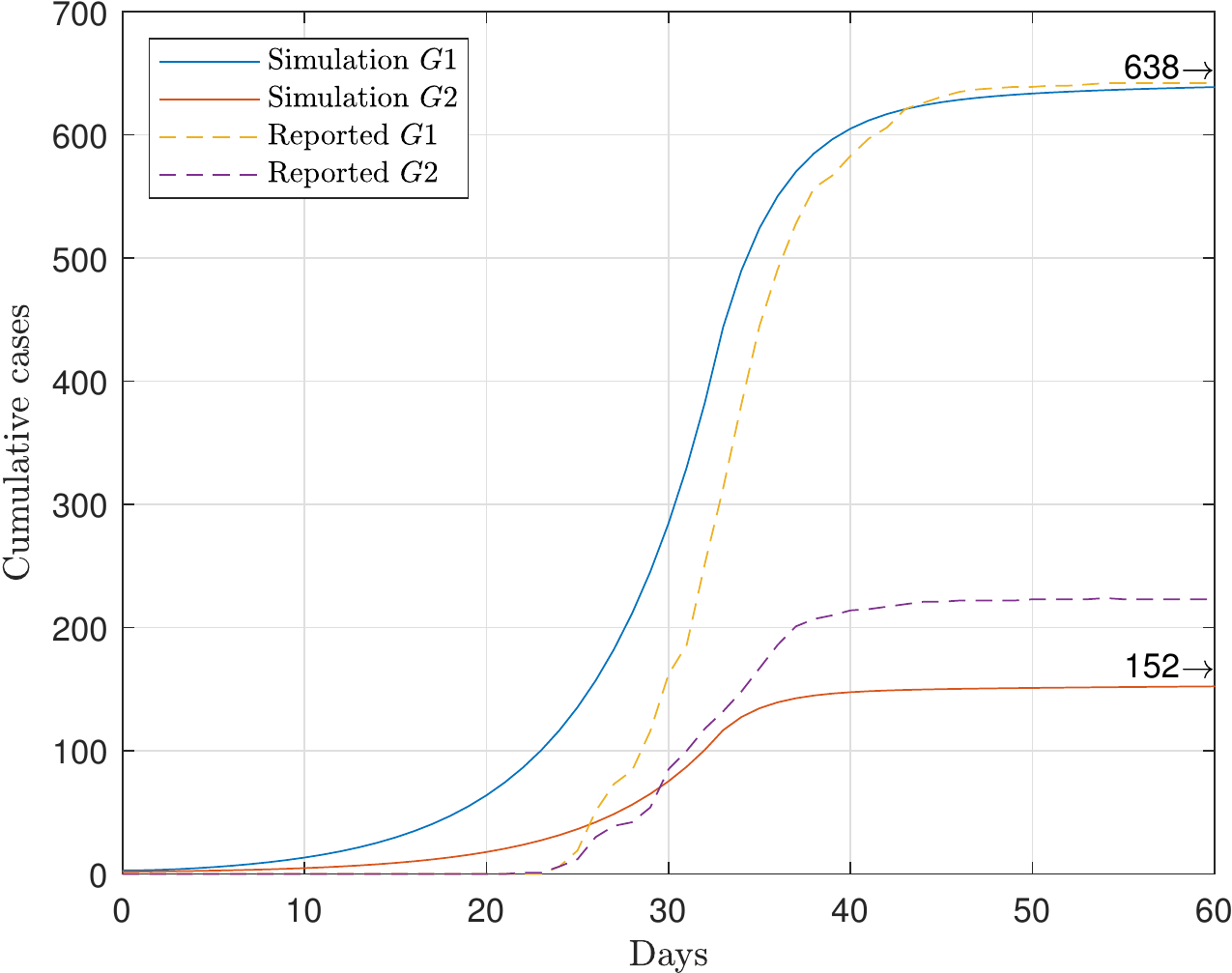}
			\caption{Low aging rate 14.38\%}
			\label{fig:p_low}
		\end{subfigure}
		\hfill
		\begin{subfigure}[b]{0.45\textwidth}
			\centering
			\includegraphics[width=\textwidth]{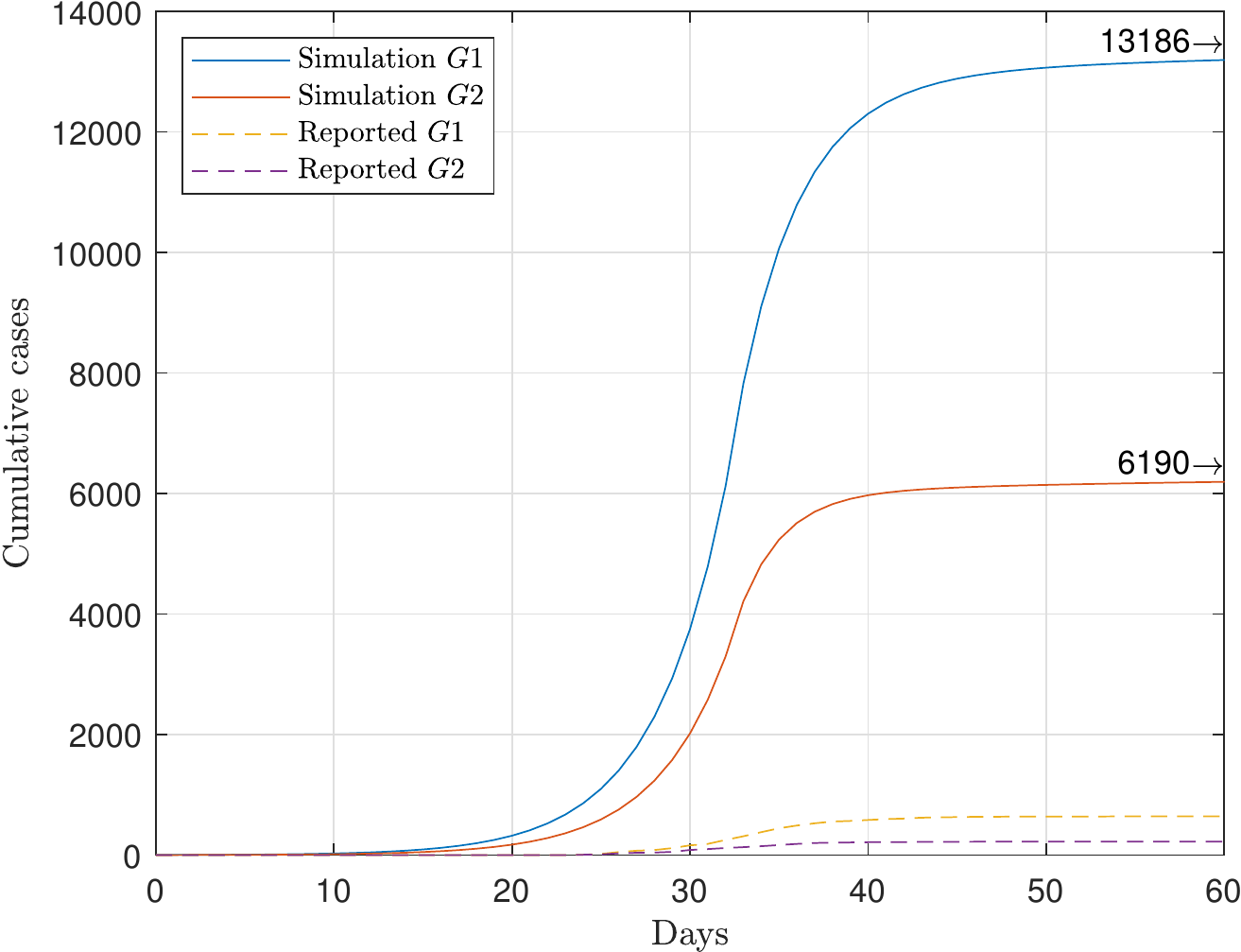}
			\caption{High aging rate 24.38\% and double contact rates}
			\label{fig:p_high_beta_double}
		\end{subfigure}
		\hfill
		\begin{subfigure}[b]{0.45\textwidth}
			\centering
			\includegraphics[width=\textwidth]{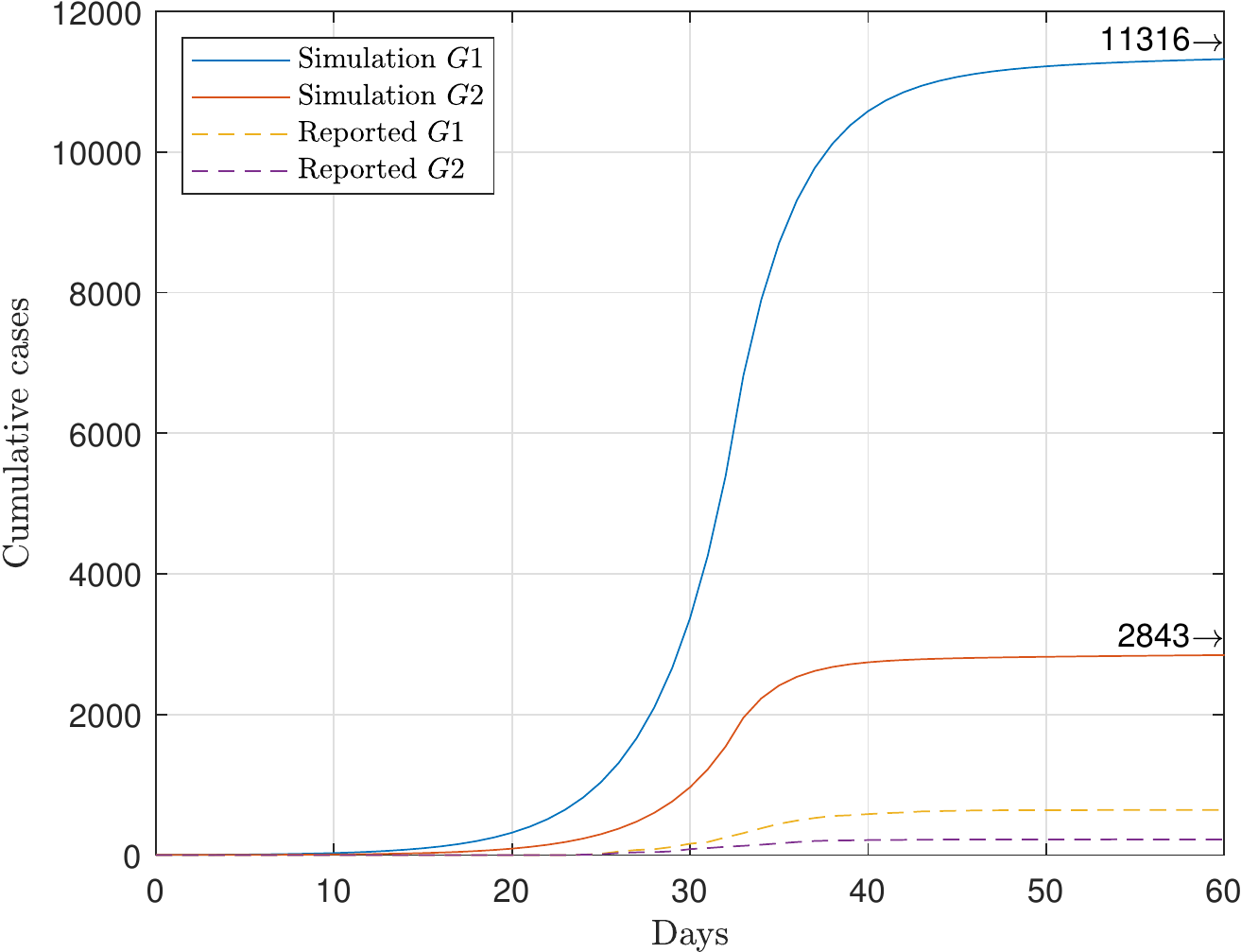}
			\caption{Low aging rate 14.38\% and double contact rates}
			\label{fig:p_low_beta_double}
		\end{subfigure}
		\caption{The impacts of the aging rate}
		\label{fig:Change_aging_rate}
	\end{figure}

	We consider the aging rates of the whole population in
	Shijiazhuang to compare the different impacts to the final size of
	the cumulative cases. Figure \ref{fig:p_high} and Figure
	\ref{fig:p_low} show that the aging rate makes the slight
	impact on the number of G1 $(< 60 \text{ yr})$, and makes the main
	impact on the number of G2 $(\geqslant 60 \text{ yr})$. The
	research results reveal that the government should pay more
	attention on the protection of the elder individuals, which can
	significantly reduce the number of confirmed cases in the elderly
	population.

	\section{Discussion} \label{sec:discussion}

	A finer age structure can reveal the more precise dynamics of COVID-19, but at the same time, more levels of age structure also implies a linear increase in the dimensionality of the equations, which make the model much more complex. As a result, the model is not suitable for the analysis of COVID-19 in the real world. A good idea is to choose the model just right, not only enough finer but also can be performed theoretically.

	We had no detailed information about individuals, such as exposure interval, date of symptom onset and so on. Without these information, we can only get the incubation period and serial interval by estimating rather than by statistical analysis. For example, incubation period might be overestimated.\cite{Data_incubation_2, Data_incubation_3} But this inaccuracy is undercontrol by reference to others studies.

	The people in the rural area frequently attended at large gatherings, and when they find theirselves had symptoms, they choosed to not to see a doctor or buy medicines in village hospital or individual clinic. This caused the outbreak to spread quickly and hard to consult early cases. The COVID-19 outbreak in Shijiazhuang is a outbreak about the rural area.\cite{Data_incubation_1} We also find this character in the simulation by our model, so the key to control the outbreak is to control the rural area. The
	governance capacity of the government is different between town and rural area. If the goverment didnot inversitigate the character mentioned above of this outbreak, they will lose the focus.

	\section{Conclusion} \label{sec:conclusion}

	In summary, we construct an age-structured model for COVID-19
	with vaccination and analyze it from multiple perspectives. We
	derive the unique disease-free equilibrium point and the basic
	reproduction number $ \mathscr{R}_0 $, then we show that the
	disease-free equilibrium is locally asymptotically stable when $
	\mathscr{R}_0 <1 $, while is unstable when $ \mathscr{R}_0 >1 $.
	We work out endemic equilibrium points and reveal their
	stability. We use sensitivity analysis to find out the influence of
	parameters to $ \mathscr{R}_0 $ which can
	help us develop more targeted strategies to control epidemics.
	Finally, this model is used to discuss the cases in Shijiazhuang,
	Hebei Province at the beginning of 2021. our study suggests the lack of anti-epidemic awareness in rural areas need to pay more attendtion to. The targeted and effective intervention is the key to control COVID-19.

	\backmatter

	\bmhead{Acknowledgments}

	The research is supported by National Natural Science Foundation
	of China (61911530398), Special Projects of the Central
	Government Guiding Local Science and Technology Development
	(2021L3018), the Natural Science Foundation of Fujian Province of
	China (2021J01621) and Scientific Research Training Program in Fuzhou University (No. 26040).

	%

	\begin{appendices}

		\section{Block matrices} \label{sec:Block matrices}
		We define
		\begin{equation}
			\begin{array}{l}
				\eta_1
				=I_1 \,\beta_{11} +I_2 \,\beta_{12}
				\text{, }
				\\
				\eta_2
				=I_1 \,\beta_{21} +I_2 \,\beta_{22}
				\text{. }
			\end{array}
		\end{equation}

		Now the block matrices of Jacobian matrix can be written as follows:
		\begin{equation}
			\begin{array}{l}
				J_{EE}
				=\left(\begin{array}{cc}
						   \vspace{1ex}
						   -\alpha_1 -g-\mu_1 -\displaystyle \frac{S_1 \,\eta_1}{N^2 } & -\displaystyle \frac{S_1 \,\eta_1 }{N^2 }                 \\ \vspace{1ex}
						   g-\displaystyle \frac{S_2 \,\eta_2}{N^2 }                         & -\alpha_2 -\mu_2 -\displaystyle \frac{S_2 \,\eta_2}{N^2 }
				\end{array}\right)
				\text{, }
			\end{array}
		\end{equation}

		\begin{equation}
			\begin{array}{l}
				J_{EI}
				=\begin{array}{l}
					 \left(\begin{array}{cc}
							   \vspace{1ex}
							   \displaystyle \frac{S_1 \,\beta_{11} }{N}-\displaystyle \frac{S_1 \,\eta_1}{N^2 } & \displaystyle \frac{S_1 \,\beta_{12} }{N}-\displaystyle \frac{S_1 \,\eta_1}{N^2 } \\\vspace{1ex}
							   \displaystyle \frac{S_2 \,\beta_{21} }{N}-\displaystyle \frac{S_2 \,\eta_2}{N^2 } & \displaystyle \frac{S_2 \,\beta_{22} }{N}-\displaystyle \frac{S_2 \,\eta_2}{N^2 }
					 \end{array}\right)
				\end{array}
				\text{, }
			\end{array}
		\end{equation}

		\begin{equation}
			\begin{array}{l}
				J_{ES}
				=-\eta_1
				\left(\begin{array}{cc}
						  \vspace{1ex}
						  \displaystyle \frac{S_1 }{N^2 }-\displaystyle \frac{1}{N} & \displaystyle \frac{S_1}{N^2 }                            \\\vspace{1ex}
						  \displaystyle \frac{S_2}{N^2 }                                  & \displaystyle \frac{S_2 }{N^2 }-\displaystyle \frac{1}{N}
				\end{array}\right)
				\text{, }
			\end{array}
		\end{equation}

		\begin{equation}
			\begin{array}{l}
				J_{ER}=J_{EV}
				=-\displaystyle \frac{1}{N^2 }
				\left(
				\begin{array}{cc}
					\vspace{1ex}
					\eta_1 & \eta_1 \\\vspace{1ex}
					\eta_2      & \eta_2
				\end{array}
				\right)
				\text{, }
			\end{array}
		\end{equation}

		\begin{equation}
			\begin{array}{l}
				J_{IE}
				=\left(\begin{array}{cc}
						   \alpha_1 & 0        \\
						   0        & \alpha_2
				\end{array}\right)
				\text{, }
			\end{array}
		\end{equation}

		\begin{equation}
			\begin{array}{l}
				J_{II}
				=\left(\begin{array}{cc}
						   -d_1 -g-\gamma_1 -\mu_1 & 0                     \\
						   g                       & -d_2 -\gamma_2 -\mu_2
				\end{array}\right)
				\text{, }
			\end{array}
		\end{equation}

		\begin{equation}
			\begin{array}{l}
				J_{SE}
				=-\displaystyle \frac{1}{N^2 }
				\left(
				\begin{array}{cc}
					\vspace{1ex}
					S_1 \,\eta_1 & S_1 \,\eta_1 \\\vspace{1ex}
					S_2 \,\eta_2      & S_2 \,\eta_2
				\end{array}
				\right)
				\text{, }
			\end{array}
		\end{equation}

		\begin{equation}
			\begin{array}{l}
				J_{SI}=-J_{EI}
				\text{, }
			\end{array}
		\end{equation}

		\begin{equation}
			\begin{array}{l}
				J_{SS}
				=\begin{array}{l}
					 \left(
					 \begin{array}{cc}
						 \vspace{1ex}
						 \displaystyle \frac{S_1 \,\eta_1}{N^2 } -\displaystyle \frac{\eta_1}{N}-\left(v_1+\mu_1+g\right) & 0                                                                                              \\\vspace{1ex}
						 g                                                                                                       & \displaystyle \frac{S_2 \,\eta_2}{N^2 } -\displaystyle \frac{\eta_2}{N}-\left(v_2+\mu_2\right)
					 \end{array}
					 \right)
				\end{array}
				\text{, }
			\end{array}
		\end{equation}

		\begin{equation}
			\begin{array}{l}
				J_{SR}=J_{SV}=-J_{ER}
				\text{, }
			\end{array}
		\end{equation}

		\begin{equation}
			\begin{array}{l}
				J_{RI}
				=\left(\begin{array}{cc}
						   \gamma_1 & 0        \\
						   0        & \gamma_2
				\end{array}\right)
				\text{, }
				J_{RR}
				=\left(\begin{array}{cc}
						   -g-\mu_1 & 0      \\
						   g        & -\mu_2
				\end{array}\right)
				\text{, }
			\end{array}
		\end{equation}

		\begin{equation}
			\begin{array}{l}
				J_{VS}
				=\left(\begin{array}{cc}
						   v_1 & 0   \\
						   0   & v_2
				\end{array}\right)
				\text{, }
				J_{VV}
				=\left(\begin{array}{cc}
						   -g-\mu_1 & 0      \\
						   g        & -\mu_2
				\end{array}\right)
			\end{array}
			\text{. }
		\end{equation}

		While the other matrices $ J_{IS}$, $J_{IR}$, $J_{IV}$, $J_{RE}$, $J_{RS}$, $J_{RV}$, $J_{VE}$, $J_{VI}$, and $J_{VR}$ are equal to zeros.

		\section{Simulation Result of Sensitivity Analysis} \label{sec:simulation result of sensitivity analysis}

		\begin{figure}[h]
			\centering
			\begin{subfigure}[b]{0.23\textwidth}
				\centering
				\includegraphics[width=\textwidth]{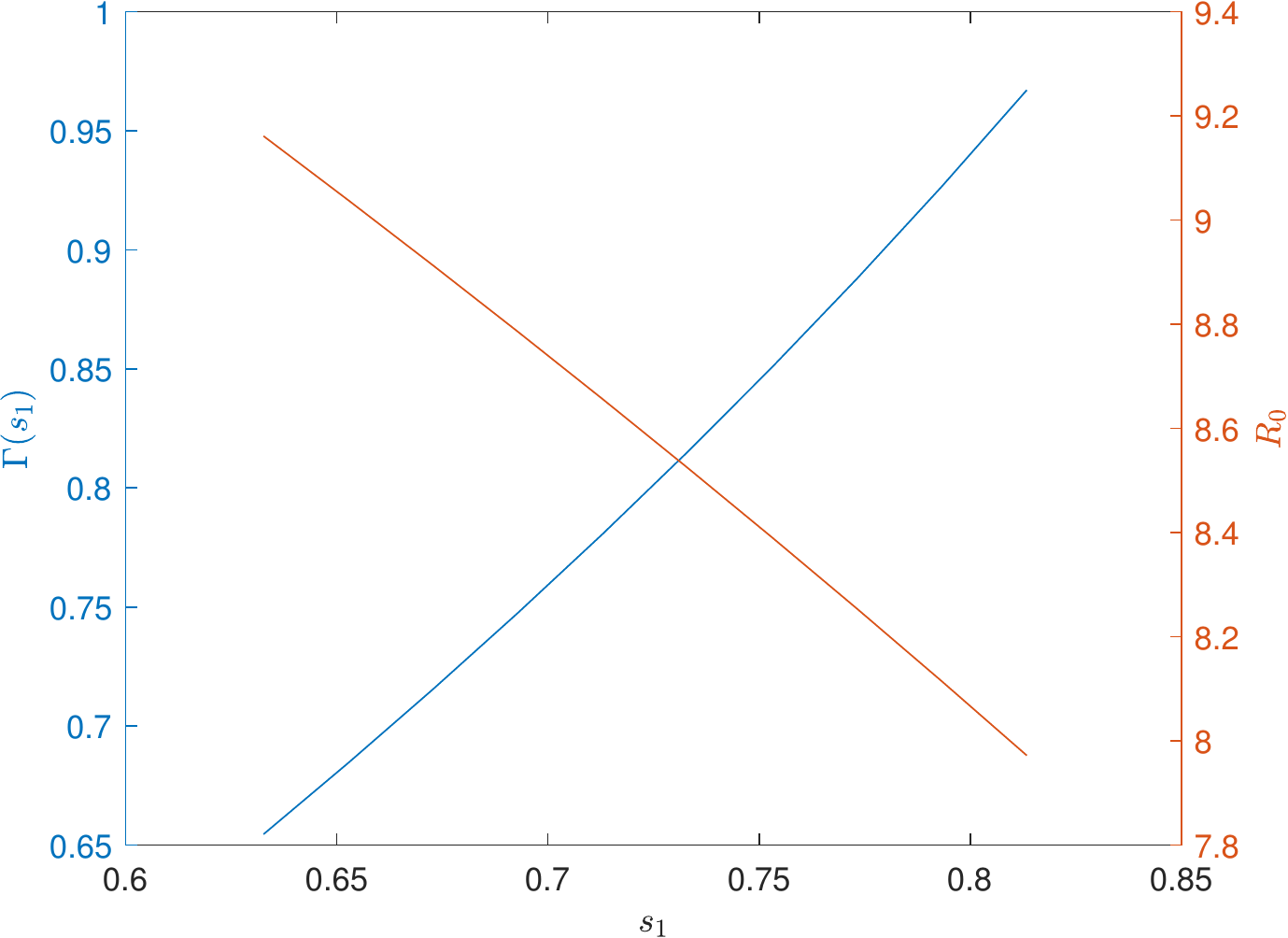}
				\caption{$ S_1 / N $}
				\label{fig:Sensitivity_s_1}
			\end{subfigure}
			\hfill
			\begin{subfigure}[b]{0.23\textwidth}
				\centering
				\includegraphics[width=\textwidth]{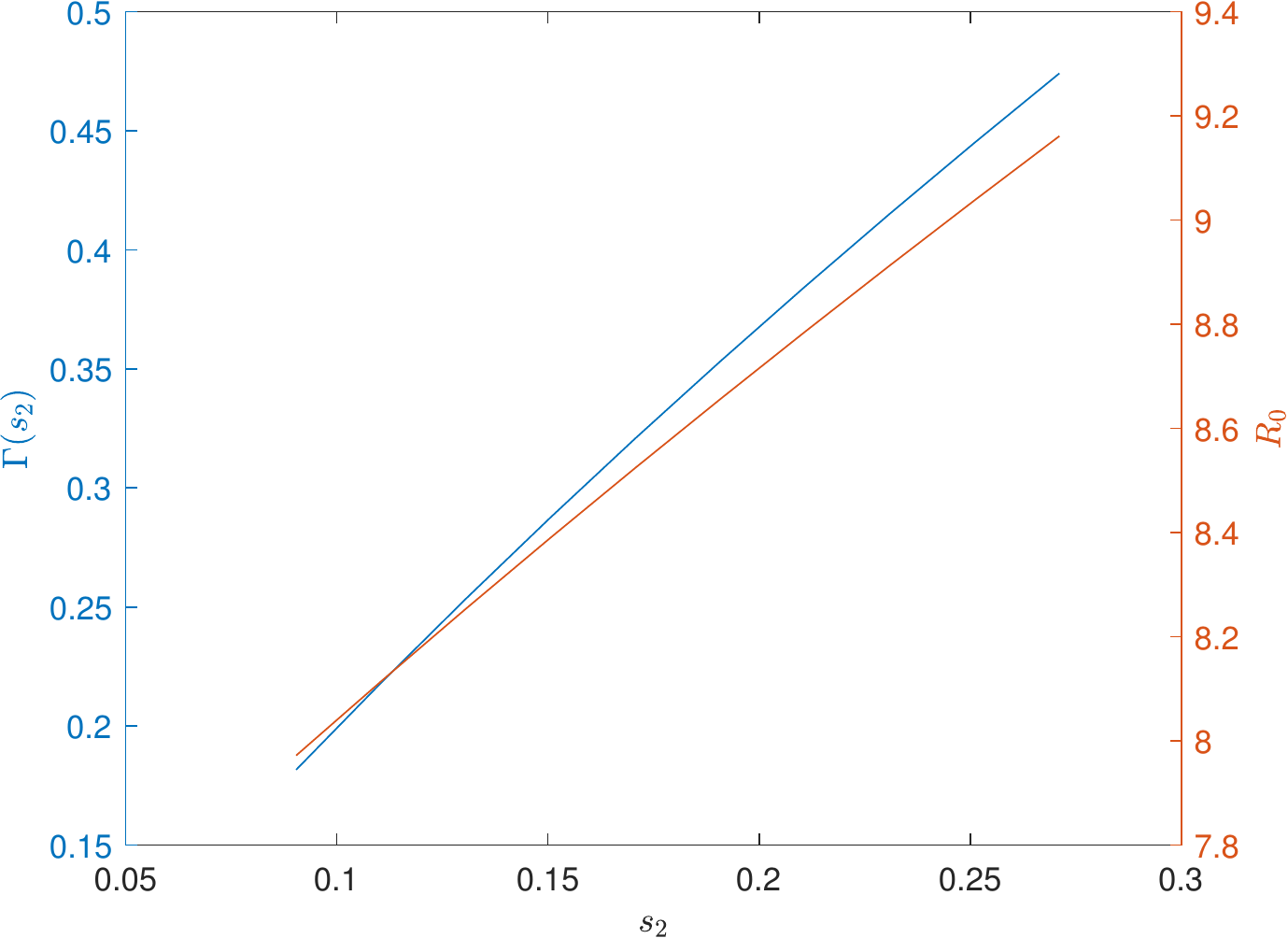}
				\caption{$ S_2 / N $}
				\label{fig:Sensitivity_s_2}
			\end{subfigure}
			\hfill
			\begin{subfigure}[b]{0.23\textwidth}
				\centering
				\includegraphics[width=\textwidth]{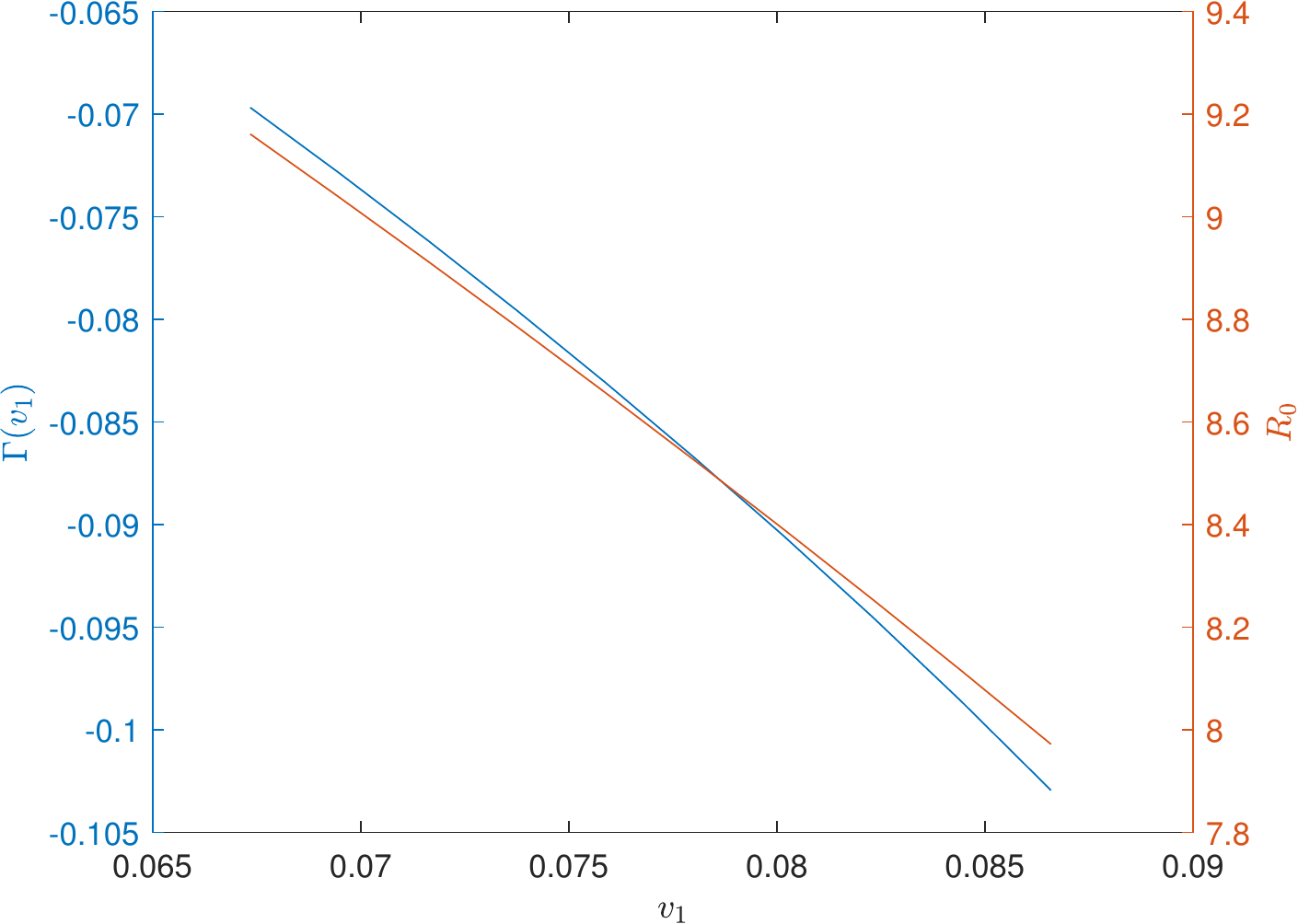}
				\caption{$ V_1 / N $}
				\label{fig:Sensitivity_v_1}
			\end{subfigure}
			\hfill
			\begin{subfigure}[b]{0.23\textwidth}
				\centering
				\includegraphics[width=\textwidth]{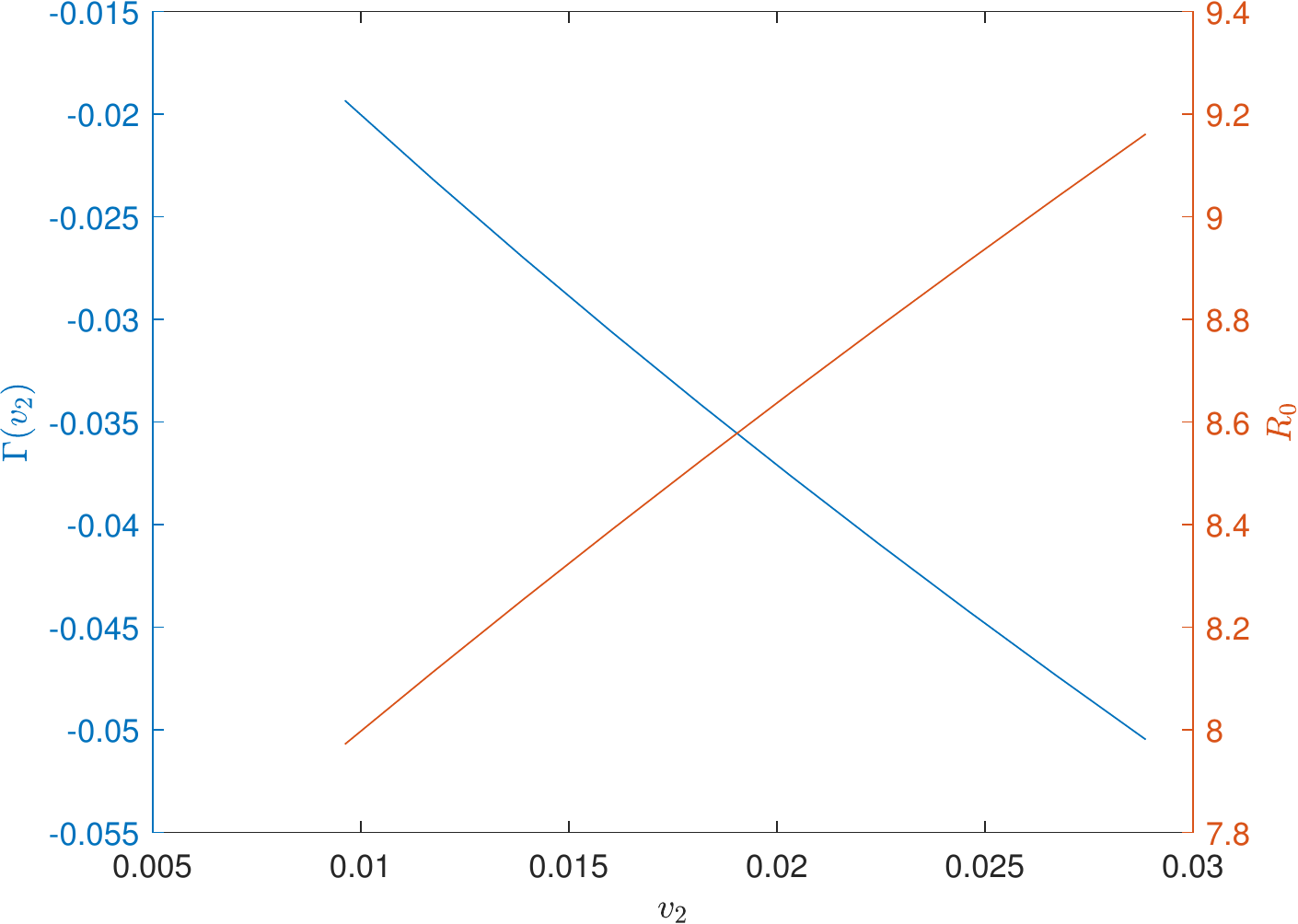}
				\caption{$ V_2 / N $}
				\label{fig:Sensitivity_v_2}
			\end{subfigure}
			\hfill
			\begin{subfigure}[b]{0.23\textwidth}
				\centering
				\includegraphics[width=\textwidth]{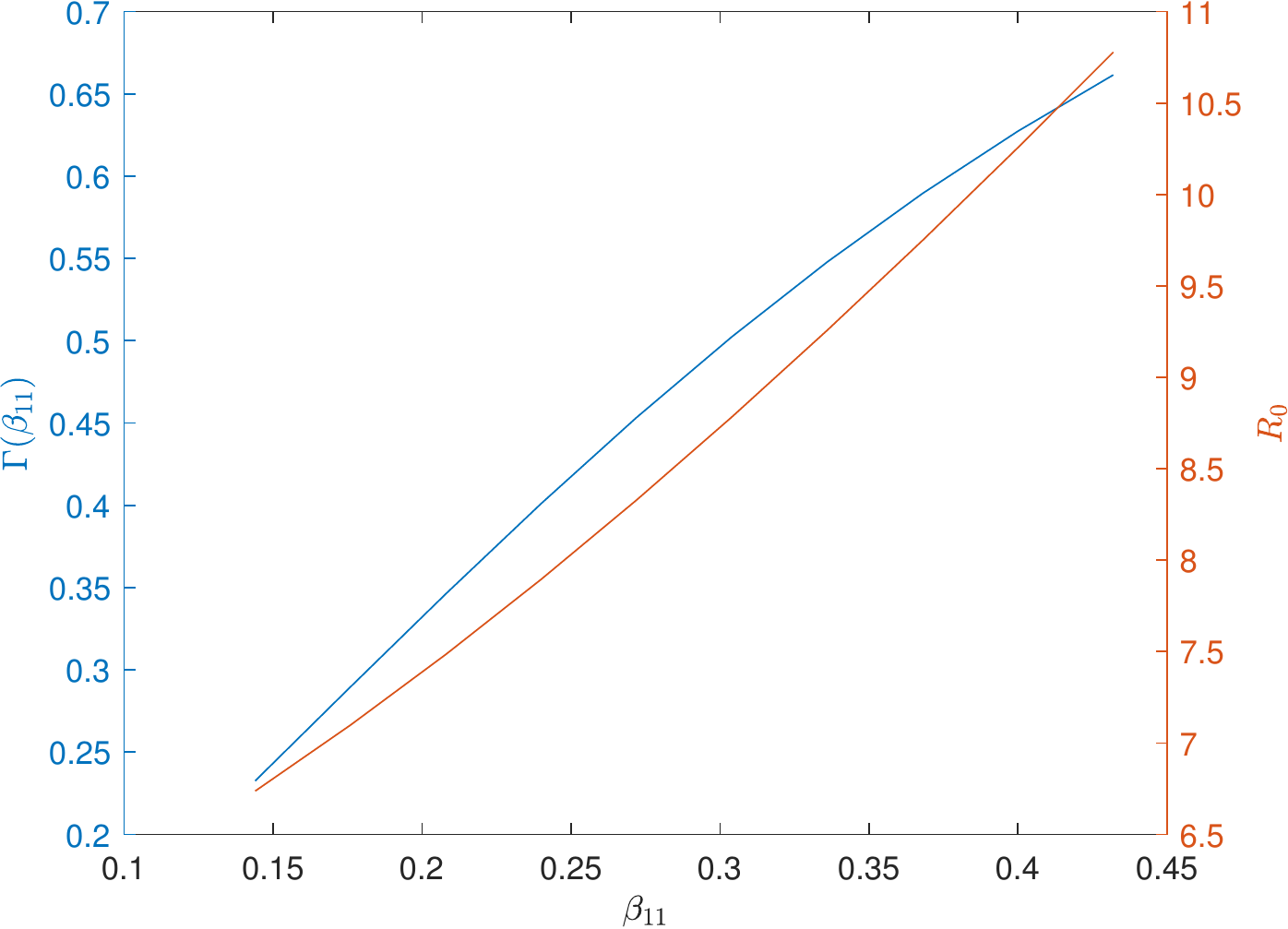}
				\caption{$ \beta_{11} $}
				\label{fig:Sensitivity_beta_{11}}
			\end{subfigure}
			\hfill
			\begin{subfigure}[b]{0.23\textwidth}
				\centering
				\includegraphics[width=\textwidth]{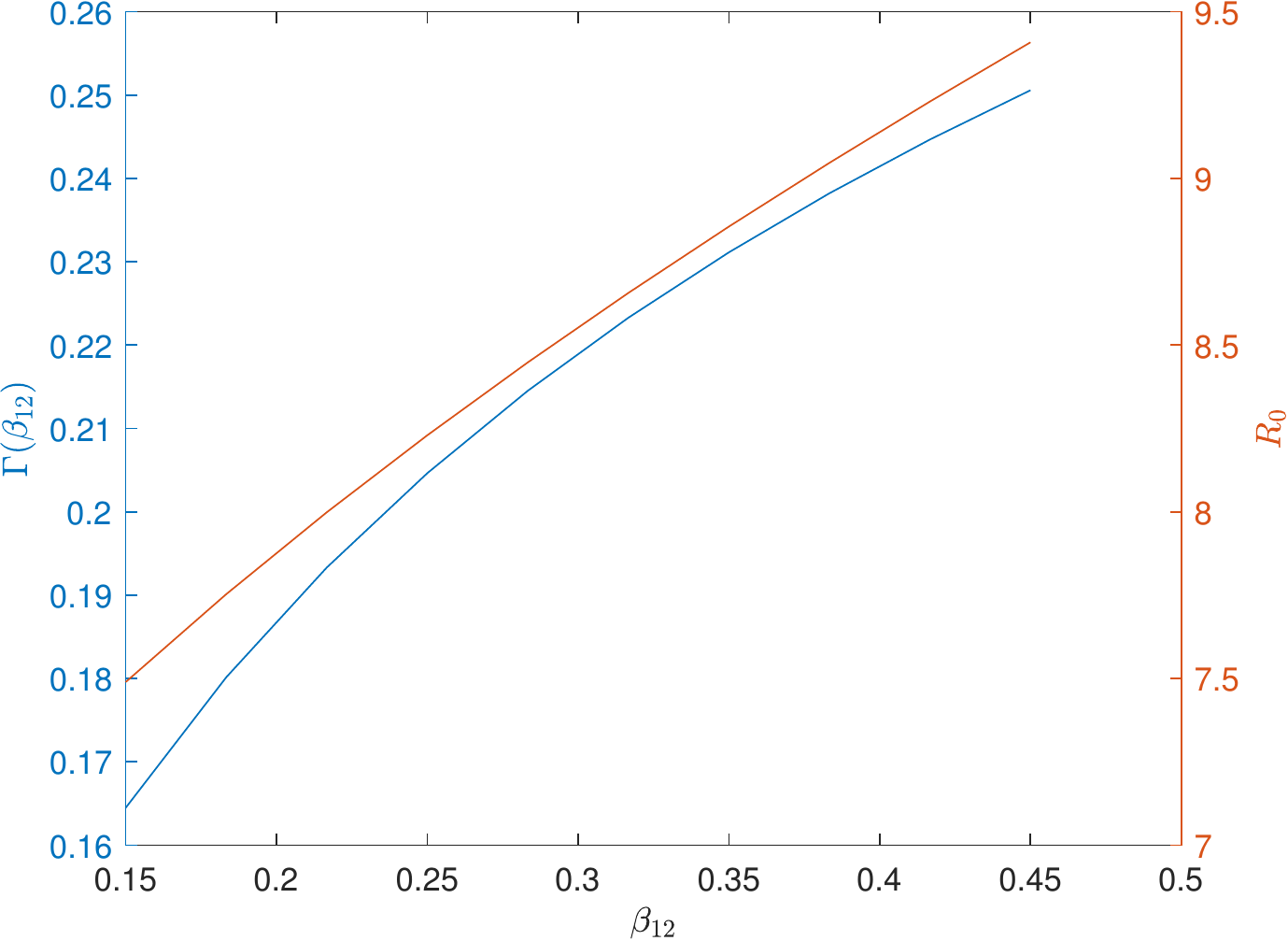}
				\caption{$ \beta_{12} $}
				\label{fig:Sensitivity_beta_{12}}
			\end{subfigure}
			\hfill
			\begin{subfigure}[b]{0.23\textwidth}
				\centering
				\includegraphics[width=\textwidth]{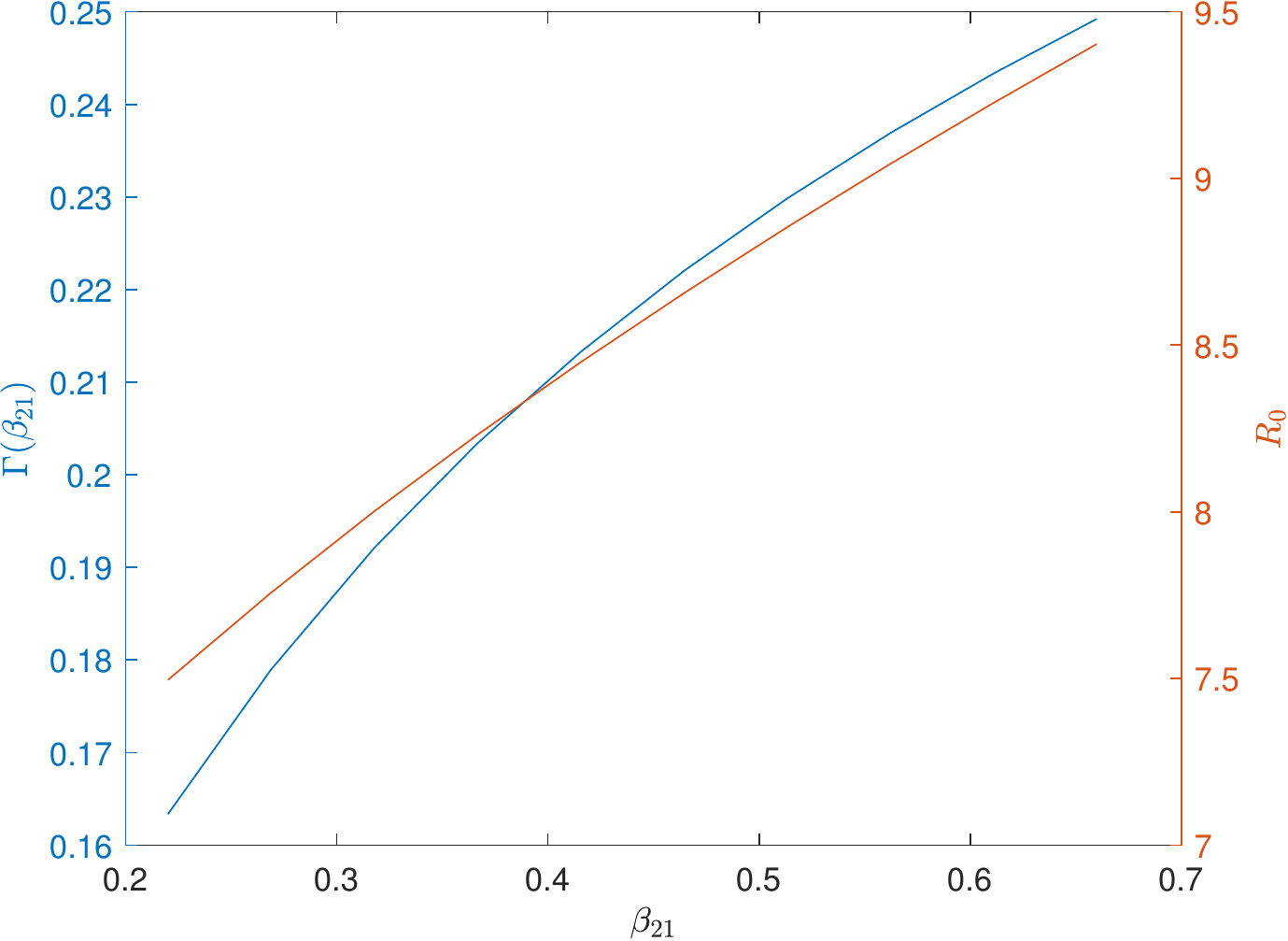}
				\caption{$ \beta_{21} $}
				\label{fig:Sensitivity_beta_{21}}
			\end{subfigure}
			\hfill
			\begin{subfigure}[b]{0.23\textwidth}
				\centering
				\includegraphics[width=\textwidth]{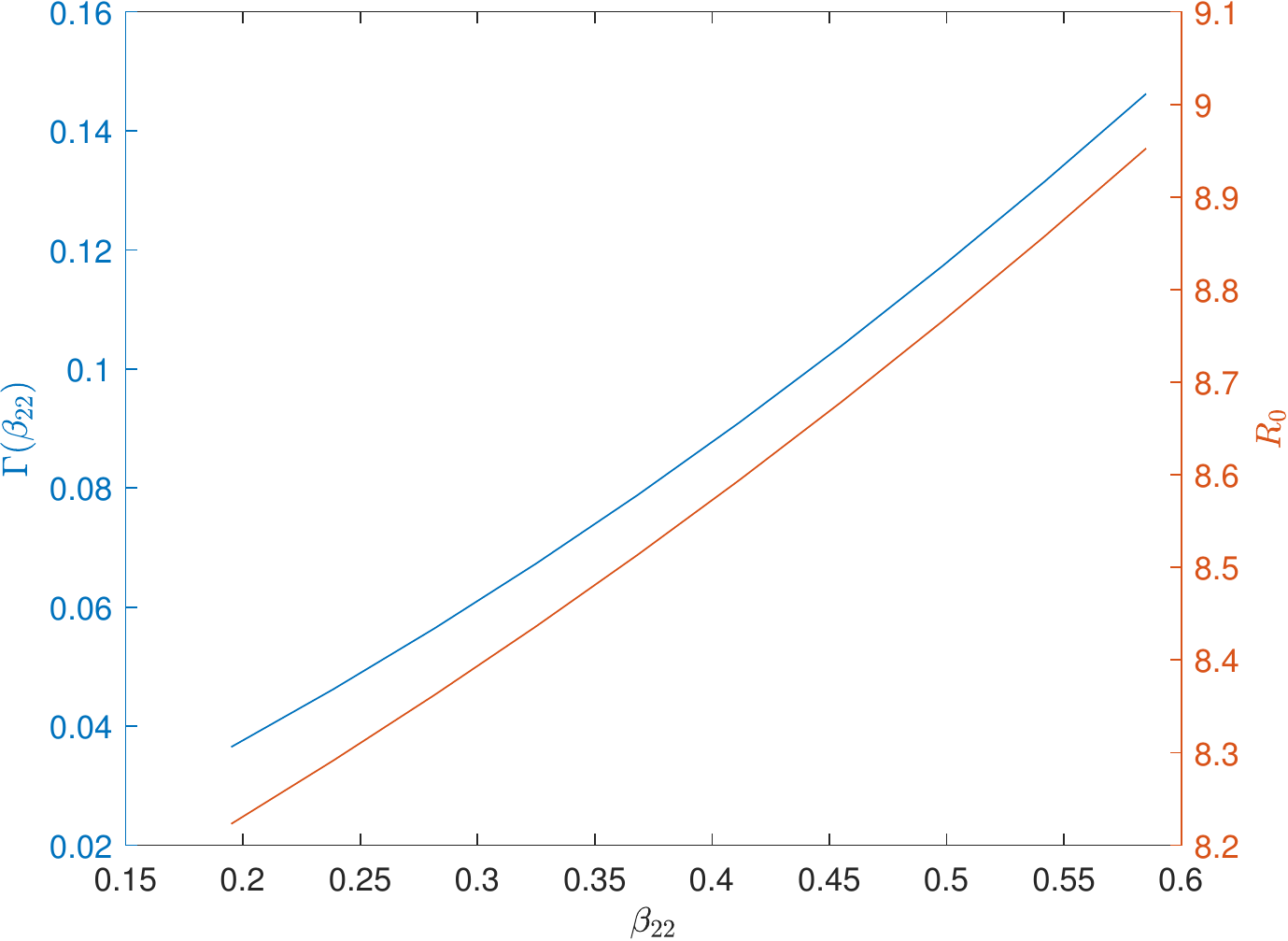}
				\caption{$ \beta_{22} $}
				\label{fig:Sensitivity_beta_{22}}
			\end{subfigure}
			\hfill
			\begin{subfigure}[b]{0.23\textwidth}
				\centering
				\includegraphics[width=\textwidth]{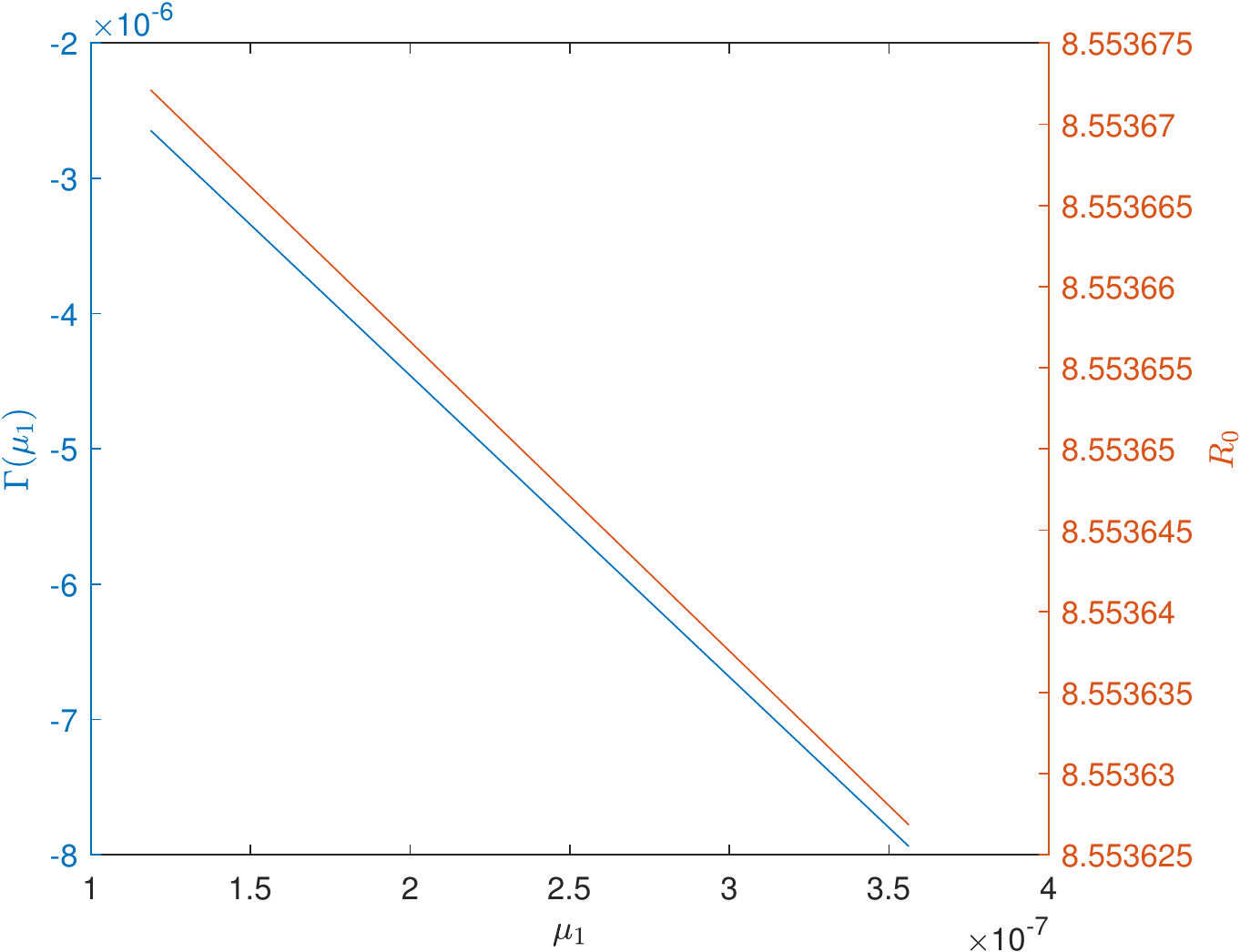}
				\caption{$ \mu_1 $}
				\label{fig:Sensitivity_mu_1}
			\end{subfigure}
			\hfill
			\begin{subfigure}[b]{0.23\textwidth}
				\centering
				\includegraphics[width=\textwidth]{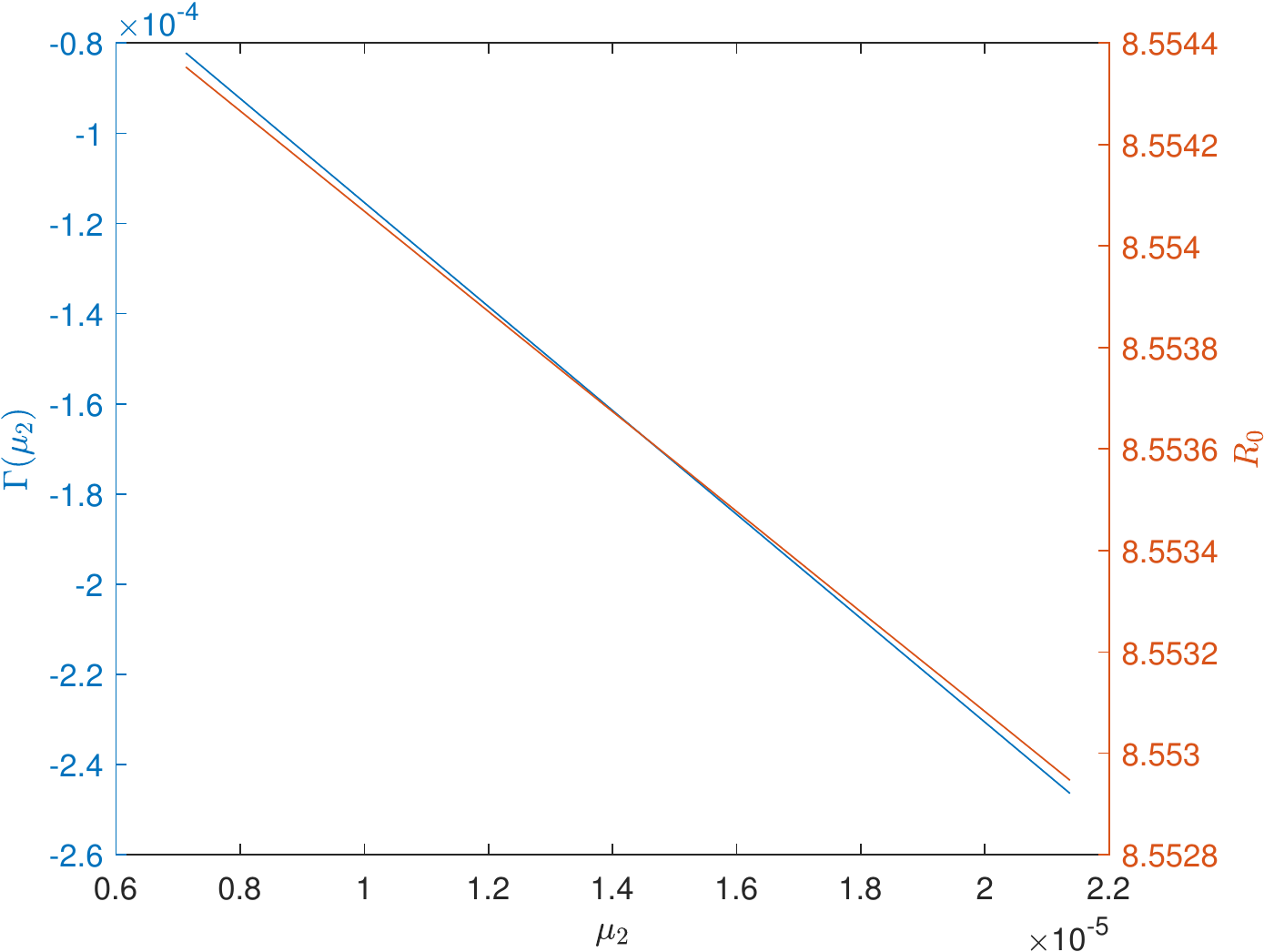}
				\caption{$ \mu_2 $}
				\label{fig:Sensitivity_mu_2}
			\end{subfigure}
			\hfill
			\begin{subfigure}[b]{0.23\textwidth}
				\centering
				\includegraphics[width=\textwidth]{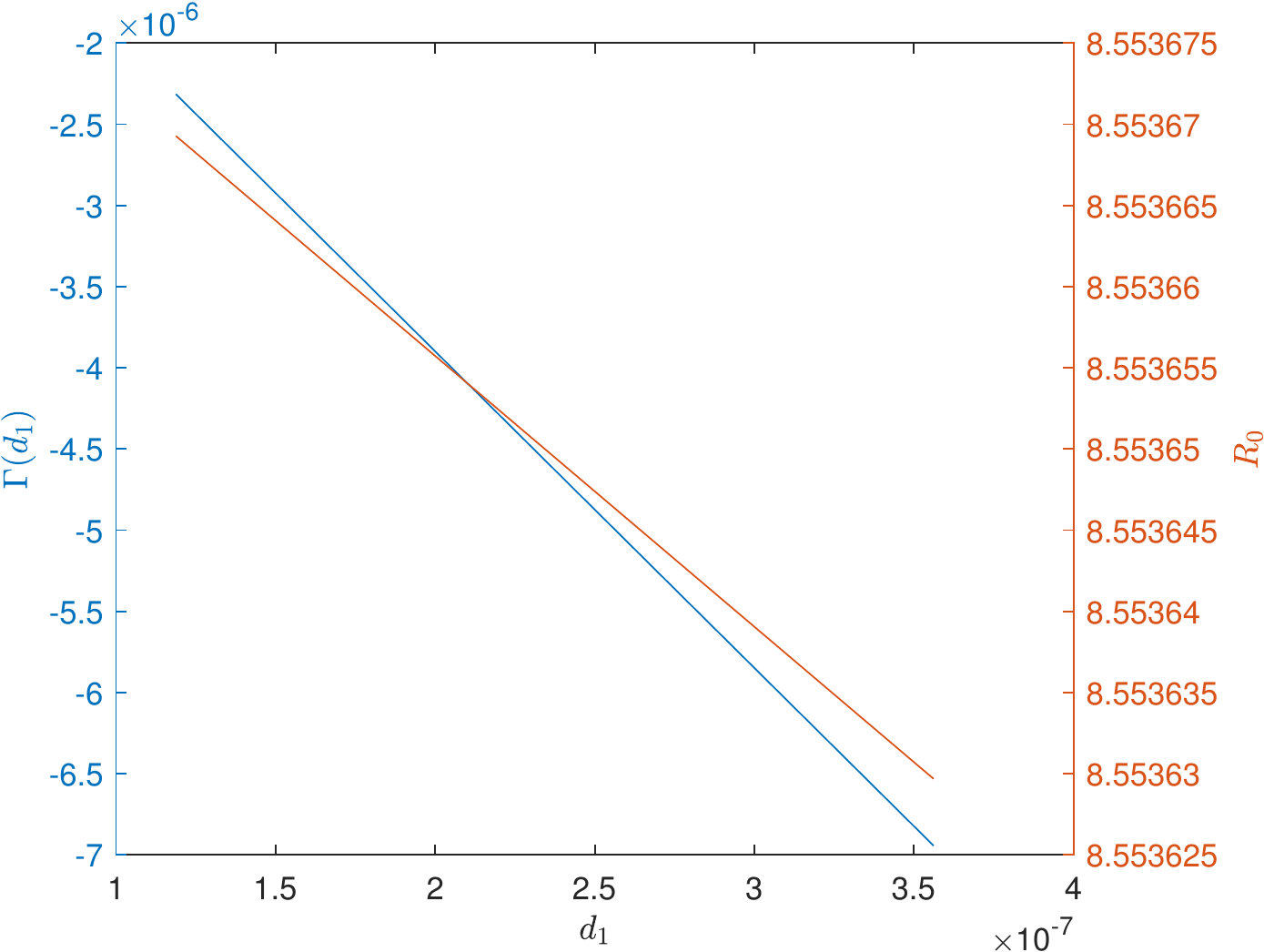}
				\caption{$ d_1 $}
				\label{fig:Sensitivity_d_1}
			\end{subfigure}
			\hfill
			\begin{subfigure}[b]{0.23\textwidth}
				\centering
				\includegraphics[width=\textwidth]{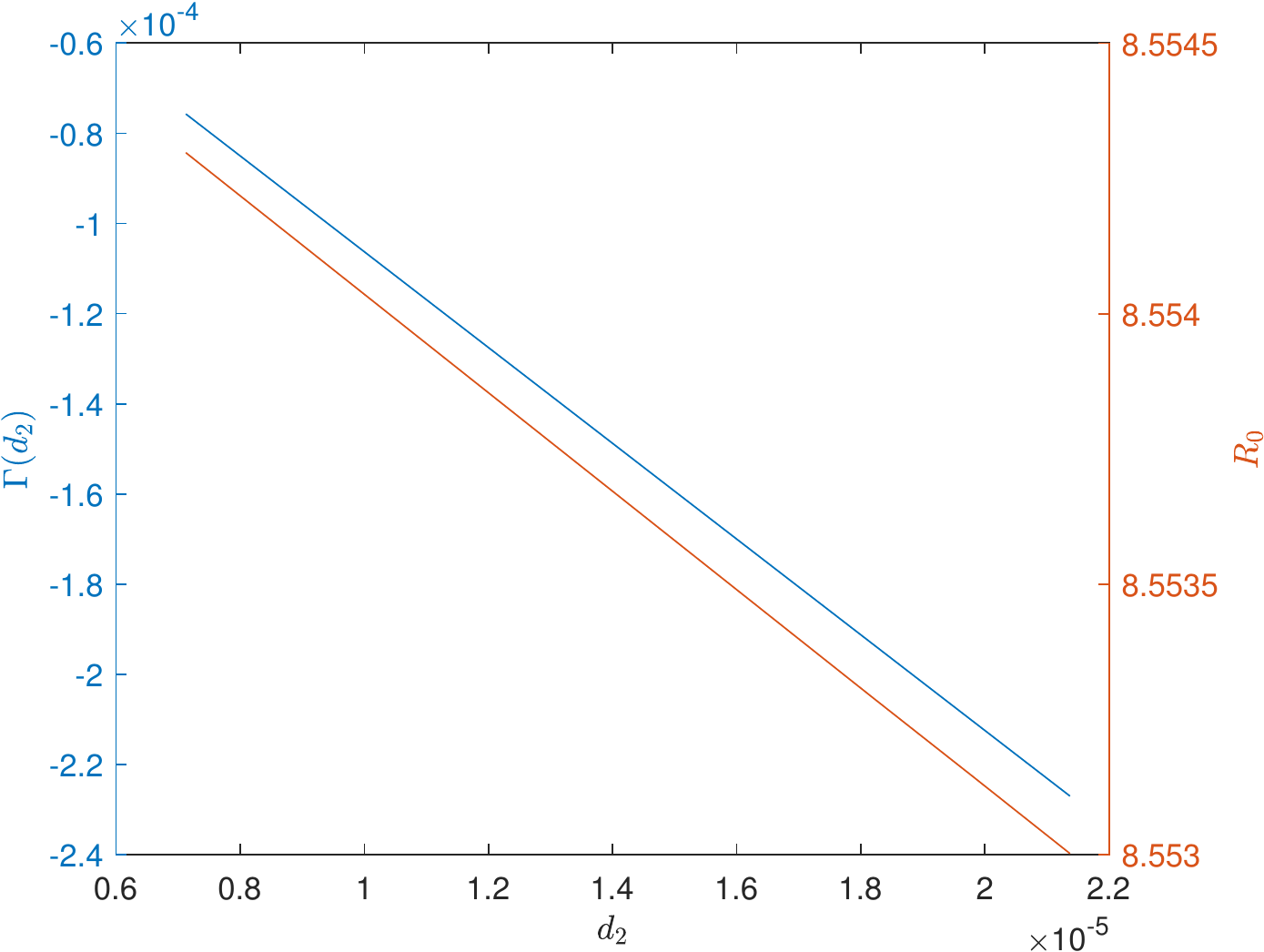}
				\caption{$ d_2 $}
				\label{fig:Sensitivity_d_2}
			\end{subfigure}
			\hfill
			\begin{subfigure}[b]{0.23\textwidth}
				\centering
				\includegraphics[width=\textwidth]{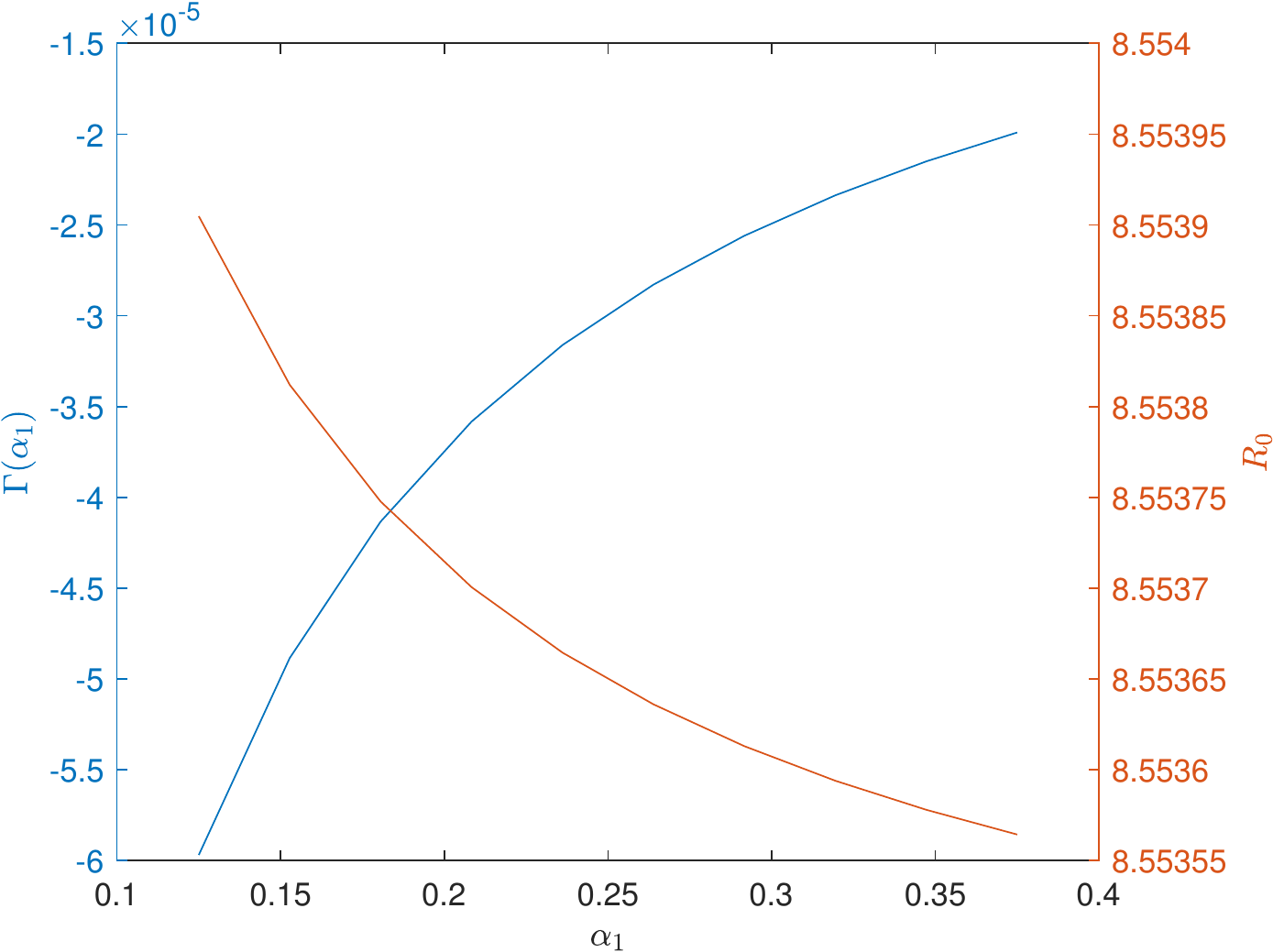}
				\caption{$ \alpha_1 $}
				\label{fig:Sensitivity_alpha_1}
			\end{subfigure}
			\hfill
			\begin{subfigure}[b]{0.23\textwidth}
				\centering
				\includegraphics[width=\textwidth]{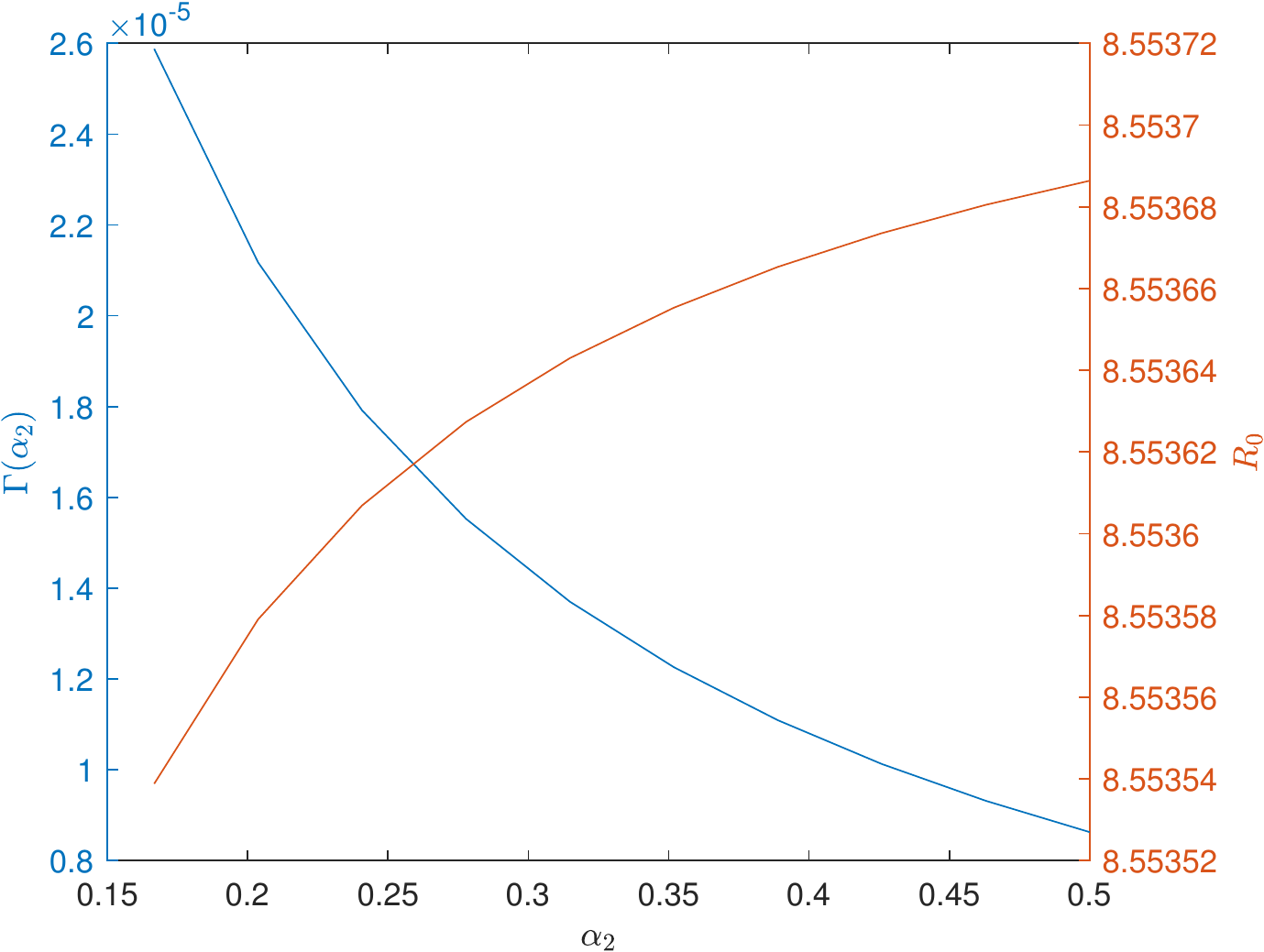}
				\caption{$ \alpha_2 $}
				\label{fig:Sensitivity_alpha_2}
			\end{subfigure}
			\hfill
			\begin{subfigure}[b]{0.23\textwidth}
				\centering
				\includegraphics[width=\textwidth]{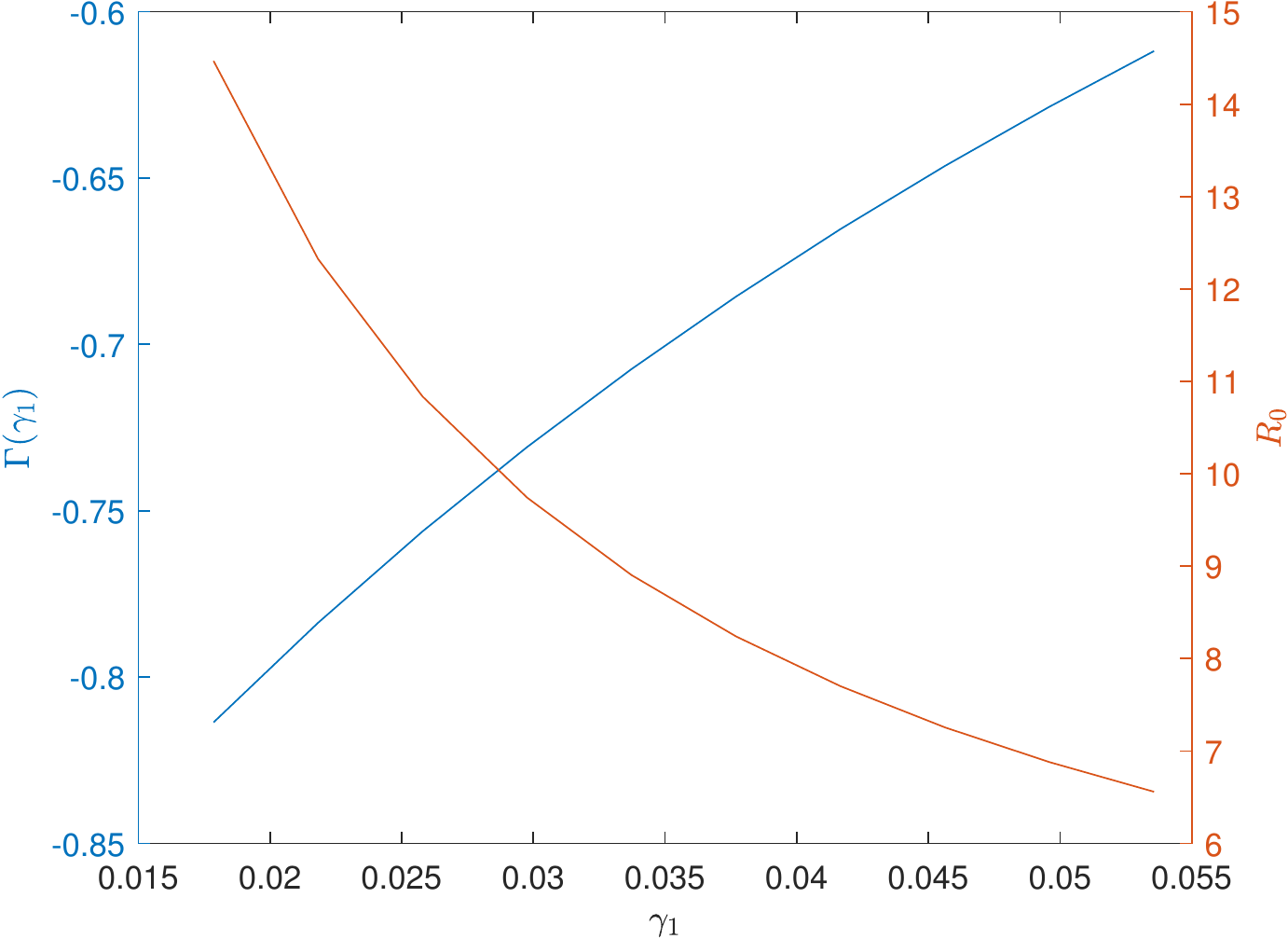}
				\caption{$ \gamma_1 $}
				\label{fig:Sensitivity_gamma_1}
			\end{subfigure}
			\hfill
			\begin{subfigure}[b]{0.23\textwidth}
				\centering
				\includegraphics[width=\textwidth]{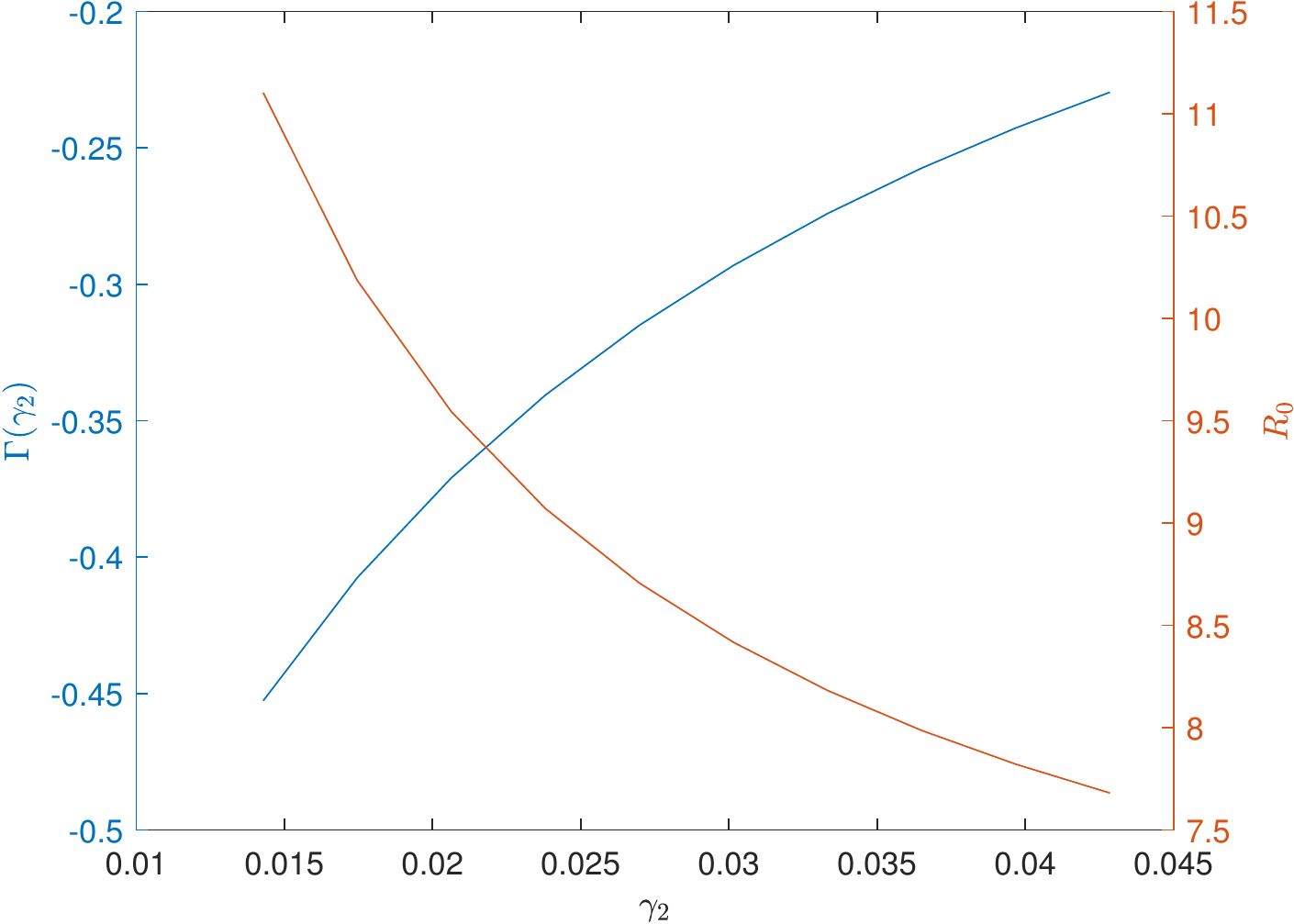}
				\caption{$ \gamma_2 $}
				\label{fig:Sensitivity_gamma_2}
			\end{subfigure}
			\caption{The sensitivity indices values}
			\label{fig:Sensitivity}
		\end{figure}




	\end{appendices}


	\bibliography{Ruiyang_Zhou_Wei}


\begin{thebibliography}{32}
\ifx \bisbn   \undefined \def \bisbn  #1{ISBN #1}\fi
\ifx \binits  \undefined \def \binits#1{#1}\fi
\ifx \bauthor  \undefined \def \bauthor#1{#1}\fi
\ifx \batitle  \undefined \def \batitle#1{#1}\fi
\ifx \bjtitle  \undefined \def \bjtitle#1{#1}\fi
\ifx \bvolume  \undefined \def \bvolume#1{\textbf{#1}}\fi
\ifx \byear  \undefined \def \byear#1{#1}\fi
\ifx \bissue  \undefined \def \bissue#1{#1}\fi
\ifx \bfpage  \undefined \def \bfpage#1{#1}\fi
\ifx \blpage  \undefined \def \blpage #1{#1}\fi
\ifx \burl  \undefined \def \burl#1{\textsf{#1}}\fi
\ifx \doiurl  \undefined \def \doiurl#1{\url{https://doi.org/#1}}\fi
\ifx \betal  \undefined \def \betal{\textit{et al.}}\fi
\ifx \binstitute  \undefined \def \binstitute#1{#1}\fi
\ifx \binstitutionaled  \undefined \def \binstitutionaled#1{#1}\fi
\ifx \bctitle  \undefined \def \bctitle#1{#1}\fi
\ifx \beditor  \undefined \def \beditor#1{#1}\fi
\ifx \bpublisher  \undefined \def \bpublisher#1{#1}\fi
\ifx \bbtitle  \undefined \def \bbtitle#1{#1}\fi
\ifx \bedition  \undefined \def \bedition#1{#1}\fi
\ifx \bseriesno  \undefined \def \bseriesno#1{#1}\fi
\ifx \blocation  \undefined \def \blocation#1{#1}\fi
\ifx \bsertitle  \undefined \def \bsertitle#1{#1}\fi
\ifx \bsnm \undefined \def \bsnm#1{#1}\fi
\ifx \bsuffix \undefined \def \bsuffix#1{#1}\fi
\ifx \bparticle \undefined \def \bparticle#1{#1}\fi
\ifx \barticle \undefined \def \barticle#1{#1}\fi
\bibcommenthead
\ifx \bconfdate \undefined \def \bconfdate #1{#1}\fi
\ifx \botherref \undefined \def \botherref #1{#1}\fi
\ifx \url \undefined \def \url#1{\textsf{#1}}\fi
\ifx \bchapter \undefined \def \bchapter#1{#1}\fi
\ifx \bbook \undefined \def \bbook#1{#1}\fi
\ifx \bcomment \undefined \def \bcomment#1{#1}\fi
\ifx \oauthor \undefined \def \oauthor#1{#1}\fi
\ifx \citeauthoryear \undefined \def \citeauthoryear#1{#1}\fi
\ifx \endbibitem  \undefined \def \endbibitem {}\fi
\ifx \bconflocation  \undefined \def \bconflocation#1{#1}\fi
\ifx \arxivurl  \undefined \def \arxivurl#1{\textsf{#1}}\fi
\csname PreBibitemsHook\endcsname

\bibitem{Characteristics01}
\begin{barticle}
\bauthor{\bsnm{Guan}, \binits{W.-J.}},
\bauthor{\bsnm{Ni}, \binits{Z.-Y.}},
\bauthor{\bsnm{Hu}, \binits{Y.}}, \betal:
\batitle{Clinical characteristics of coronavirus disease 2019 in china}.
\bjtitle{New England Journal of Medicine}
\bvolume{382}(\bissue{18}),
\bfpage{1708}--\blpage{1720}
(\byear{2020})
\end{barticle}
\endbibitem

\bibitem{Age-structured1}
\begin{botherref}
\oauthor{\bsnm{Deforche}, \binits{K.}}:
An age-structured epidemiological model of the belgian covid-19 epidemic
(2020)
\end{botherref}
\endbibitem

\bibitem{Age-structured2}
\begin{botherref}
\oauthor{\bsnm{Mark~Kimathi}, \binits{V.O.} \bsuffix{Samuel M~Musili}},
\oauthor{\bsnm{Gathungu}, \binits{D.K.}}:
Age-structured model for covid-19:effectiveness of social distancing and
  contact reduction
(2020)
\end{botherref}
\endbibitem

\bibitem{Age-structured3}
\begin{barticle}
\bauthor{\bsnm{Fatmawati}},
\bauthor{\bsnm{Dyah~Purwati}, \binits{U.}},
\bauthor{\bsnm{Riyudha}, \binits{F.}}, \betal:
\batitle{Optimal control of a discrete age-structured model for tuberculosis
  transmission}.
\bjtitle{Heliyon}
\bvolume{6}(\bissue{1}),
\bfpage{03030}
(\byear{2020})
\end{barticle}
\endbibitem

\bibitem{Age-structured4}
\begin{barticle}
\bauthor{\bsnm{Zhou}, \binits{L.}},
\bauthor{\bsnm{Wang}, \binits{Y.}},
\bauthor{\bsnm{Xiao}, \binits{Y.}}, \betal:
\batitle{Global dynamics of a discrete age-structured sir epidemic model with
  applications to measles vaccination strategies}.
\bjtitle{Math Biosci}
\bvolume{308},
\bfpage{27}--\blpage{37}
(\byear{2019})
\end{barticle}
\endbibitem

\bibitem{vaccination1}
\begin{botherref}
\oauthor{\bsnm{Cot}, \binits{C.}},
\oauthor{\bsnm{Cacciapaglia}, \binits{G.}},
\oauthor{\bsnm{Islind}, \binits{A.S.}}, et al.:
Impact of us vaccination strategy on covid-19 wave dynamics.
Scientific Reports
\textbf{11}(1)
(2021)
\end{botherref}
\endbibitem

\bibitem{vaccination2}
\begin{barticle}
\bauthor{\bsnm{Makhoul}, \binits{M.}},
\bauthor{\bsnm{Chemaitelly}, \binits{H.}},
\bauthor{\bsnm{Ayoub}, \binits{H.H.}}, \betal:
\batitle{Epidemiological differences in the impact of covid-19 vaccination in
  the united states and china}.
\bjtitle{Vaccines}
\bvolume{9}(\bissue{3}),
\bfpage{223}
(\byear{2021})
\end{barticle}
\endbibitem

\bibitem{Data_AD}
\begin{barticle}
\bauthor{\bsnm{Huang}, \binits{S.-Z.}},
\bauthor{\bsnm{Wei}, \binits{F.}},
\bauthor{\bsnm{Peng}, \binits{Z.}}, \betal:
\batitle{Assessment method of coronavirus disease 2019 outbreaks under normal
  prevention and control}.
\bjtitle{Disease Surveillance}
\bvolume{35}(\bissue{8}),
\bfpage{679}--\blpage{686}
(\byear{2020})
\end{barticle}
\endbibitem

\bibitem{Data_people}
\begin{botherref}
\oauthor{\bparticle{of} \bsnm{Statistics}, \binits{S.M.B.}}:
Shijiazhuang City, the seventh national population census bulletin (No. 1).
\url{https://www.sjz.gov.cn/col/1596018184396/2021/05/31/1622426640444.html}.
[Online; accessed 21-October-2021]
(2021)
\end{botherref}
\endbibitem

\bibitem{Data_age_rate}
\begin{botherref}
\oauthor{\bparticle{of} \bsnm{Statistics}, \binits{S.M.B.}}:
Shijiazhuang City, the seventh national population census bulletin (No. 2).
\url{https://www.sjz.gov.cn/col/1596018184396/2021/05/31/1622426985480.html}.
[Online; accessed 21-October-2021]
(2021)
\end{botherref}
\endbibitem

\bibitem{Data_incubation}
\begin{barticle}
\bauthor{\bsnm{Wu~Yu}, \binits{L.M.}}:
\batitle{The incubation period of covid-19 caused by different sars-cov-2
  variants}.
\bjtitle{Chinese General Practice}
\bvolume{25}(\bissue{11}),
\bfpage{1309}
(\byear{2022})
\end{barticle}
\endbibitem

\bibitem{Data_incubation_1}
\begin{barticle}
\bauthor{\bsnm{Guo}, \binits{J.H.}},
\bauthor{\bsnm{Zhang}, \binits{S.Y.}},
\bauthor{\bsnm{Liu}, \binits{X.S.}}, \betal:
\batitle{Epidemiological characteristics of covid-19 outbreak in gaocheng
  district of shijiazhuan}.
\bjtitle{Zhonghua Liu Xing Bing Xue Za Zhi}
\bvolume{42}(\bissue{10}),
\bfpage{1769}--\blpage{1773}
(\byear{2021})
\end{barticle}
\endbibitem

\bibitem{Data_incubation_2}
\begin{botherref}
\oauthor{\bsnm{Zhu}, \binits{W.}},
\oauthor{\bsnm{Zhang}, \binits{M.}},
\oauthor{\bsnm{Pan}, \binits{J.}}, et al.:
Effects of prolonged incubation period and centralized quarantine on the
  covid-19 outbreak in shijiazhuang, china: a modeling study.
BMC Medicine
\textbf{19}(1)
(2021)
\end{botherref}
\endbibitem

\bibitem{Data_gamma_0}
\begin{barticle}
\bauthor{\bsnm{Wu}, \binits{S.}},
\bauthor{\bsnm{Xue}, \binits{L.}},
\bauthor{\bsnm{Legido-Quigley}, \binits{H.}}, \betal:
\batitle{Understanding factors influencing the length of hospital stay among
  non-severe covid-19 patients: A retrospective cohort study in a fangcang
  shelter hospital}.
\bjtitle{PLOS ONE}
\bvolume{15}(\bissue{10}),
\bfpage{0240959}
(\byear{2020})
\end{barticle}
\endbibitem

\bibitem{Data_gamma_1}
\begin{botherref}
\oauthor{\bsnm{Mizrahi}, \binits{B.}},
\oauthor{\bsnm{Shilo}, \binits{S.}},
\oauthor{\bsnm{Rossman}, \binits{H.}}, et al.:
Longitudinal symptom dynamics of covid-19 infection.
Nature Communications
\textbf{11}(1)
(2020)
\end{botherref}
\endbibitem

\bibitem{Data_clinic}
\begin{barticle}
\bauthor{\bsnm{Guan}, \binits{W.-J.}},
\bauthor{\bsnm{Ni}, \binits{Z.-Y.}},
\bauthor{\bsnm{Hu}, \binits{Y.}}, \betal:
\batitle{Clinical characteristics of coronavirus disease 2019 in china}.
\bjtitle{New England Journal of Medicine}
\bvolume{382}(\bissue{18}),
\bfpage{1708}--\blpage{1720}
(\byear{2020})
\end{barticle}
\endbibitem

\bibitem{Data_death}
\begin{barticle}
\bauthor{\bsnm{Fanelli}, \binits{D.}},
\bauthor{\bsnm{Piazza}, \binits{F.}}:
\batitle{Analysis and forecast of covid-19 spreading in china, italy and
  france}.
\bjtitle{Chaos Solitons Fractals}
\bvolume{134},
\bfpage{109761}
(\byear{2020})
\end{barticle}
\endbibitem

\bibitem{Data_death1}
\begin{barticle}
\bauthor{\bsnm{Ioannidis}, \binits{J.P.A.}}:
\batitle{Infection fatality rate of covid-19 inferred from seroprevalence
  data}.
\bjtitle{Bulletin of the World Health Organization}
\bvolume{99}(\bissue{1}),
\bfpage{19}--\blpage{33}
(\byear{2021})
\end{barticle}
\endbibitem

\bibitem{Data_mu}
\begin{botherref}
\oauthor{\bparticle{of} \bsnm{Statistics}, \binits{S.M.B.}}:
2020 Shijiazhuang Statistical Yearbook.
\url{http://tjj.sjz.gov.cn/col/1584345186126/2021/07/21/1626855580453.html}.
[Online; accessed 21-July-2021]
(2021)
\end{botherref}
\endbibitem

\bibitem{reproduction1}
\begin{botherref}
\oauthor{\bsnm{Driessche}, \binits{P.v.d.}},
\oauthor{\bsnm{Watmough}, \binits{J.}}:
Reproduction numbers and sub-threshold endemic equilibria for compartmental
  models of disease transmission.
Mathematical Biosciences
(2002)
\end{botherref}
\endbibitem

\bibitem{Descartes01}
\begin{botherref}
\oauthor{\bsnm{Curtiss}, \binits{D.R.}}:
Recent extensions of descartes' rule of signs.
Annals of Mathematics
\textbf{19}
(1918)
\end{botherref}
\endbibitem

\bibitem{Descartes02}
\begin{bbook}
\bauthor{\bsnm{Meserve}, \binits{B.E.}}:
\bbtitle{Fundamental Concepts of Algebra}.
\bpublisher{Dover Publications},
\blocation{New {Y}ork}
(\byear{1982})
\end{bbook}
\endbibitem

\bibitem{Descartes03}
\begin{bbook}
\bauthor{\bsnm{Latham}, \binits{D.E.S.}},
\bauthor{\bsnm{M.L.}}:
\bbtitle{The Geometry of René Descartes with a Facsimile of the First
  Edition}.
\bpublisher{Dover Publications},
\blocation{New {Y}ork}
(\byear{1954})
\end{bbook}
\endbibitem

\bibitem{sensitivity2}
\begin{barticle}
\bauthor{\bsnm{Chitnis}, \binits{N.}},
\bauthor{\bsnm{Hyman}, \binits{J.M.}},
\bauthor{\bsnm{Cushing}, \binits{J.M.}}:
\batitle{Determining important parameters in the spread of malaria through the
  sensitivity analysis of a mathematical model}.
\bjtitle{Bulletin of Mathematical Biology}
\bvolume{70}(\bissue{5}),
\bfpage{1272}--\blpage{1296}
(\byear{2008})
\end{barticle}
\endbibitem

\bibitem{sensitivity3}
\begin{barticle}
\bauthor{\bsnm{Augeraud}, \binits{E.}},
\bauthor{\bsnm{Yang}, \binits{J.}},
\bauthor{\bsnm{Wang}, \binits{G.}}, \betal:
\batitle{Analysis of the age-structured epidemiological characteristics of
  sars-cov-2 transmission in mainland china: An aggregated approach}.
\bjtitle{Mathematical Modelling of Natural Phenomena}
\bvolume{15},
\bfpage{39}
(\byear{2020})
\end{barticle}
\endbibitem

\bibitem{sjz01}
\begin{botherref}
\oauthor{\bsnm{Cheng}, \binits{X.M.}},
\oauthor{\bsnm{Li}, \binits{Y.F.}},
\oauthor{\bsnm{Zhang}, \binits{Y.L.}}, et al.:
The characteristics of and responses to the two covid-19 outbreak waves in
  hebei province of china, january 2020 to february 2021.
Epidemiology and Infection,
1--22
(2021)
\end{botherref}
\endbibitem

\bibitem{sjz02}
\begin{barticle}
\bauthor{\bsnm{Wang}, \binits{B.}},
\bauthor{\bsnm{Zheng}, \binits{H.}}:
\batitle{From blanket quarantine in wuhan to distant centralized quarantine in
  shijiazhuang: the evolution of china's covid-19 quarantine approach}.
\bjtitle{Infection}
\bvolume{49}(\bissue{4}),
\bfpage{765}--\blpage{767}
(\byear{2021})
\end{barticle}
\endbibitem

\bibitem{Data_stage2_0}
\begin{botherref}
\oauthor{\bsnm{News}, \binits{S.}}:
The city held the second new crown pneumonia outbreak prevention and control
  conference.
\url{https://www.sjz.gov.cn/col/1596014213243/2021/01/07/1610070300288.html}.
[Online; accessed 7-Jaunary-2021]
(2021)
\end{botherref}
\endbibitem

\bibitem{Data_stage2_1}
\begin{botherref}
\oauthor{\bsnm{Network}, \binits{H.N.}}:
Hebei is in full state of emergency.
\url{http://wsjkw.hebei.gov.cn/html/zwyw/20210107/375266.html}.
[Online; accessed 7-Jaunary-2021]
(2021)
\end{botherref}
\endbibitem

\bibitem{Data_stage3}
\begin{botherref}
\oauthor{\bsnm{News}, \binits{S.}}:
The city held the ninth new crown pneumonia outbreak prevention and control
  conference.
\url{https://www.sjz.gov.cn/col/1596014213243/2021/01/14/1610589345799.html}.
[Online; accessed 14-Jaunary-2021]
(2021)
\end{botherref}
\endbibitem

\bibitem{Data_stage4}
\begin{botherref}
\oauthor{\bsnm{Office}, \binits{H.P.P.G.I.}}:
Hebei held a press conference on the prevention and control of the new crown
  pneumonia outbreak (ninth).
\url{http://www.scio.gov.cn/xwfbh/gssxwfbh/xwfbh/hebei/Document/1698137/1698137.htm}.
[Online; accessed 28-Jaunary-2021]
(2021)
\end{botherref}
\endbibitem

\bibitem{Data_incubation_3}
\begin{barticle}
\bauthor{\bsnm{Sun}, \binits{J.W.}},
\bauthor{\bsnm{Cui}, \binits{Z.F.}},
\bauthor{\bsnm{Li}, \binits{F.J.}}, \betal:
\batitle{Epidemiological characteristics of novel coronavirus pneumonia in
  shijiazhuang, china}.
\bjtitle{Zhonghua Jie He He Hu Xi Za Zhi}
\bvolume{44}(\bissue{11}),
\bfpage{961}--\blpage{965}
(\byear{2021})
\end{barticle}
\endbibitem

\end{thebibliography}



\end{document}